\PassOptionsToPackage{svgnames}{xcolor}

\documentclass[sigconf,10pt,nonacm]{acmart}

\AtBeginDocument{%
  }

\newcommand{\para}[1]{\smallskip\noindent\textbf{#1}~}

\usepackage[frozencache,cachedir=.]{minted}

\usepackage{cuted}
\usepackage{multicol}
\usepackage{lipsum}
\usepackage[many]{tcolorbox}
\usepackage{subcaption}
\usepackage{graphicx}
\usepackage{enumitem}
\usepackage{booktabs}

\usepackage[T1]{fontenc}
\usepackage[utf8]{inputenc}
\usepackage{fancyvrb}
\usepackage{upquote}

\makeatletter
\renewcommand{\@titlefont}{\LARGE\sffamily\bfseries}
\makeatother

\begin{document}

\title[]{Congestion Control System Optimization with Large Language Models}

\author{\large Zhiyuan He$^1$, Aashish Gottipati$^{2}$, Lili Qiu$^{12}$, Yuqing Yang$^1$, Francis Y. Yan$^{3}$}

\affiliation{
  \vspace{5pt}
  \institution{$^1$Microsoft Research, $^2$UT Austin, $^3$UIUC}
  \country{}
  \vspace{15pt}
}

\thanks{$^*$Lili Qiu and Francis Y. Yan are the corresponding authors.}

\makeatletter
\renewcommand{\@titlefont}{\LARGE\sffamily\bfseries}
\makeatother
\renewcommand{\authors}{Zhiyuan He, Aashish Gottipati, Lili Qiu, Yuqing Yang, and Francis Y. Yan}
\renewcommand{\shortauthors}{Z. He, A. Gottipati, L. Qiu, Y. Yang, F. Y. Yan}

\begin{abstract}
Congestion control is a fundamental component of Internet infrastructure, and researchers have dedicated considerable effort to developing improved congestion control algorithms. However, despite extensive study, existing algorithms continue to exhibit suboptimal performance across diverse network environments. In this paper, we introduce a novel approach that automatically optimizes congestion control algorithms using large language models (LLMs). Our framework consists of a structured algorithm generation process, an emulation-based evaluation pipeline covering a broad range of network conditions, and a statistically guided method to substantially reduce evaluation time. Empirical results from four distinct LLMs validate the effectiveness of our approach. We successfully identify algorithms that achieve up to 27\% performance improvements over the original BBR algorithm in a production QUIC implementation. Our work demonstrates the potential of LLMs to accelerate the design of high-performance network algorithms and paves the way for broader applications in networking systems.
\end{abstract}

\maketitle

\section{Introduction}

Congestion control \cite{brakmo1994tcp, ha2008cubic, cardwell2016bbr, arun2018copa, dong2018pcc} algorithms play a crucial role in maintaining the stability and efficiency of modern networks, yet optimizing them remains an exceptionally challenging task. These algorithms must operate within highly dynamic network environments, where factors such as fluctuating bandwidth, varying latency, and unpredictable traffic patterns create complex interactions that are difficult to model and manage.

Recently, the rapid advancements in large language models (LLMs) have demonstrated their ability to generate code across various domains \cite{hurst2024gpt, jaech2024openai, grattafiori2024llama, yang2024qwen2}. Building on this progress, we explore the use of LLMs for designing networking algorithms via code generation, with a particular focus on congestion control. For the first time, we demonstrate that state-of-the-art LLMs can propose substantial algorithmic modifications that lead to significant improvement of end-to-end network performance.

Leveraging LLMs for network system optimization presents several key challenges. First, congestion control algorithms are inherently complex and difficult to optimize. For instance, the implementation of the BBR algorithm we enhance in this study consists of over 30 functions and more than 1,000 lines of code. Fully rewriting such a system with LLMs is impractical---not only due to its complexity but also because of token limits when using API-based LLMs. Previous LLM-driven system optimizations typically focus on modifying a single function rather than handling intricate systems like congestion control \cite{ma2023eureka,romera2024mathematical,chen2023evoprompting}. Second, network environments are highly dynamic, requiring precise evaluation of algorithmic modifications.  Direct deployment of LLM-generated algorithms on real networks is infeasible due to potential inefficiencies and correctness concerns. Third, exhaustive network emulation to validate the proposed algorithms is computationally expensive, limiting the feasibility and scalability of evaluating numerous LLM-generated candidates.

To address these challenges, we propose a framework that integrates LLM-driven algorithm design and efficient evaluation techniques based on statistical methods.

First, to address the complexity of generating congestion control algorithms from scratch, we use LLMs to generate modifications to an existing congestion control algorithm, with BBR~\cite{cardwell2016bbr} as a case study throughout the paper. We provide LLMs with the complete source code and prompt each LLM to output only the functions and variables that require updates. We also incorporate chain-of-thought \cite{wei2022chain} prompting techniques to enrich the diversity of generated algorithms.

To ensure real-world applicability, we validate generated algorithms in trace-driven emulation environments that capture realistic network conditions. Using real-world trace datasets from broadband, Starlink, 4G, and 5G networks, we emulate diverse network scenarios to verify that the improvements are robust across various conditions.

Furthermore, we observe that the generation-evaluation pipeline based on LLMs introduces a distinct challenge: generating new algorithms is typically fast, but evaluating them requires significantly longer time, thereby constraining the scalability of the entire framework. In our work, generating an algorithm takes less than one minute, whereas conducting emulation experiments under diverse network conditions requires approximately seven hours. To overcome this bottleneck, we propose a statistically guided, efficient evaluation method that substantially reduces the number of required network emulation runs while maintaining reliable performance assessments. Our method builds on two key insights: (1) congestion control algorithms exhibit higher performance \textit{variance} under certain network conditions, and (2) algorithm performance across some network conditions exhibits strong \textit{correlations}. We model these relationships using a multivariate Gaussian distribution, allowing us to accurately estimate the true performance with a small subset of carefully selected network conditions.

Specifically, we apply our methodology to enhance the BBR algorithm implemented in MsQuic \cite{msquic}, a production QUIC implementation shown to have strong performance across a variety of network settings \cite{mishra2022understanding,bunstorf2023msquic}. Unlike TCP, whose congestion control operates in the kernel space, QUIC's congestion control runs in the user space, making it easier to modify and experiment with new algorithms.
As noted in \cite{mishra2024keeping}, BBR is widely deployed on QUIC servers across the Internet.

We conduct experiments with four different LLMs, including both closed-source and open-source models. Our results demonstrate that LLM-generated modifications to BBR may significantly outperform the default BBR implementation, showcasing the potential of LLMs to enhance congestion control mechanisms. Our approach achieves up to a 27.28\% increase in throughput with negligible impact on latency. Additionally, our efficient evaluation method reduces emulation time by up to 87.6\%, making large-scale testing substantially more feasible.

It is important to note that, in this study, we do not aim to identify the optimal, one-size-fits-all congestion control algorithm. Instead, our objective is to demonstrate that LLMs can automatically enhance existing congestion control methods. Our approach is naturally extensible to other network algorithms, such as routing, adaptive bitrate streaming, and wireless resource scheduling.

By integrating LLM-driven innovation with rigorous validation via network emulation and efficient evaluation through statistical sampling, we present a scalable and effective methodology for automatically designing and optimizing network algorithms.

\section{Related Work}
\label{sec:related}

\subsection{Congestion Control}

Congestion control is a fundamental challenge in networking where network resources and packet flows must be dynamically adapted to avoid degrading network efficiency. Many congestion control algorithms have been proposed, each with their own unique optimization objectives and characteristics. Early works such as TCP Vegas~\cite{brakmo1994tcp} adjust send rates based on observed packet-round trip times to avoid packet loss. Cubic~\cite{ha2008cubic} introduces a cubic growth function for the congestion window, allowing for more aggressive probing of available bandwidth while maintaining stability in high-bandwidth, high-latency networks. BBR~\cite{cardwell2016bbr} introduces a model-based approach, estimating the network's bottleneck bandwidth and round-trip propagation time to achieve high throughput and low latency simultaneously. More recent works include Copa \cite{arun2018copa} and Vivace \cite{dong2018pcc}. Our methodology is fundamentally different from these approaches, as we achieve automate algorithmic improvements by leveraging LLMs, reducing the reliance on manual tuning by human researchers.

\subsection{LLMs for Network}
Recently, LLMs have shown promising capabilities in various networking tasks. Several works leverage LLMs to generating network configuration, either by automatically synthesizing verified router configurations~\cite{mondal2023llms} or translating high-level requirements into low-level device settings~\cite{wang2023making}. Others use LLMs to automatically extract protocol specifications from complex RFC documents, significantly reducing manual effort~\cite{sharma2023prosper}. Beyond parsing text-based configurations and protocols, LLMs are also utilized for optimizing network performance. NADA~\cite{he2024designing} and NetLLM~\cite{wu2024netllm} apply LLMs to enhance learning-based optimization algorithms across tasks such as adaptive bitrate streaming and job scheduling. \cite{shrestha2024adapting} explores combining LLMs with congestion control in WiFi environments. However, their method primarily focuses on selecting the most suitable existing congestion control algorithm using a trained LLM, rather than directly improving the congestion control algorithm itself. Moreover, deploying their method in real-world environments would require running computationally expensive LLM inference in the real-time network environment. In contrast, our methodology leverage LLMs to automatically enhance a widely-used congestion control algorithm, thus enabling direct deployment of the improved algorithm for superior network performance, providing a more essential and practical innovation.

\subsection{LLM-Based Optimization by Code Generation}

LLMs demonstrate strong code generation capabilities and are successfully applied in various optimization tasks. For NP-hard problems, FunSearch~\cite{romera2024mathematical} utilizes LLMs to generate priority scoring functions for the cap set problem, surpassing previous best-known solutions. Similarly, Evolution of Heuristic (EoH) \cite{liu2024example} uses LLM to first generate thoughts then translate them into functions, improving online bin packing heuristics. In reinforcement learning, Eureka \cite{ma2023eureka} employs LLMs to automatically create reward functions for complex robotic tasks like dexterous pen spinning, outperforming expert-designed rewards. EvoPrompting \cite{chen2023evoprompting} applies LLM-generated code mutations for neural architecture search, discovering novel convolutional and graph neural network structures that beat established architectures.  Recently, \cite{gpuKernerl} leverages LLMs to generate optimized GPU kernels, surpassing manual optimization. However, existing methods typically focus on optimizing a single function. In contrast, our work addresses a more complex scenario: optimizing 31 functions in MsQuic's BBR implementation, to achieve fundamental system optimization. We also propose an efficient evaluation method to address the unique challenge of reducing evaluating time of the generated algorithms.

\section{Our Basic Approach}
\label{sec:approach}

We present our approach to algorithm generation using LLMs in Section \ref{sec:algo-gen}, as well as our methodology for evaluating these generated algorithms in Section \ref{sec:algo-eval}. Furthermore, beyond basic generation and evaluation, we detail our strategies for efficient evaluation in Section~\ref{sec:towards-efficient-evaluation}.

\subsection{Algorithm Generation}
\label{sec:algo-gen}

Simply prompting LLMs to generate entirely new congestion control algorithms often results in vague and impractical suggestions. It is also difficult and inefficient to implement and evaluate those suggestions. To achieve concrete system optimization, we leverage the strong code generation capabilities of state-of-the-art LLMs, which are extensively trained on large-scale code repositories and proficient at creating detailed code snippets \cite{hurst2024gpt, jaech2024openai, grattafiori2024llama, yang2024qwen2}. Our method involves generating numerous specific and diverse code modifications, and then systematically evaluating these changes to identify improvements.

We also employ the following design decisions to ensure the effectiveness of our approach:

\para{Optimizing congestion control with QUIC.}  Modifying traditional TCP congestion control algorithms is challenging, as they reside in kernel space. Testing modified TCP congestion control requires complex procedures such as recompiling the Linux kernel and rebooting machines. Instead, we focus on optimizing QUIC, whose congestion control algorithms operate entirely in user space. This user-space characteristic greatly simplifies modifications and enables concurrent evaluation of multiple algorithms on a single test machine.

\para{Full input, patched output.} Current LLM systems struggle to generate a complete congestion control algorithm from scratch, as the quality deteriorates significantly when generating lengthy responses. However, they handle long inputs effectively \cite{huang2023advancing}. Therefore, we adopt a "long input, short response" approach: we provide LLMs with the full source code of the widely-used BBR algorithm and prompt them to suggest substantial modification patches. These concise patches can then be directly applied and evaluated. This framework also supports multi-iteration improvements: After one patch is accepted, LLMs can generate new patches based the new source code. Specifically, we aim to improve the BBR congestion control algorithm implemented in MsQuic \cite{msquic}. MsQuic is a production-level open-source implementation of the IETF QUIC protocol. It is written in C and is reported to have strong performance \cite{mishra2022understanding, bunstorf2023msquic}.

\para{Encouraging diversity.} We find that a diverse set of candidate algorithms is critical to achieving a fundamental improvement. To encourage such diversity, we use two strategies: employing a chain-of-thought approach \cite{wei2022chain} that explicitly prompts LLMs to propose novel ideas clearly before coding, and setting the generation temperature to a high value (temperature=1) to promote creativity. We further enhance this diversity by exploring a broad array of LLMs, including both closed-source and open-source ones.

The following provides an outline of our prompt structure. For the complete prompt structure, please refer to the Appendix \ref{sec:appendix:full-prompt}.

\begin{figure}[htbp]
\begin{tcolorbox}[title=Prompt outline:]
\begin{minted}[fontsize=\small,breaklines,breaksymbolleft=]{markdown}
Here's the implementation of the BBR congestion control algorithm in QUIC from an open source project:
```
<Full Source Code>
```
We aim to improve the performance of the above algorithm. Please design an innovative mechanism and implement it. Your code implementation should follow the "update block" format below:
UPDATE FUNCTION `<Function Name>`:
```
Rewrite the function here.
```
UPDATE VARIABLE `<Variable Name>`:
```
Rewrite the variable definition.
```
ADD MEMBER TO `<Struct Name>`:
```
Add members to the struct.
```
Your output can consist of a single update block or multiple such update blocks. Brainstorm ideas and select the best one. Then, provide the updated code modifications in the format described above.
\end{minted}
\end{tcolorbox}
\end{figure}

\subsection{Algorithm Evaluation}
\label{sec:algo-eval}

Suppose we have generated $N$ candidate algorithms, denoted as $a_1, a_2, \dots, a_N$. Our next step is to comprehensively evaluate their performance and select the best-performing algorithm. To ensure robustness and substantial improvement across diverse network conditions, we test these algorithms over a wide range of scenarios, denoted by $c_1, c_2, \dots, c_M$.

Specifically, under a given network condition $c_j$, the performance of algorithm $a_i$ is quantified by a utility function $u(a_i, c_j)$. To facilitate systematic comparison among the algorithms, we define the final utility $\overline{u}(a_i)$ of algorithm $a_i$ as the average utility across all network conditions:
\[
\overline{u}(a_i) = \frac{1}{M}\sum_{j=1}^{M} u(a_i, c_j).
\]

Finally, the algorithm achieving the highest average utility $\overline{u}(a_i)$ is selected as the optimal algorithm.

In practice, we utilize real-world traces collected from diverse network environments, such as broadband, Starlink, 4G, and 5G, to represent the various network conditions ($c_1, c_2, \dots, c_M$). To accurately quantify algorithm performance, the utility $u(a_i, c_j)$ is obtained through emulation experiments driven by these real traces. We prefer emulation over simulation because emulation involves running actual algorithms in realistic, ensuring the optimizations introduced by LLM-generated code changes yield reliable and practical improvements.

Specifically, we measure two critical performance metrics: end-to-end throughput and latency. First, we set up a server and a client within the emulated environment. The client is programmed to download a large file from the server continuously for 10 seconds, during which we measure the throughput. This throughput measurement is repeated three times, and the average value is recorded as $tput$. Next, to assess latency, we simulate realistic small web-page requests by having the client repeatedly upload 512-byte data packets and download 4KB responses from the server immediately afterward. This lasts for 30 seconds and the median latency of these requests is recorded as $lat$. Such a test is the standard benchmarking method of MsQuic and is implemented in the official benchmarking tool, \textit{secnetperf} \cite{secnetperf}. It accurately reflects real end-to-end performance metrics.

The final utility function combines these metrics into a single numerical score:
\[
u = \frac{tput - tput_0}{tput_0} - \lambda \frac{lat - lat_0}{lat_0},\]
where $tput_0$ and $lat_0$ represent the throughput and latency obtained by the original, unmodified BBR algorithm under the same emulated network conditions. The parameter $\lambda$ balances the relative importance of throughput and latency improvements. In this work, we set $\lambda = 10$, because we observed that latency changes are typically subtler compared to throughput enhancements.

\section{Towards Efficient Evaluation}
\label{sec:towards-efficient-evaluation}

\subsection{Motivation}
To ensure the reliability and robustness of LLM-generated algorithms,  we carry out comprehensive evaluation across a wide range of network conditions, as described in Section \ref{sec:algo-eval}. However, such extensive evaluation can be extremely time-consuming. Specifically, in this paper, we consider a total of 408 network conditions. Emulation experiments for each condition require approximately one minute to complete. Consequently, evaluating a single algorithm across all network conditions would take around 408 minutes, equivalent to roughly 7 hours. In contrast, generating an algorithm using an LLM typically takes less than 1 minute. Thus, a fundamental challenge in LLM-based algorithm exploration arises:

\textit{How can we efficiently and effectively evaluate the large number of generated algorithms?}

In this paper, we aim to address this challenge by accurately estimating an algorithm's overall performance through evaluating only a carefully selected subset of network conditions. This approach significantly reduces the evaluation time without compromising accuracy.

\subsection{Problem Formulation}
Given a set of algorithms $a_1, a_2, \dots, a_N$ ($N$ can be infinite owing to the generative power of LLMs) and a set of network conditions $c_1, c_2, \dots, c_M$ ($M$ is a large number, e.g., 408), our goal is to identify the optimal algorithm $a_{i^*}$ that maximizes a predefined utility function. Formally, we aim to find:
\[
i^* = \arg\max_{i} \overline{u}(a_i),
\]
where $\overline{u}(a_i)$ denotes the average utility of algorithm $a_i$ across all network conditions. However, directly optimizing this objective is %
prohibitively expensive, as computing $\overline{u}(a_i)$ for each newly generated
algorithm $a_i$ requires testing it on $M$ network conditions, which is both time-consuming and resource-demanding.

To address this challenge, we propose to estimate the average utility by testing each algorithm on only a small subset $S \subset \{1, 2, \ldots, M\}$ of size $|S| = K \ll M$ of the network conditions. Let $\hat{u}(a_i | \boldsymbol{u}_S)$ be our estimate of $\bar{u}(a_i)$ based on observing the algorithm's performance only on subset $S$, where $\boldsymbol{u}_S$ represents the observed value vectors. Now, the key question becomes:

\textit{How can we select the optimal subset $S$ to ensure that our estimate $\hat{u}(a_i | \boldsymbol{u}_S)$ is as close as possible to the true average utility $\bar{u}(a_i)$?}

One approach would be to minimize the expected squared error between our estimate and the true average: $$\min_S \mathbb{E}[(\hat{u}(a_i | \boldsymbol{u}_S) - \bar{u}(a_i))^2]$$ However, this objective is challenging to optimize directly because: 1) It depends on the specific algorithm $a_i$ being evaluated 2) It requires knowing the true average utility $\bar{u}(a_i)$, which is precisely what we are trying to estimate 3) The optimization space over all possible subsets $S$ of size $K$ is combinatorially large.

To simplify the task, we take a statistical approach to estimating the average utility. Specifically, we model the utility values $u(\cdot, c_1), u(\cdot, c_2), ..., u(\cdot, c_M)$ as random variables and assume that they follow a certain probability distribution.
Here, different generated algorithms provide the randomness in the utility function, and each evaluation of an algorithm provides one realization (observation) of these random variables.
For simplicity, we denote $u_1=u(\cdot, c_1), u_2 = u(\cdot, c_2), ..., u_M=u(\cdot, c_M)$.
The overall utility we seek to estimate is the empirical mean across these conditions:
$$\overline{u} = \frac{1}{M}\sum_{j=1}^{M} u_j.$$

When we can only observe utilities for a subset $S$ of conditions, the optimal estimator for $\bar{u}$ under squared error loss is the conditional expectation: $$\hat{u}(\boldsymbol{u}_S) = \mathbb{E}[\bar{u} | \boldsymbol{u}_S] = \frac{1}{M}\left(\sum_{j \in S} u_j + \mathbb{E}\left[\sum_{j \in U} u_j | \boldsymbol{u}_S\right]\right)$$ where $U = \{1, 2, \ldots, M\} \setminus S$ is the set of unobserved conditions. It follows the bias-variance decomposition of mean-squared error~\cite{stats-book} that for any estimator $f(\boldsymbol{u}_S)$, the expected squared error follows the general relationship: $$\mathbb{E}[(\overline{u} - f(\boldsymbol{u}_S))^2] = \mathbb{E}[\text{Var}(\overline{u}|\boldsymbol{u}_S)] + \mathbb{E}[(\mathbb{E}[\overline{u}|\boldsymbol{u}_S] - f(\boldsymbol{u}_S))^2]$$ The uncertainty in this estimation is quantified by the conditional variance $\text{Var}(\overline{u} | \boldsymbol{u}_S)$. When using the optimal estimator $\hat{u}(\boldsymbol{u}_S) = \mathbb{E}[\bar{u}|\boldsymbol{u}_S]$, the second term becomes zero, giving us: $$\mathbb{E}[(\overline{u} - \hat{u}(\boldsymbol{u}_S))^2] = \mathbb{E}[\text{Var}(\overline{u} | \boldsymbol{u}_S)]$$ This shows that the expected conditional variance represents the irreducible component of the expected squared error. Therefore, minimizing the conditional variance: $$\min_S \text{Var}(\overline{u} | \boldsymbol{u}_S)$$ directly reduces the inherent uncertainty in our estimator, transforming our original problem into a well-defined statistical objective that does not depend on specific algorithm values or knowledge of the true average utility.

Overall, the problem is formally defined as follows: Given random variables $u_1, u_2, \dots, u_M$ and an integer $K$ ($K \ll M$), we seek a subset $S \subseteq \{1,2,\dots,M\}$ of size $|S| = K$ that minimizes the conditional variance $\text{Var}(\overline{u} | \boldsymbol{u}_S)$, where $\overline{u} = \frac{1}{M}\sum_{j=1}^{M}u_j$ represents the overall mean, and $\boldsymbol{u}_S$ denotes the observed values corresponding to the subset $S$.

\subsection{Empirical Obersevations}
We empirically find important statistical properties among the random variables $u_1, u_2, \dots, u_M$: some network conditions exhibit significantly higher \textit{variance} of utility, while others show strong \textit{correlations}. Leveraging these variance and correlation properties can considerably reduce evaluation time. To illustrate this, we evaluate 500 algorithms generated by the o1-preview model \cite{jaech2024openai} on 10 network traces from the FCC dataset \cite{fccMeasuringBroadband}, with each trace representing a distinct network condition ($c_j$). Figure~\ref{fig:utility_variance_of_trace} shows the variance of utilities achieved by 500 algorithms on each trace. We make two key observations from this analysis:

\begin{figure}[tbp]
    \centering
    \includegraphics[width=0.95\linewidth]{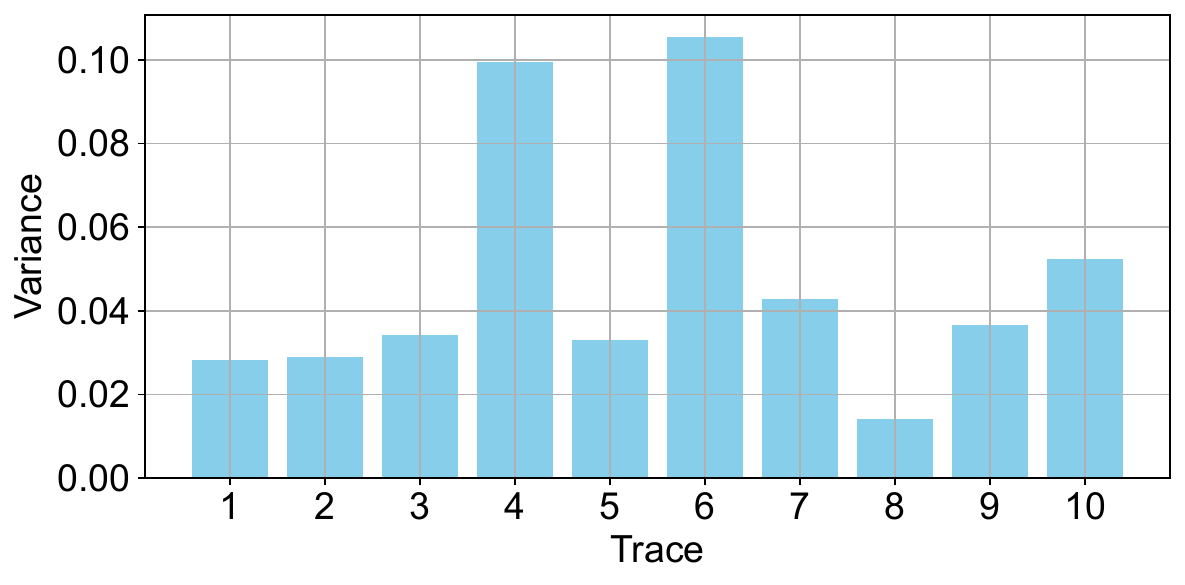}
    \caption{Utility variance of 500 algorithms on ten different traces.}
    \label{fig:utility_variance_of_trace}
\end{figure}

\para{Observation 1: Certain traces exhibit notably higher utility variance.} As depicted in Figure~\ref{fig:utility_variance_of_trace}, traces 4 and 6 demonstrate higher performance variance than others, meaning different algorithms achieve highly diverse utility values on them.
Intuitively, such traces are more discriminative, and thus prioritizing them
during testing can amplify performance differences between algorithms.
Consider a simplified scenario with two independent random variables \( u_1 \) and \( u_2 \). If we observe the random variable \( u_1 \), the conditional variance of the average becomes $\text{Var}(\overline{u}|u_1) = \text{Var}((u_1+u_2)/2|u_1) =\text{Var}(u_2)/4$. Similarly, if we observe the random variable \( u_2 \), the conditional variance becomes $\text{Var}(\overline{u}|u_2) = \text{Var}(u_1)/4$. Therefore, to minimize the conditional variance by observing only one of the variables, we should choose to observe the variable with the higher variance.

\begin{figure}[tbp]
    \centering
    \includegraphics[width=0.95\linewidth]{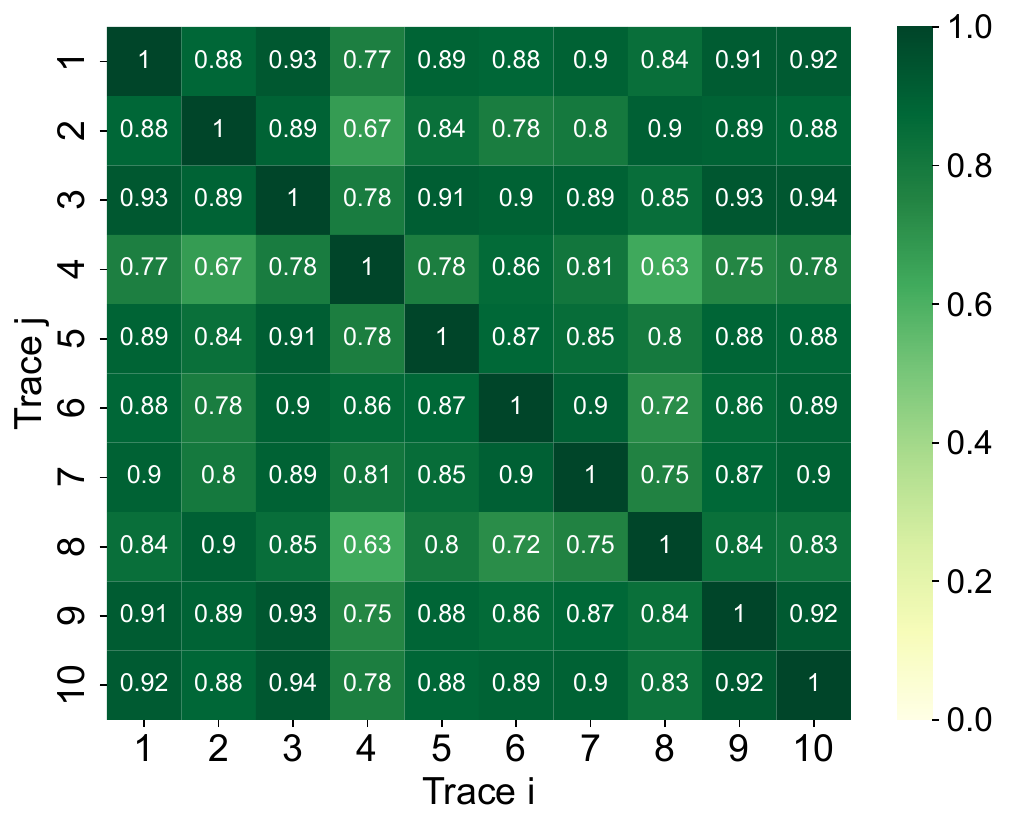}
    \caption{Utility correlation between different traces}
    \label{fig:utility_correlation}
\end{figure}

\para{Observation 2: Some traces exhibit strong correlations.} \\
Figure~\ref{fig:utility_correlation} presents the Pearson correlation coefficients among random variables \( u_1, u_2, \dots, u_{10} \). Several pairs of traces display notably high correlations; for instance, the correlation coefficient between trace~3 and trace~10 is \(0.94\). High correlations indicate that observing one variable provides substantial information to reduce uncertainty about its correlated counterpart. For instance, if two traces have a strong positive correlation, a high utility in one suggests a similarly high utility in the other, reducing uncertainty even without direct observation of the other variable.

\subsection{Our Selection Method}
To reflect the observed statistical properties, we choose to model the random variables \( u_1, u_2, \dots, u_M \) with a multivariate Gaussian distribution---a reasonable (and common) assumption given that we do not have knowledge about the underlying distribution of LLM-generated algorithms.

Formally, we assume:
\[
(u_1, u_2, \dots, u_M)^\top \sim \mathcal{N}(\boldsymbol{\mu}, \boldsymbol{\Sigma}),
\]
where the mean vector \(\boldsymbol{\mu}\) and covariance matrix \(\boldsymbol{\Sigma}\) are defined as:
\begin{equation}
\boldsymbol{\mu} = (\mu_1, \mu_2, \dots, \mu_M)^\top,\quad
\boldsymbol{\Sigma} = [\sigma_{ij}]_{M \times M}.
\label{eq:para}
\end{equation}

Recall that our goal is to select a subset \( \mathcal{S} \subseteq \{1,2,\dots,M\} \) of size \( |\mathcal{S}|=K \) to observe such that $\text{Var}(\overline{u}|\boldsymbol{u}_S)$ is minimized. We denote the unobserved set as \( \mathcal{U} = \{1,2,\dots,M\} \setminus \mathcal{S} \). To calculate $\text{Var}(\overline{u}|\boldsymbol{u}_S)$ in the setting of multivariate Gaussian distribution, we partition \( \boldsymbol{\mu} \) and \( \boldsymbol{\Sigma} \) as follows:
\[
\boldsymbol{\mu} = \begin{pmatrix}
\boldsymbol{\mu}_S \\
\boldsymbol{\mu}_U
\end{pmatrix}, \quad
\boldsymbol{\Sigma} = \begin{pmatrix}
\boldsymbol{\Sigma}_{SS} & \boldsymbol{\Sigma}_{SU} \\
\boldsymbol{\Sigma}_{US} & \boldsymbol{\Sigma}_{UU}
\end{pmatrix},
\]
where \(\boldsymbol{\mu}_S\) and \(\boldsymbol{\mu}_U\) are mean vectors of the observed and unobserved variables, and \(\boldsymbol{\Sigma}_{SS}, \boldsymbol{\Sigma}_{SU}, \boldsymbol{\Sigma}_{US}, \boldsymbol{\Sigma}_{UU}\) are corresponding covariance sub-matrices.

Given observations \(\boldsymbol{u}_S\), the conditional distribution of the unobserved variables \(\boldsymbol{u}_U\) is also Gaussian, expressed as:
\[
\boldsymbol{u}_U | \boldsymbol{u}_S \sim \mathcal{N}(\boldsymbol{\mu}_{U|S}, \boldsymbol{\Sigma}_{U|S}),
\]
with the conditional mean and covariance given by:
\begin{align}
\boldsymbol{\mu}_{U|S} &= \boldsymbol{\mu}_U + \boldsymbol{\Sigma}_{US}\boldsymbol{\Sigma}_{SS}^{-1}(\boldsymbol{u}_S - \boldsymbol{\mu}_S), \label{eq:mean} \\
\boldsymbol{\Sigma}_{U|S} &= \boldsymbol{\Sigma}_{UU} - \boldsymbol{\Sigma}_{US}\boldsymbol{\Sigma}_{SS}^{-1}\boldsymbol{\Sigma}_{SU}. \label{eq:var}
\end{align}

Our optimization target, the conditional variance of the average utility can be formulated as
\[
\text{Var}(\overline{u}|\boldsymbol{u}_S)
= \text{Var}\left(\frac{1}{M}\sum_{i \in \mathcal{S}}u_i + \frac{1}{M}\sum_{j \in \mathcal{U}} u_j | \boldsymbol{u}_S\right).
\]

Since the observed utilities $\boldsymbol{u}_S$ are known,
their variance is zero. Thus,
\begin{equation}
\text{Var}(\overline{u}|\boldsymbol{u}_S)
= \frac{1}{M^2}\text{Var}\left(\sum_{j \in \mathcal{U}} u_j | \boldsymbol{u}_S\right).
\label{eq:cv_sum}
\end{equation}

For any random variables $x_1, x_2, ..., x_M$, $\text{Var}(\sum^{M}_{i=1}{x_i})$ equals $\sum_{i=1}^{M} \sum_{j=1}^{M} \text{Cov}(x_i, x_j)$, which is the sum of elements of the covariance matrix. Thus equation (\ref{eq:cv_sum}) can be simplifies into
\begin{equation}
\label{eq:final}
\text{Var}(\overline{u}|\boldsymbol{u}_S)
= \frac{1}{M^2}\mathbf{1}_{M-K}^\top \boldsymbol{\Sigma}_{U|S}\mathbf{1}_{M-K},
\end{equation}
where $\mathbf{1}_{M-K}^\top \boldsymbol{\Sigma}_{U|S}\mathbf{1}_{M-K}$ stands for the element sum of the conditional covariance matrix $\boldsymbol{\Sigma}_{U|S}$ shown in Equation \ref{eq:var}.

Given $K$, selecting the optimal $K$ random variables is NP-hard because it can be reduced to a subset selection problem, where the objective function is given by Equation~\ref{eq:final}. To address this complexity, we adopt a greedy algorithm approach. Specifically, we iteratively select one variable at a time that minimizes the conditional variance, %
and repeat this selection process $K$ times.

In practice, we first estimate the mean vector and covariance matrix in Equation \(\ref{eq:para}\), by evaluating a randomly selected subset of \(L\) algorithms (\(L \ll N\)) from the total \(N\) algorithms under all \(M\) network conditions.  Next, we select \(K\) network conditions (\(K \ll M\)) using the greedy algorithm mentioned before. For the remaining \(N - L\) algorithms, we evaluate them only under these selected \(K\) network conditions. Consequently, for each algorithm, we obtain an observed performance vector \(\boldsymbol{u}_S\). The expectation of the unobserved variables, \(\boldsymbol{\mu}_{U|S}\), can be estimated using Equation \(\ref{eq:mean}\). Then, the average utility $\overline{u}$ is given by averaging \(\boldsymbol{\mu}_{U|S}\) and \(\boldsymbol{u}_S\).

Consider the first step of the greedy algorithm. For each candidate $k \in \{1,2,\dots,M\}$, the variable set to observe is $\mathcal{S}=\{k\}$, and the corresponding unobserved variable set $\mathcal{U}=\{1, 2,\dots,M\} \setminus \{k\}$. Selecting the optimal $k^*$ involves minimizing the Equation \ref{eq:final}, which translates to minimizing the element sum of $\boldsymbol{\Sigma}_{U|S} = \boldsymbol{\Sigma}_{UU} - \boldsymbol{\Sigma}_{US}\boldsymbol{\Sigma}_{SS}^{-1}\boldsymbol{\Sigma}_{SU}$. We can prove that the optimal $k^*$ is determined by:
$$
k^{*} = \arg\max_k {\frac{(\text{Var}(u_k) + \sum_{j=1, j \neq k }^{M}{\text{Cov}(u_j, u_k)})^2}{\text{Var}(u_k)}}.
$$

Notably, if all variables are independent, the covariance terms $\sum_{j \neq k}\text{Cov}(u_j, u_k)$ become zero. In this scenario, the selection simplifies to choosing variables with the highest variance, aligning directly with Observation 1. However, when correlations between variables exist, the selection must consider both high individual variance and substantial mutual correlations among variables, aligning with Observation 2.

\section{Evaluation}
\label{sec:eval}

This section presents a comprehensive evaluation of our approach. We utilize four different LLMs to generate algorithms and conduct emulation-based experiments on real-world traces to identify improvements.

Details of the experimental setup and main findings are provided in Sections~\ref{sec:settings}, \ref{sec:utility-distribution}, and \ref{sec:best-algos}. Section~\ref{sec:efficient-evaluation} examines the effectiveness of our ``efficient evaluation'' methods (Section~\ref{sec:towards-efficient-evaluation}). Section~\ref{sec:domain-specific} analyzes whether a domain-specific or general congestion control algorithm is more advantageous. Finally, Section~\ref{sec:iterative-pipeline} investigates the impact of an iterative pipeline on further performance gains.

\subsection{Settings}
\label{sec:settings}

\begin{table*}[t]
    \centering
    \renewcommand{\arraystretch}{1.3} %
    \begin{tabular}{l l c c c}
        \toprule
        \textbf{Parameter} & \textbf{Meaning} & \textbf{Conservative} & \textbf{Default} & \textbf{Aggressive} \\
        \midrule
        InitialWindowPackets & Initial congestion window (packets) & 10 & 16 & 32 \\
        kHighGain & Pacing gain during the Startup state & \(2400/1000 + 1\) & \(2885/1000 + 1\) & \(3500/1000 + 1\) \\
        kStartupGrowthTarget & Bandwidth growth per RTT during Startup & \(6/5\) & \(5/4\) & \(3/2\) \\
        kDrainGain & Pacing gain during the Drain state & \(1000/3500\) & \(1000/2885\) & \(1000/2400\) \\
        kCwndGain & Congestion window gain during ProbeBw & \(3/2\) & \(2\) & \(5/2\) \\
        \bottomrule
    \end{tabular}
    \vspace{5pt}
    \caption{We tune five key BBR parameters in emulated networks. Each parameter can take its default value, a more conservative setting, or a more aggressive setting, resulting in 243 possible configurations. The configuration that achieves the best performance serves as the configuration-optimized BBR baseline.}
    \vspace{-10pt}
    \label{tab:bbr_parameters}
\end{table*}

\begin{table}[t]
\begin{tabular}{cccc}
\toprule
\textbf{Dataset} & \textbf{\# of traces} & \textbf{Avg BW (Mbps)} & \textbf{BDP (bytes)} \\
\midrule
FCC              & 85                 & 1.3                        & 16,000                       \\
Starlink         & 19                 & 1.9                        & 24,000                       \\
4G               & 50                 & 18.1                       & 230,000                      \\
5G               & 50                 & 31.7                       & 400,000                      \\
\bottomrule
\end{tabular}
\vspace{5pt}
\caption{We employ four real-world network trace sets for emulation. This table summarizes the number of traces and the average bandwidth for each dataset. We also compute the bandwidth-delay product (BDP) assuming a 100 ms RTT, and configure
the queue size in \texttt{mahimahi} to both 1$\times$ BDP and 0.5$\times$ BDP,
resulting in a total of %
408 distinct network conditions.}
\label{tab:evaluation-setting}
\vspace{-8pt}
\end{table}

\para{LLMs.} We examine the following LLMs: (1) GPT-4o, a flagship general-purpose LLMs from OpenAI known for its strong capabilities across various tasks~\cite{hurst2024gpt}; (2) o1-preview, a specialized reasoning model from OpenAI for mathematical problem-solving and complex coding tasks~\cite{jaech2024openai}; (3) Llama-3.1-70B, a widely-used open-source LLM developed by Meta~\cite{grattafiori2024llama};  (4) Qwen-2.5-72B, a popular open-source model developed by Alibaba, whose coding capability is reported to be better than Llama-3.1-70B~\cite{yang2024qwen2}.

\para{Algorithm generation.} We utilize MsQuic v2.4.5 and provide the LLMs with the full source code in \textit{src/core/bbr.h} and \textit{src/core/bbr.c}, where the BBR algorithm is implemented. The prompt is detailed in Section~\ref{sec:algo-gen}. The LLMs will suggest function updates, global variable updates, and C struct updates in a predefined syntax. We will extract these updates and apply them to the source code. Finally, we will compile the modified source code. If the compilation succeeds, we will conduct further emulation tests using real-world traces. We generate 3,000 samples using each LLM.

\para{Datasets.} Our goal is to develop a general-purpose improved BBR algorithm that performs well across various network scenarios. To achieve this, we leverage the following real-world traces and conduct emulation experiments:

\begin{itemize}[itemsep=1pt,topsep=5pt,leftmargin=*]
    \item \textbf{FCC:} The FCC dataset provides broadband network measurements in the U.S.~\cite{fccMeasuringBroadband}.
    \item \textbf{Starlink:} We collect throughput traces from a Starlink RV terminal in the U.S. Although Starlink maintains sufficient throughput during off-peak hours, its bandwidth decreases significantly during peak periods due to the shared usage of satellite links. To simulate network congestion, we adjust the throughput in Starlink traces to one-eighth of the original value.
    \item \textbf{4G:} We measure downlink throughput using {\em iperf} from a cloud server located in the central U.S. to a local client over the internet, where the wireless hop utilizes 4G networks.
    \item \textbf{5G:} We collect 5G downlink throughput data using the same method as for 4G.
\end{itemize}

These datasets represent the mainstream network types today. An improved BBR algorithm should achieve consistently better performance across all of these traces. Detailed information of these datasets is provided in Table~\ref{tab:evaluation-setting}.

\para{Emulation settings.} We use \texttt{mahimahi}~\cite{mahimahi} to set up emulated network environments and test the generated congestion control algorithms. In our experiments, we set the round-trip time (RTT) to 100 ms in \texttt{mahimahi}. To better evaluate the performance of the congestion control algorithm, we configure a fixed queue size for the \texttt{mahimahi} network. Packets are buffered in the queue, and any exceeding packets are dropped when the queue reaches its capacity. The queue size is determined based on the bandwidth-delay product (BDP) for each dataset. The specific BDP settings are provided in Table \ref{tab:evaluation-setting}. For each network trace, we evaluate two queue size conditions: 1$\times$ BDP and 0.5$\times$ BDP. Consequently, a total number of %
408 network conditions are tested. For each condition, we measure the end-to-end throughput and latency using the test tool \textit{secnetperf} provided in MsQuic. The utility of each congestion control algorithm is then quantified based on its relative improvement compared with the default BBR algorithm. 
Specifically, under each network condition, the utility is computed as \( u = \frac{\text{tput} - \text{tput}_0}{\text{tput}_0} - \lambda \frac{\text{lat} - \text{lat}_0}{\text{lat}_0} \), where \( \text{tput} \) and \( \text{lat} \) denote the measured end-to-end throughput and latency of the generated algorithm, while \( \text{tput}_0 \) and \( \text{lat}_0 \) represent the corresponding values for the default BBR under the same conditions. The utility of the generated algorithm is the average utility across all network conditions.
The detailed testing process and utility formulation are presented in
Section~\ref{sec:algo-eval}.

\para{Baselines.} We compare LLM-generated algorithms with the default BBR
algorithm in MsQuic~\cite{msquic}.
We adopted MsQuic as it is reported to be one of the most performant QUIC
implementation in production environments, and its BBR algorithm is shown to deliver
strong performance~\cite{mishra2022understanding,bunstorf2023msquic}.
To further strengthen the default BBR, we also perform parameter tuning on our
traces. We identify five key parameters in BBR and test three values for each
parameter: the default, a more conservative option, and a more aggressive option.
This results in 243 parameter combinations, which we evaluate across all network conditions. The combination with the highest average utility serves as a configuration-optimized BBR baseline. Details of these parameters are
in Table~\ref{tab:bbr_parameters}.

We emphasize that this study does not aim to identify the optimal, one-size-fits-all congestion control algorithm. Instead, our objective is to demonstrate that LLMs can automatically enhance existing congestion control methods.
To this end, we focus on BBR as a case study due to its widespread adoption and practical significance on today's Internet~\cite{mishra2024keeping}, while
our methodology is broadly applicable to other congestion control methods
(which we defer to future work due to the high inference costs of LLMs).

In total, we employ four LLMs---GPT-4o, o1-preview, Llama-3.1-70B, and Qwen-2.5-72B---to generate 3,000 candidate algorithms each. These algorithms are then compiled and evaluated in emulated networks. Each algorithm undergoes evaluation across 408 network conditions, covering broadband, Starlink, 4G, and 5G networks. The utility of the generated algorithms is measured and compared against both the default BBR and a configuration-optimized BBR.

\subsection{Utility Distribution}
\label{sec:utility-distribution}

\begin{table*}[t]
\begin{tabular}{cccccccc}
\toprule
\textbf{Method}             & \textbf{Generated} & \multicolumn{1}{l}{\textbf{Code Length}} & \textbf{Compiled} & \textbf{Simlarity} & \textbf{Best Utility} & \textbf{Tput} & \textbf{Latency} \\ \midrule
BBR (default configuration) & --                  & --                                   & --                 & --                  & 0.000                 & 0.00\%              & 0.00\%           \\
BBR (best configuration)    & --                  & --                                   & --               & --                  & 0.121                 & +9.66\%              & -0.26\%          \\
o1-preview                 & 3,000              & 8,252                               & 726               & 2.27               & 0.257                 & +25.63\%             & -0.01\%          \\
GPT-4o                     & 3,000              & 3,905                               & 1,663             & 2.83               & 0.261                 & +26.03\%             & 0.00\%           \\
Llama-3.1-70B              & 3,000              & 2,428                               & 138               & 1.61               & 0.206                 & +22.45\%             & +0.18\%           \\
Qwen-2.5-72B               & 3,000              & 9,536                               & 1,028             & 1.82               & 0.271                 & +27.28\%             & +0.02\%           \\ \bottomrule
\end{tabular}
\vspace{5pt}
\caption{Main results. Code Length refers to the average character length (not token count) of code over all samples from each LLM, excluding any non-code text. Similarity represents the average similarity score, computed by randomly selecting output pairs and querying GPT-4 for a similarity score with a scale from 1 (least similar) to 5 (most similar). Best Utility represents the best average utility achieved among all samples from each LLM. Tput and Latency indicate the throughput and latency changes of the best sample relative to the default BBR.}
\label{tab:best-sample}
\vspace{-8pt}
\end{table*}

\begin{figure}[tbp]
    \centering
    \includegraphics[width=0.8\linewidth]{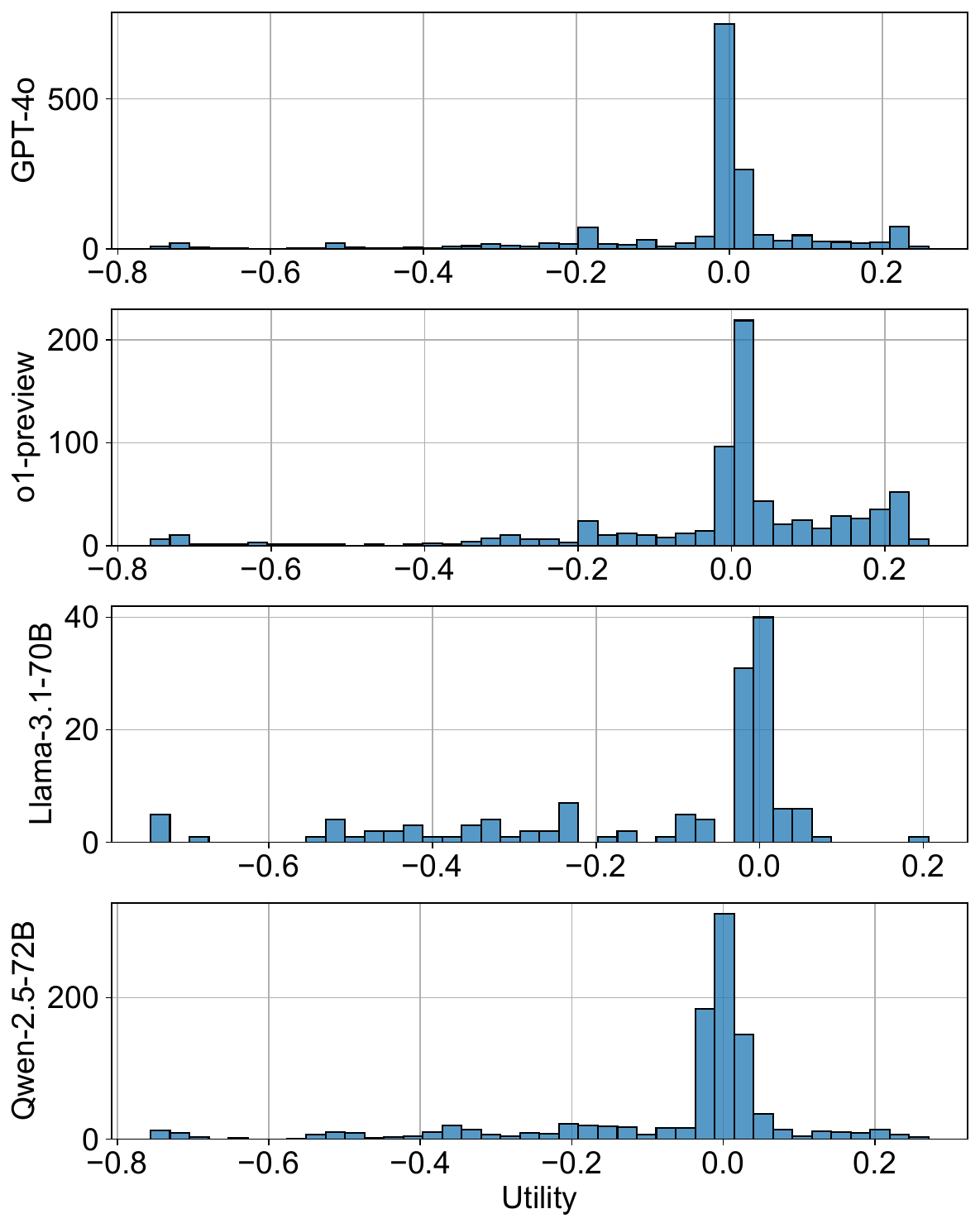}
    \caption{The utility distribution of compiled samples for all models. Samples from different models are shown in separate subfigures. In each subfigure, the X-axis represents the utility, while the Y-axis denotes the sample count corresponding to the utility. Specifically, for better visualization, we exclude a sample with a minimum utility of $-2.07$ from the first subfigure (GPT-4o). For other subfigures, no samples are removed.
    }
    \label{fig:utility_distribution}
\end{figure}

\begin{figure}[tbp]
    \centering
    \includegraphics[width=0.8\linewidth]{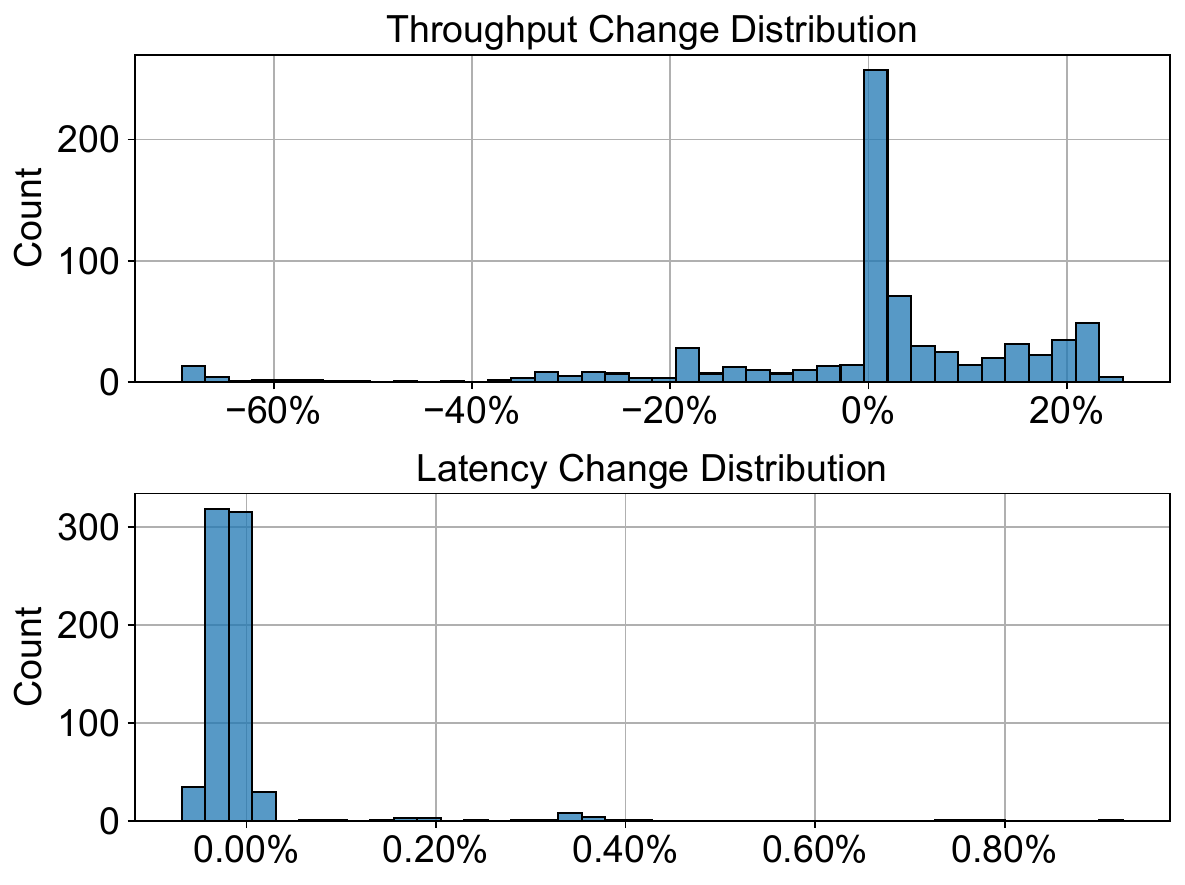}
    \caption{Distribution of throughput and latency relative changes for all compiled samples from the o1-preview model, measured against the default BBR. }
    \label{fig:detailed_distribution_o1}
\end{figure}

We present evaluation results for all network conditions in Table~\ref{tab:best-sample} and Figure~\ref{fig:utility_distribution}. This section examines the distribution of algorithms generated by various LLMs, focusing on their average quality and similarity.

All models are found to produce reasonable code modifications, but their compilation success rates vary. As shown in Table~\ref{tab:best-sample}, GPT-4o achieves the highest compilation rate, with 1,663 out of 3,000 generations compiling successfully. In contrast, Llama-3.1 has the lowest rate, with only 138 compilable generations.
We employ the same prompt across LLMs for a fair comparison,
while customizing the prompt for Llama might be able to improve its compilation success rate.

We also analyze the average character length of generated code, excluding any non-code text from the response. o1-preview and Qwen-2.5-72B generate significantly longer code than GPT-4o and Llama-3.1-70B. In particular, Llama-3.1-70B produces the shortest code modifications.

Furthermore, we assess the similarity of generated samples by randomly selecting 100 pairs from the compiled outputs of each model. We use GPT-4 to classify each pair into five similarity levels, ranging from 1 (least similar) and 5 (most similar).
As shown in Table~\ref{tab:best-sample}, %
GPT-4o tends to produce the most similar algorithms overall, whereas Qwen-2.5-72B appears
more creative (Llama-3.1-70B generates significantly fewer compilable outputs).

Figure~\ref{fig:utility_distribution} presents the utility distribution of compiled samples from all models. The utility is measured as the relative improvement over default BBR. The distribution follows a consistent pattern: (1) Most compiled modifications have minimal impact, forming a sharp peak around 0 (utility remains unchanged); (2) Some modifications significantly degrade performance, with utilities ranging from $-0.8$ to $-0.2$ (indicating a 20\% to 80\% reduction in performance); (3) Crucially, a small fraction of modifications consistently enhance performance, appearing as short but distinct bars between 0.1 and 0.25 (10\% to 25\% improvement). The o1-preview model generates a notable percentage of modifications with utility above 0.1. This suggests that while LLMs generally struggle to produce effective algorithmic improvements, generating a large number of samples and applying proper evaluation can reveal valuable system optimizations.

As an illustrative example, Figure~\ref{fig:detailed_distribution_o1} presents the distribution of throughput and latency changes for the o1-preview model. We observe that most modifications result in minimal latency changes, whereas throughput is substantially affected, with several modifications producing significant variations. Notably, a substantial portion of samples exhibit an average throughput improvement of approximately $20\%$ across the $408$ evaluated network conditions. This trend is consistent across the other models as well: latency variations remain negligible at the most of time, while throughput variations are more pronounced, indicating that the observed changes in overall utility primarily stem from changes in throughput.

\begin{figure*}[t]
\centering
\begin{subfigure}[t]{0.49\textwidth}
    \centering
    \includegraphics[height=120pt]{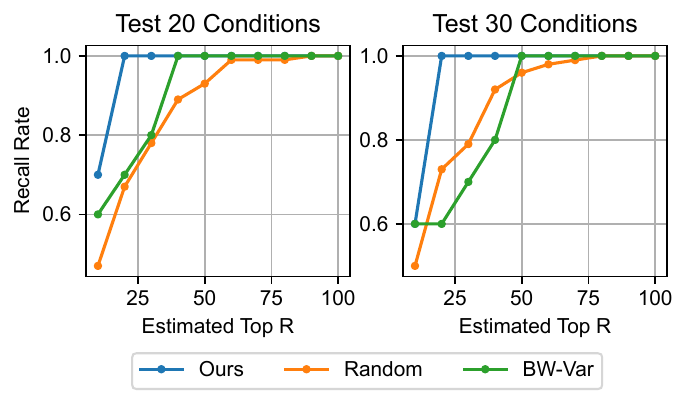}
    \vspace{-5pt}
    \caption{o1-preview}
    \label{fig:recall_o1}
\end{subfigure}
\begin{subfigure}[t]{0.49\textwidth}
    \centering
    \includegraphics[height=120pt]{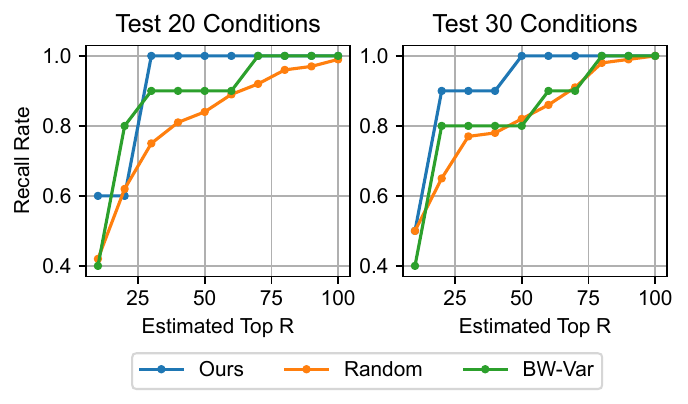}
    \vspace{-5pt}
    \caption{GPT-4o}
    \label{fig:recall_gpt4o}
\end{subfigure}
\vspace{-10pt}
\caption{Effectiveness of our trace sampling strategy
(\S\ref{sec:towards-efficient-evaluation}) vs.
two baselines (\S\ref{sec:efficient-evaluation}),
when evaluating the remaining \(N-L\) algorithms generated by o1-preview and GPT-4o.
Each figure shows the \textit{recall rate} (y-axis) of the ground-truth top 10
algorithms, measured as the fraction appearing in our estimated top $R$ algorithms
(x-axis) ranked based on the conditional expectation of average utility given observed
utilities in $K = 20$ or $30$ network conditions.
}
\end{figure*}

\subsection{The Best Algorithms}
\label{sec:best-algos}

In this section, we focus on analyzing the best-performing samples generated by each model. As detailed in Table \ref{tab:best-sample}, all LLMs can generate algorithms that surpass both the default BBR algorithm and its configuration-optimized variant. Specifically, optimizing BBR's parameters alone results in only a 9.66\% throughput improvement, whereas the LLM-based approaches achieve substantially greater gains ranging from 22.45\% to 27.28\%---with negligible latency changes.

Upon closer inspection, we find that the top-performing samples involve non-trivial algorithmic modifications:

\begin{itemize}[itemsep=1pt,topsep=5pt,leftmargin=*]
    \item \textbf{o1-preview}: This implementation modifies one struct and four functions to introduce a loss-rate-aware mechanism. The key modification involves scaling \texttt{pacing\_gain} and \texttt{cwnd\_gain} by ($1 - $\texttt{loss\_rate}), thereby making BBR sensitive to packet loss events. Consequently, whenever a loss event occurs, the corresponding gain values decrease proportionally, enabling the algorithm to adapt more responsively to changing network conditions. Although Table~\ref{tab:bbr_parameters} also includes static adjustments to \texttt{pacing\_gain} and \texttt{cwnd\_gain} to explore optimal configurations, the improvements obtained through these static adjustments are smaller compared to the dynamic, loss-aware mechanism proposed here.

    \item \textbf{GPT-4o}: This best sample from this model proposes to compute $\texttt{ACK\_Bytes}/\min(\texttt{RTT})$ as an instant maximum bandwidth estimate. Here, the bandwidth estimation uses the minimum observed RTT, thereby reflecting an instant upper bound on network capacity. If this instant bandwidth is lower than BBR's current bandwidth estimate, the model interprets this as an opportunity to slightly increase throughput. Accordingly, it incrementally raises both \texttt{pacing\_gain} and \texttt{cwnd\_gain} by a factor of $\frac{1}{16}$ under such conditions.

    \item \textbf{Llama-3.1-70B}: The sample from this model introduces a dual-threshold control mechanism, defining high and low thresholds for bytes in flight. When the bytes in flight exceed the high threshold or fall below the low threshold, the congestion window is actively adjusted; otherwise, the algorithm maintains standard behavior.

    \item \textbf{Qwen-2.5-72B}: Two modifications are introduced to improve BBR's responsiveness upon detecting packet loss. First, the congestion window is reduced by subtracting the total bytes of lost packets. Second, if the current RTT exceeds the historical minimum and the bytes in flight are fewer than the congestion window, it proactively reduces the congestion window. Together, these adjustments enable BBR to react more quickly and sensitively to network changes. Among all tested samples, this approach achieves the highest overall utility.
\end{itemize}

Detailed code changes corresponding to each of these best-performing samples are provided in Appendix \ref{sec:appendix:best-sample}. These modifications can be directly integrated into MsQuic v2.4.5 to reproduce our reported results. Beyond the top-performing samples, LLMs have also generated a diverse set of algorithmic variants, many of which achieve throughput gains exceeding 20\% (with nonsignificant changes to latency). These variants could further inspire novel optimizations in congestion control systems.

\subsection{Efficient Evaluation}
\label{sec:efficient-evaluation}

In this section, we empirically validate the efficient evaluation method described in Section \ref{sec:towards-efficient-evaluation}. Suppose we have $N$ algorithms and $M$ network conditions; our goal is to efficiently identify the top-performing algorithms. Naive exhaustive evaluation would require $M\times N$ evaluations, which is computationally expensive. In our experiments, $M = 409$, and $N$ represents the number of successfully compiled algorithms.

Our proposed efficient evaluation approach consists of the following steps:

\begin{enumerate}[itemsep=1pt,topsep=5pt,leftmargin=*]
    \item Randomly select a subset of $L$ algorithms and evaluate their performance under all $M$ network conditions. We set $L=100$. Using these results, we estimate the parameters of the Gaussian distribution in Equation \ref{eq:para}.
    \item Based on these estimated parameters, we select a subset of $K$ network conditions using a greedy method to minimize the variance of overall utility. We test $K=20$ and $K=30$.
    \item Evaluate the remaining $N - L$ algorithms only under these $K$ selected network conditions.
    \item Rank the remaining $N - L$ algorithms based on a combination of observed utilities (in the selected $K$ conditions) and estimated utilities (from Equation \ref{eq:mean}) in the untested conditions.
\end{enumerate}

To evaluate the quality of our method, we measure the recall rate of the top-performing algorithms. Specifically, we focus on recalling the top 10 algorithms among the remaining $N - L$ candidates. We test our method on both the o1-preview and GPT-4o models. Two baseline approaches are compared:

\begin{itemize}[itemsep=1pt,topsep=5pt,leftmargin=*]
    \item ``Random'': Random selection of $K$ network conditions.
    \item ``BW-Var'': Selection of $K$ network conditions with the highest bandwidth variance. Fluctuating network bandwidth
    tends to influence algorithm behavior differently, leading to diverse performance outcomes.
\end{itemize}

Figures \ref{fig:recall_o1} and \ref{fig:recall_gpt4o} illustrate the recall rate results for the o1-preview and GPT-4o models, respectively. Our proposed method consistently achieves higher recall rates compared with both baselines. Moreover, it always requires the fewest ``estimated top $R$'' to reach a 100\% recall rate. Specifically:

\begin{itemize}[itemsep=1pt,topsep=5pt,leftmargin=*]
    \item For the o1-preview model, testing 20 network conditions and then recalling 20 algorithms has a 100\% recall rate for the top 10 algorithms.
    \item For the GPT-4o model, testing 20 network conditions and then recalling 30 algorithms achieves a 100\% recall rate.
\end{itemize}

Overall, the efficiency gains are substantial. Without this method, we would need $M\times N$ evaluations. In contrast, our efficient method reduces the required evaluations to only $LM + (N - L)K$, plus $M \times R$ additional evaluations to confirm the actual top algorithms. In our experiments, with $M=408$, $L=100$, and $K=20$, the required evaluations are reduced by approximately \textbf{79.2\%} for the o1-preview model ($N=726$, $R=20$) and \textbf{87.6\%} for the GPT-4o model ($N=1663$, $R=30$), representing a significant reduction in computational overhead.

\subsection{Domain-Specific vs. General Algorithms}
\label{sec:domain-specific}

Our previous discussions focus on leveraging LLMs to optimize a single, general-purpose congestion control algorithm applicable to diverse network environments, such as broadband, Starlink, 4G, and 5G. In this section, we explore whether employing LLMs for optimizing domain-specific congestion control, tailored explicitly for certain network types, yields advantages over the generalized approach.

Intuitively, a domain-specific congestion control algorithm should outperform a general-purpose algorithm within its respective network environment. To evaluate the extent of these potential benefits, we analyze all congestion control algorithms generated by the o1-preview model as a case study. First, we identify the algorithm with the highest average utility across all network scenarios. Subsequently, we measure its domain-specific performance separately on the FCC, Starlink, 4G, and 5G datasets, comparing it against the algorithm achieving the highest utility specifically within each domain. The results of this analysis are presented in Table~\ref{tab:domain1}.

Our findings indicate that designing domain-specific algorithms can provide performance gains for all network types, though the magnitude of these gains varies. For FCC, 4G, and 5G networks, the improvements are modest. The best-performing algorithm for 4G and 5G is the same, likely due to their similar wireless network characteristics. In contrast, the Starlink network exhibited notably larger improvements when using a specialized algorithm. We manually inspect the traces in Starlink and find specific behaviors such as abrupt throughput drops linked to low Earth orbit (LEO) satellite handovers. Such distinct characteristics likely contribute to the greater effectiveness of domain-specific congestion control in satellite networks.

Furthermore, we conduct cross-domain evaluations by applying the best-performing algorithm from one network scenario to the others, with results summarized in Table~\ref{tab:domain2}. Our analysis demonstrates that an algorithm optimized specifically for one network type does not necessarily generalize equally well to other scenarios. For instance, the best-performing algorithm for Starlink networks exhibits poor performance when applied to 4G and 5G networks.

In summary, our results support the benefit of developing domain-specific congestion control algorithms, as they consistently offer performance gains compared to generalized algorithms, though the extent of improvement varies by network type. Network scenarios sharing similar dynamics, such as 4G and 5G, can effectively leverage similar algorithmic designs.
\begin{table*}[t]
\begin{tabular}{ccccc}
\toprule
                  & \textbf{Test in FCC} & \textbf{Test in Starlink} & \textbf{Test in 4G} & \textbf{Test in 5G} \\ \midrule
\textbf{Best in FCC}      & 0.324                & 0.395                     & 0.051               & -0.104              \\
\textbf{Best in Starlink} & 0.314                & 0.440                     & -0.411              & -0.608              \\
\textbf{Best in 4G}       & 0.219                & 0.233                     & 0.203               & 0.227               \\
\textbf{Best in 5G}       & 0.219                & 0.233                     & 0.203               & 0.227               \\ \bottomrule
\end{tabular}
\vspace{5pt}
\caption{The utility of evaluating the best sample from one network type in a different network type.}
\label{tab:domain2}
\vspace{-8pt}
\end{table*}

\begin{table}[htbp]
\begin{tabular}{lccccc}
\toprule
\textbf{Iteration}    & \textbf{0} & \textbf{1} & \textbf{2} & \textbf{3} & \textbf{4} \\ \midrule
\textbf{Utility} & 0.257      & 0.247      & 0.274      & 0.274      & 0.277      \\
\textbf{Tput}   & +25.6\%   & +25.5\%   & +27.4\%   & +27.5\%   & +27.5\%   \\
\textbf{Latency}      & -0.01\%    & +0.07\%    & 0.00\%     & +0.01\%    & -0.02\%    \\ \bottomrule
\end{tabular}
\vspace{5pt}
\caption{``Utility'' represents the highest utility in each iteration. ``Tput'' and ``Latency'' indicate the relative change compared with the default BBR for the sample with the highest utility in each iteration.}
\label{tab:iterative}
\vspace{-8pt}
\end{table}

\begin{table}[]
\begin{tabular}{ccc}
\toprule
\textbf{Scenario} & \textbf{Best (specific)} & \textbf{Best (general)} \\
\midrule
FCC               & 0.324                              & 0.309                             \\
Starlink          & 0.440                              & 0.380                             \\
4G                & 0.203                              & 0.162                             \\
5G                & 0.227                              & 0.217                             \\
\bottomrule
\end{tabular}
\vspace{5pt}
\caption{Utility achieved in each specific network type. ``Best (general)'' refers to the single sample with the highest average utility across all network types, while ``Best (specific)'' refers to the domain-specific sample with the highest utility in each network type.}
\label{tab:domain1}
\vspace{-8pt}
\end{table}

\subsection{Iterative Pipeline}
\label{sec:iterative-pipeline}

Previously, our methodology only involves a single round of generating and testing algorithms with LLMs. To further improve performance, we introduce a multi-round iterative pipeline. This approach involves repeatedly refining the algorithms based on their performance in each round. We opt to use the o1-preview model to experiment with iterative refinement. The pipeline works as follows:

\begin{itemize}[itemsep=1pt,topsep=5pt,leftmargin=*]
    \item \textbf{Iteration 0}: We start by generating 3,000 algorithm samples with the o1-preview model. We then select 20 representative network conditions using the efficient evaluation method described in Section~\ref{sec:towards-efficient-evaluation}. After evaluation on these conditions, we rank the algorithms and select the top 100 performers.

    \item \textbf{Iteration 1}: For each of the 100 top algorithms from Iteration 0, we generate 100 new variants using the o1-preview model, resulting in a total of 10,000 new algorithms. These algorithms are evaluated and ranked using the same 20 conditions, and again we select the top 100 performers.

    \item \textbf{Iteration 2}: We repeat this process, generating another 10,000 variants from the top 100 algorithms selected in Iteration 1, followed by evaluation and selection.

    \item This iterative refinement continues similarly in further iterations.
\end{itemize}

We evaluate the selected top 100 algorithms from each iteration across all 408 network conditions to confirm their improvements. Table~\ref{tab:iterative} shows the best utility achieved in each iteration. Through multiple iterations, we observe a relative improvement of approximately 7.8\%, with utility increasing from 0.257 in Iteration 0 to 0.277 by Iteration 4.

The best-performing algorithm from Iteration 4, presented in Appendix~\ref{sec:appendix:best-sample-iterative}, combines several enhancements. These improvements include adaptive gain adjustments based on changes in bandwidth and RTT, and congestion window tuning informed by observed packet loss rates.

One possible reason for the relatively modest improvement is the limitation of our dataset—certain network conditions exhibit stable throughput, providing limited room for further optimization. Improvements stay the same for these stable network conditions in all iterations. Future work may be conducted to explore whether greater improvements are achievable in more dynamic network environments.

\section{Conclusion}
\label{sec:conclusion}

In this paper, we propose a framework that leverages LLMs to automatically optimize congestion control algorithms. The framework first generates a diverse set of high-quality algorithm candidates and then identifies the best ones through emulation experiments. A key challenge in this process is the excessively long emulation time required to evaluate each candidate, which we effectively address through statistically guided sampling. Our approach achieves a performance improvement of 27\% compared with the default BBR algorithm, while reducing emulation time by up to 87.6\%. %
As part of our future work, we plan to apply our approach to optimize other network systems, such as routing, adaptive video bitrate control, and wireless resource allocation.

\bibliographystyle{ACM-Reference-Format}
\bibliography{main}

\appendix

\onecolumn
\section{Appendix}
\label{sec:appendix}

\subsection{The Best Samples}
\label{sec:appendix:best-sample}

\makeatletter
\def\PY@reset{\let\PY@it=\relax \let\PY@bf=\relax%
    \let\PY@ul=\relax \let\PY@tc=\relax%
    \let\PY@bc=\relax \let\PY@ff=\relax}
\def\PY@tok#1{\csname PY@tok@#1\endcsname}
\def\PY@toks#1+{\ifx\relax#1\empty\else%
    \PY@tok{#1}\expandafter\PY@toks\fi}
\def\PY@do#1{\PY@bc{\PY@tc{\PY@ul{%
    \PY@it{\PY@bf{\PY@ff{#1}}}}}}}
\def\PY#1#2{\PY@reset\PY@toks#1+\relax+\PY@do{#2}}

\@namedef{PY@tok@w}{\def\PY@tc##1{\textcolor[rgb]{0.73,0.73,0.73}{##1}}}
\@namedef{PY@tok@c}{\let\PY@it=\textit\def\PY@tc##1{\textcolor[rgb]{0.24,0.48,0.48}{##1}}}
\@namedef{PY@tok@cp}{\def\PY@tc##1{\textcolor[rgb]{0.61,0.40,0.00}{##1}}}
\@namedef{PY@tok@k}{\let\PY@bf=\textbf\def\PY@tc##1{\textcolor[rgb]{0.00,0.50,0.00}{##1}}}
\@namedef{PY@tok@kp}{\def\PY@tc##1{\textcolor[rgb]{0.00,0.50,0.00}{##1}}}
\@namedef{PY@tok@kt}{\def\PY@tc##1{\textcolor[rgb]{0.69,0.00,0.25}{##1}}}
\@namedef{PY@tok@o}{\def\PY@tc##1{\textcolor[rgb]{0.40,0.40,0.40}{##1}}}
\@namedef{PY@tok@ow}{\let\PY@bf=\textbf\def\PY@tc##1{\textcolor[rgb]{0.67,0.13,1.00}{##1}}}
\@namedef{PY@tok@nb}{\def\PY@tc##1{\textcolor[rgb]{0.00,0.50,0.00}{##1}}}
\@namedef{PY@tok@nf}{\def\PY@tc##1{\textcolor[rgb]{0.00,0.00,1.00}{##1}}}
\@namedef{PY@tok@nc}{\let\PY@bf=\textbf\def\PY@tc##1{\textcolor[rgb]{0.00,0.00,1.00}{##1}}}
\@namedef{PY@tok@nn}{\let\PY@bf=\textbf\def\PY@tc##1{\textcolor[rgb]{0.00,0.00,1.00}{##1}}}
\@namedef{PY@tok@ne}{\let\PY@bf=\textbf\def\PY@tc##1{\textcolor[rgb]{0.80,0.25,0.22}{##1}}}
\@namedef{PY@tok@nv}{\def\PY@tc##1{\textcolor[rgb]{0.10,0.09,0.49}{##1}}}
\@namedef{PY@tok@no}{\def\PY@tc##1{\textcolor[rgb]{0.53,0.00,0.00}{##1}}}
\@namedef{PY@tok@nl}{\def\PY@tc##1{\textcolor[rgb]{0.46,0.46,0.00}{##1}}}
\@namedef{PY@tok@ni}{\let\PY@bf=\textbf\def\PY@tc##1{\textcolor[rgb]{0.44,0.44,0.44}{##1}}}
\@namedef{PY@tok@na}{\def\PY@tc##1{\textcolor[rgb]{0.41,0.47,0.13}{##1}}}
\@namedef{PY@tok@nt}{\let\PY@bf=\textbf\def\PY@tc##1{\textcolor[rgb]{0.00,0.50,0.00}{##1}}}
\@namedef{PY@tok@nd}{\def\PY@tc##1{\textcolor[rgb]{0.67,0.13,1.00}{##1}}}
\@namedef{PY@tok@s}{\def\PY@tc##1{\textcolor[rgb]{0.73,0.13,0.13}{##1}}}
\@namedef{PY@tok@sd}{\let\PY@it=\textit\def\PY@tc##1{\textcolor[rgb]{0.73,0.13,0.13}{##1}}}
\@namedef{PY@tok@si}{\let\PY@bf=\textbf\def\PY@tc##1{\textcolor[rgb]{0.64,0.35,0.47}{##1}}}
\@namedef{PY@tok@se}{\let\PY@bf=\textbf\def\PY@tc##1{\textcolor[rgb]{0.67,0.36,0.12}{##1}}}
\@namedef{PY@tok@sr}{\def\PY@tc##1{\textcolor[rgb]{0.64,0.35,0.47}{##1}}}
\@namedef{PY@tok@ss}{\def\PY@tc##1{\textcolor[rgb]{0.10,0.09,0.49}{##1}}}
\@namedef{PY@tok@sx}{\def\PY@tc##1{\textcolor[rgb]{0.00,0.50,0.00}{##1}}}
\@namedef{PY@tok@m}{\def\PY@tc##1{\textcolor[rgb]{0.40,0.40,0.40}{##1}}}
\@namedef{PY@tok@gh}{\let\PY@bf=\textbf\def\PY@tc##1{\textcolor[rgb]{0.00,0.00,0.50}{##1}}}
\@namedef{PY@tok@gu}{\let\PY@bf=\textbf\def\PY@tc##1{\textcolor[rgb]{0.50,0.00,0.50}{##1}}}
\@namedef{PY@tok@gd}{\def\PY@tc##1{\textcolor[rgb]{0.63,0.00,0.00}{##1}}}
\@namedef{PY@tok@gi}{\def\PY@tc##1{\textcolor[rgb]{0.00,0.52,0.00}{##1}}}
\@namedef{PY@tok@gr}{\def\PY@tc##1{\textcolor[rgb]{0.89,0.00,0.00}{##1}}}
\@namedef{PY@tok@ge}{\let\PY@it=\textit}
\@namedef{PY@tok@gs}{\let\PY@bf=\textbf}
\@namedef{PY@tok@gp}{\let\PY@bf=\textbf\def\PY@tc##1{\textcolor[rgb]{0.00,0.00,0.50}{##1}}}
\@namedef{PY@tok@go}{\def\PY@tc##1{\textcolor[rgb]{0.44,0.44,0.44}{##1}}}
\@namedef{PY@tok@gt}{\def\PY@tc##1{\textcolor[rgb]{0.00,0.27,0.87}{##1}}}
\@namedef{PY@tok@err}{\def\PY@bc##1{{\setlength{\fboxsep}{\string -\fboxrule}\fcolorbox[rgb]{1.00,0.00,0.00}{1,1,1}{\strut ##1}}}}
\@namedef{PY@tok@kc}{\let\PY@bf=\textbf\def\PY@tc##1{\textcolor[rgb]{0.00,0.50,0.00}{##1}}}
\@namedef{PY@tok@kd}{\let\PY@bf=\textbf\def\PY@tc##1{\textcolor[rgb]{0.00,0.50,0.00}{##1}}}
\@namedef{PY@tok@kn}{\let\PY@bf=\textbf\def\PY@tc##1{\textcolor[rgb]{0.00,0.50,0.00}{##1}}}
\@namedef{PY@tok@kr}{\let\PY@bf=\textbf\def\PY@tc##1{\textcolor[rgb]{0.00,0.50,0.00}{##1}}}
\@namedef{PY@tok@bp}{\def\PY@tc##1{\textcolor[rgb]{0.00,0.50,0.00}{##1}}}
\@namedef{PY@tok@fm}{\def\PY@tc##1{\textcolor[rgb]{0.00,0.00,1.00}{##1}}}
\@namedef{PY@tok@vc}{\def\PY@tc##1{\textcolor[rgb]{0.10,0.09,0.49}{##1}}}
\@namedef{PY@tok@vg}{\def\PY@tc##1{\textcolor[rgb]{0.10,0.09,0.49}{##1}}}
\@namedef{PY@tok@vi}{\def\PY@tc##1{\textcolor[rgb]{0.10,0.09,0.49}{##1}}}
\@namedef{PY@tok@vm}{\def\PY@tc##1{\textcolor[rgb]{0.10,0.09,0.49}{##1}}}
\@namedef{PY@tok@sa}{\def\PY@tc##1{\textcolor[rgb]{0.73,0.13,0.13}{##1}}}
\@namedef{PY@tok@sb}{\def\PY@tc##1{\textcolor[rgb]{0.73,0.13,0.13}{##1}}}
\@namedef{PY@tok@sc}{\def\PY@tc##1{\textcolor[rgb]{0.73,0.13,0.13}{##1}}}
\@namedef{PY@tok@dl}{\def\PY@tc##1{\textcolor[rgb]{0.73,0.13,0.13}{##1}}}
\@namedef{PY@tok@s2}{\def\PY@tc##1{\textcolor[rgb]{0.73,0.13,0.13}{##1}}}
\@namedef{PY@tok@sh}{\def\PY@tc##1{\textcolor[rgb]{0.73,0.13,0.13}{##1}}}
\@namedef{PY@tok@s1}{\def\PY@tc##1{\textcolor[rgb]{0.73,0.13,0.13}{##1}}}
\@namedef{PY@tok@mb}{\def\PY@tc##1{\textcolor[rgb]{0.40,0.40,0.40}{##1}}}
\@namedef{PY@tok@mf}{\def\PY@tc##1{\textcolor[rgb]{0.40,0.40,0.40}{##1}}}
\@namedef{PY@tok@mh}{\def\PY@tc##1{\textcolor[rgb]{0.40,0.40,0.40}{##1}}}
\@namedef{PY@tok@mi}{\def\PY@tc##1{\textcolor[rgb]{0.40,0.40,0.40}{##1}}}
\@namedef{PY@tok@il}{\def\PY@tc##1{\textcolor[rgb]{0.40,0.40,0.40}{##1}}}
\@namedef{PY@tok@mo}{\def\PY@tc##1{\textcolor[rgb]{0.40,0.40,0.40}{##1}}}
\@namedef{PY@tok@ch}{\let\PY@it=\textit\def\PY@tc##1{\textcolor[rgb]{0.24,0.48,0.48}{##1}}}
\@namedef{PY@tok@cm}{\let\PY@it=\textit\def\PY@tc##1{\textcolor[rgb]{0.24,0.48,0.48}{##1}}}
\@namedef{PY@tok@cpf}{\let\PY@it=\textit\def\PY@tc##1{\textcolor[rgb]{0.24,0.48,0.48}{##1}}}
\@namedef{PY@tok@c1}{\let\PY@it=\textit\def\PY@tc##1{\textcolor[rgb]{0.24,0.48,0.48}{##1}}}
\@namedef{PY@tok@cs}{\let\PY@it=\textit\def\PY@tc##1{\textcolor[rgb]{0.24,0.48,0.48}{##1}}}

\def\PYZbs{\char`\\}
\def\PYZus{\char`\_}
\def\PYZob{\char`\{}
\def\PYZcb{\char`\}}
\def\PYZca{\char`\^}
\def\PYZam{\char`\&}
\def\PYZlt{\char`\<}
\def\PYZgt{\char`\>}
\def\PYZsh{\char`\#}
\def\PYZpc{\char`\%}
\def\PYZdl{\char`\$}
\def\PYZhy{\char`\-}
\def\PYZsq{\char`\'}
\def\PYZdq{\char`\"}
\def\PYZti{\char`\~}
\def\PYZat{@}
\def\PYZlb{[}
\def\PYZrb{]}
\makeatother

\subsubsection{The Best Sample from o1-preview} 

\hfill\\
The struct definition of \textit{QUIC\_CONGESTION\_CONTROL\_BBR} in bbr.h is updated. Besides, the function \textit{BbrCongestionControlOnDataSent}, \textit{BbrCongestionControlOnDataLost}, \textit{BbrCongestionControlUpdateCongestionWindow}, 
 and \textit{BbrCongestionControlSetSendQuantum} are updated.

\begin{Verbatim}[commandchars=\\\{\},numbers=left,firstnumber=1,stepnumber=1]
\colorbox{white}{\parbox{\linewidth}{\PY{o}{\PYZhy{}}\PY{o}{\PYZhy{}}\PY{o}{\PYZhy{}}\PY{+w}{ }\PY{n}{bbr}\PY{p}{.}\PY{n}{h}\vphantom{fg}}}
\colorbox{white}{\parbox{\linewidth}{\PY{o}{+}\PY{o}{+}\PY{o}{+}\PY{+w}{ }\PY{n}{msquic}\PY{o}{/}\PY{n}{src}\PY{o}{/}\PY{n}{core}\PY{o}{/}\PY{n}{bbr}\PY{p}{.}\PY{n}{h}\vphantom{fg}}}
\colorbox{white}{\parbox{\linewidth}{\PY{err}{@}\PY{err}{@}\PY{+w}{ }\PY{l+m+mi}{\PYZhy{}212}\PY{p}{,}\PY{l+m+mi}{6}\PY{+w}{ }\PY{o}{+}\PY{l+m+mi}{212}\PY{p}{,}\PY{l+m+mi}{13}\PY{+w}{ }\PY{err}{@}\PY{err}{@}\vphantom{fg}}}
\colorbox{white}{\parbox{\linewidth}{\PY{+w}{   }\PY{c+c1}{// BBR estimates maximum bandwidth by the maximum recent bandwidth}\vphantom{fg}}}
\colorbox{white}{\parbox{\linewidth}{\PY{+w}{   }\PY{c+c1}{//}\vphantom{fg}}}
\colorbox{white}{\parbox{\linewidth}{\PY{+w}{   }\PY{n}{BBR\PYZus{}BANDWIDTH\PYZus{}FILTER}\PY{+w}{ }\PY{n}{BandwidthFilter}\PY{p}{;}\vphantom{fg}}}
\colorbox{PaleGreen}{\parbox{\linewidth}{\vphantom{fg}}}
\colorbox{PaleGreen}{\parbox{\linewidth}{\PY{+w}{    }\PY{c+c1}{//}\vphantom{fg}}}
\colorbox{PaleGreen}{\parbox{\linewidth}{\PY{+w}{    }\PY{c+c1}{// Variables for loss\PYZhy{}aware gain adjustment}\vphantom{fg}}}
\colorbox{PaleGreen}{\parbox{\linewidth}{\PY{+w}{    }\PY{c+c1}{//}\vphantom{fg}}}
\colorbox{PaleGreen}{\parbox{\linewidth}{\PY{+w}{    }\PY{k+kt}{uint64\PYZus{}t}\PY{+w}{ }\PY{n}{TotalLostBytes}\PY{p}{;}\vphantom{fg}}}
\colorbox{PaleGreen}{\parbox{\linewidth}{\PY{+w}{    }\PY{k+kt}{uint64\PYZus{}t}\PY{+w}{ }\PY{n}{TotalSentBytes}\PY{p}{;}\vphantom{fg}}}
\colorbox{PaleGreen}{\parbox{\linewidth}{\PY{+w}{    }\PY{k+kt}{double}\PY{+w}{ }\PY{n}{LossRate}\PY{p}{;}\vphantom{fg}}}
\colorbox{white}{\parbox{\linewidth}{\PY{+w}{ }\vphantom{fg}}}
\colorbox{white}{\parbox{\linewidth}{\PY{+w}{ }\PY{p}{\PYZcb{}}\PY{+w}{ }\PY{n}{QUIC\PYZus{}CONGESTION\PYZus{}CONTROL\PYZus{}BBR}\PY{p}{;}\vphantom{fg}}}
\colorbox{white}{\parbox{\linewidth}{\PY{+w}{ }\vphantom{fg}}}
\end{Verbatim}

\begin{Verbatim}[commandchars=\\\{\},numbers=left,firstnumber=1,stepnumber=1]
\colorbox{white}{\parbox{\linewidth}{\PY{o}{\PYZhy{}}\PY{o}{\PYZhy{}}\PY{o}{\PYZhy{}}\PY{+w}{ }\PY{n}{bbr}\PY{p}{.}\PY{n}{c}\vphantom{fg}}}
\colorbox{white}{\parbox{\linewidth}{\PY{o}{+}\PY{o}{+}\PY{o}{+}\PY{+w}{ }\PY{n}{msquic}\PY{o}{/}\PY{n}{src}\PY{o}{/}\PY{n}{core}\PY{o}{/}\PY{n}{bbr}\PY{p}{.}\PY{n}{c}\vphantom{fg}}}
\colorbox{white}{\parbox{\linewidth}{\PY{err}{@}\PY{err}{@}\PY{+w}{ }\PY{l+m+mi}{\PYZhy{}443}\PY{p}{,}\PY{l+m+mi}{6}\PY{+w}{ }\PY{o}{+}\PY{l+m+mi}{443}\PY{p}{,}\PY{l+m+mi}{11}\PY{+w}{ }\PY{err}{@}\PY{err}{@}\vphantom{fg}}}
\colorbox{white}{\parbox{\linewidth}{\PY{+w}{       }\PY{o}{\PYZhy{}}\PY{o}{\PYZhy{}}\PY{n}{Bbr}\PY{o}{\PYZhy{}}\PY{o}{\PYZgt{}}\PY{n}{Exemptions}\PY{p}{;}\vphantom{fg}}}
\colorbox{white}{\parbox{\linewidth}{\PY{+w}{   }\PY{p}{\PYZcb{}}\vphantom{fg}}}
\colorbox{white}{\parbox{\linewidth}{\PY{+w}{ }\vphantom{fg}}}
\colorbox{PaleGreen}{\parbox{\linewidth}{\PY{+w}{    }\PY{c+c1}{//}\vphantom{fg}}}
\colorbox{PaleGreen}{\parbox{\linewidth}{\PY{+w}{    }\PY{c+c1}{// Update TotalSentBytes for loss\PYZhy{}aware gain adjustment}\vphantom{fg}}}
\colorbox{PaleGreen}{\parbox{\linewidth}{\PY{+w}{    }\PY{c+c1}{//}\vphantom{fg}}}
\colorbox{PaleGreen}{\parbox{\linewidth}{\PY{+w}{    }\PY{n}{Bbr}\PY{o}{\PYZhy{}}\PY{o}{\PYZgt{}}\PY{n}{TotalSentBytes}\PY{+w}{ }\PY{o}{=}\PY{+w}{ }\PY{n}{NumRetransmittableBytes}\PY{p}{;}\vphantom{fg}}}
\colorbox{PaleGreen}{\parbox{\linewidth}{\vphantom{fg}}}
\colorbox{white}{\parbox{\linewidth}{\PY{+w}{   }\PY{n}{BbrCongestionControlUpdateBlockedState}\PY{p}{(}\PY{n}{Cc}\PY{p}{,}\PY{+w}{ }\PY{n}{PreviousCanSendState}\PY{p}{)}\PY{p}{;}\vphantom{fg}}}
\colorbox{white}{\parbox{\linewidth}{\PY{+w}{ }\PY{p}{\PYZcb{}}\vphantom{fg}}}
\colorbox{white}{\parbox{\linewidth}{\PY{+w}{ }\vphantom{fg}}}
\colorbox{white}{\parbox{\linewidth}{\PY{err}{@}\PY{err}{@}\PY{+w}{ }\PY{l+m+mi}{\PYZhy{}690}\PY{p}{,}\PY{l+m+mi}{12}\PY{+w}{ }\PY{o}{+}\PY{l+m+mi}{695}\PY{p}{,}\PY{l+m+mi}{27}\PY{+w}{ }\PY{err}{@}\PY{err}{@}\vphantom{fg}}}
\colorbox{white}{\parbox{\linewidth}{\PY{+w}{ }\PY{k+kt}{void}\vphantom{fg}}}
\colorbox{white}{\parbox{\linewidth}{\PY{+w}{ }\PY{n}{BbrCongestionControlSetSendQuantum}\PY{p}{(}\vphantom{fg}}}
\colorbox{white}{\parbox{\linewidth}{\PY{+w}{   }\PY{n}{\PYZus{}In\PYZus{}}\PY{+w}{ }\PY{n}{QUIC\PYZus{}CONGESTION\PYZus{}CONTROL}\PY{o}{*}\PY{+w}{ }\PY{n}{Cc}\vphantom{fg}}}
\colorbox{LightPink}{\parbox{\linewidth}{\PY{p}{)}\vphantom{fg}}}
\colorbox{PaleGreen}{\parbox{\linewidth}{\PY{+w}{    }\PY{p}{)}\vphantom{fg}}}
\colorbox{white}{\parbox{\linewidth}{\PY{+w}{ }\PY{p}{\PYZob{}}\vphantom{fg}}}
\colorbox{white}{\parbox{\linewidth}{\PY{+w}{   }\PY{n}{QUIC\PYZus{}CONGESTION\PYZus{}CONTROL\PYZus{}BBR}\PY{+w}{ }\PY{o}{*}\PY{n}{Bbr}\PY{+w}{ }\PY{o}{=}\PY{+w}{ }\PY{o}{\PYZam{}}\PY{n}{Cc}\PY{o}{\PYZhy{}}\PY{o}{\PYZgt{}}\PY{n}{Bbr}\PY{p}{;}\vphantom{fg}}}
\colorbox{white}{\parbox{\linewidth}{\PY{+w}{   }\PY{n}{QUIC\PYZus{}CONNECTION}\PY{o}{*}\PY{+w}{ }\PY{n}{Connection}\PY{+w}{ }\PY{o}{=}\PY{+w}{ }\PY{n}{QuicCongestionControlGetConnection}\PY{p}{(}\PY{n}{Cc}\PY{p}{)}\PY{p}{;}\vphantom{fg}}}
\colorbox{white}{\parbox{\linewidth}{\PY{+w}{ }\vphantom{fg}}}
\colorbox{white}{\parbox{\linewidth}{\PY{+w}{   }\PY{k+kt}{uint64\PYZus{}t}\PY{+w}{ }\PY{n}{Bandwidth}\PY{+w}{ }\PY{o}{=}\PY{+w}{ }\PY{n}{BbrCongestionControlGetBandwidth}\PY{p}{(}\PY{n}{Cc}\PY{p}{)}\PY{p}{;}\vphantom{fg}}}
\colorbox{PaleGreen}{\parbox{\linewidth}{\vphantom{fg}}}
\colorbox{PaleGreen}{\parbox{\linewidth}{\PY{+w}{    }\PY{c+c1}{// Update LossRate}\vphantom{fg}}}
\colorbox{PaleGreen}{\parbox{\linewidth}{\PY{+w}{    }\PY{k}{if}\PY{+w}{ }\PY{p}{(}\PY{n}{Bbr}\PY{o}{\PYZhy{}}\PY{o}{\PYZgt{}}\PY{n}{TotalSentBytes}\PY{+w}{ }\PY{o}{\PYZgt{}}\PY{+w}{ }\PY{l+m+mi}{0}\PY{p}{)}\PY{+w}{ }\PY{p}{\PYZob{}}\vphantom{fg}}}
\colorbox{PaleGreen}{\parbox{\linewidth}{\PY{+w}{        }\PY{n}{Bbr}\PY{o}{\PYZhy{}}\PY{o}{\PYZgt{}}\PY{n}{LossRate}\PY{+w}{ }\PY{o}{=}\PY{+w}{ }\PY{p}{(}\PY{k+kt}{double}\PY{p}{)}\PY{n}{Bbr}\PY{o}{\PYZhy{}}\PY{o}{\PYZgt{}}\PY{n}{TotalLostBytes}\PY{+w}{ }\PY{o}{/}\PY{+w}{ }\PY{p}{(}\PY{k+kt}{double}\PY{p}{)}\PY{n}{Bbr}\PY{o}{\PYZhy{}}\PY{o}{\PYZgt{}}\PY{n}{TotalSentBytes}\PY{p}{;}\vphantom{fg}}}
\colorbox{PaleGreen}{\parbox{\linewidth}{\PY{+w}{    }\PY{p}{\PYZcb{}}\PY{+w}{ }\PY{k}{else}\PY{+w}{ }\PY{p}{\PYZob{}}\vphantom{fg}}}
\colorbox{PaleGreen}{\parbox{\linewidth}{\PY{+w}{        }\PY{n}{Bbr}\PY{o}{\PYZhy{}}\PY{o}{\PYZgt{}}\PY{n}{LossRate}\PY{+w}{ }\PY{o}{=}\PY{+w}{ }\PY{l+m+mf}{0.0}\PY{p}{;}\vphantom{fg}}}
\colorbox{PaleGreen}{\parbox{\linewidth}{\PY{+w}{    }\PY{p}{\PYZcb{}}\vphantom{fg}}}
\colorbox{PaleGreen}{\parbox{\linewidth}{\vphantom{fg}}}
\colorbox{PaleGreen}{\parbox{\linewidth}{\PY{+w}{    }\PY{c+c1}{// Adjust PacingGain based on LossRate}\vphantom{fg}}}
\colorbox{PaleGreen}{\parbox{\linewidth}{\PY{+w}{    }\PY{k}{if}\PY{+w}{ }\PY{p}{(}\PY{n}{Bbr}\PY{o}{\PYZhy{}}\PY{o}{\PYZgt{}}\PY{n}{LossRate}\PY{+w}{ }\PY{o}{\PYZgt{}}\PY{+w}{ }\PY{l+m+mf}{0.0}\PY{p}{)}\PY{+w}{ }\PY{p}{\PYZob{}}\vphantom{fg}}}
\colorbox{PaleGreen}{\parbox{\linewidth}{\PY{+w}{        }\PY{n}{Bbr}\PY{o}{\PYZhy{}}\PY{o}{\PYZgt{}}\PY{n}{PacingGain}\PY{+w}{ }\PY{o}{=}\PY{+w}{ }\PY{p}{(}\PY{k+kt}{uint32\PYZus{}t}\PY{p}{)}\PY{p}{(}\PY{n}{Bbr}\PY{o}{\PYZhy{}}\PY{o}{\PYZgt{}}\PY{n}{PacingGain}\PY{+w}{ }\PY{o}{*}\PY{+w}{ }\PY{p}{(}\PY{l+m+mf}{1.0}\PY{+w}{ }\PY{o}{\PYZhy{}}\PY{+w}{ }\PY{n}{Bbr}\PY{o}{\PYZhy{}}\PY{o}{\PYZgt{}}\PY{n}{LossRate}\PY{p}{)}\PY{p}{)}\PY{p}{;}\vphantom{fg}}}
\colorbox{PaleGreen}{\parbox{\linewidth}{\PY{+w}{        }\PY{k}{if}\PY{+w}{ }\PY{p}{(}\PY{n}{Bbr}\PY{o}{\PYZhy{}}\PY{o}{\PYZgt{}}\PY{n}{PacingGain}\PY{+w}{ }\PY{o}{\PYZlt{}}\PY{+w}{ }\PY{n}{GAIN\PYZus{}UNIT}\PY{+w}{ }\PY{o}{/}\PY{+w}{ }\PY{l+m+mi}{2}\PY{p}{)}\PY{+w}{ }\PY{p}{\PYZob{}}\vphantom{fg}}}
\colorbox{PaleGreen}{\parbox{\linewidth}{\PY{+w}{            }\PY{n}{Bbr}\PY{o}{\PYZhy{}}\PY{o}{\PYZgt{}}\PY{n}{PacingGain}\PY{+w}{ }\PY{o}{=}\PY{+w}{ }\PY{n}{GAIN\PYZus{}UNIT}\PY{+w}{ }\PY{o}{/}\PY{+w}{ }\PY{l+m+mi}{2}\PY{p}{;}\PY{+w}{ }\PY{c+c1}{// Minimum PacingGain}\vphantom{fg}}}
\colorbox{PaleGreen}{\parbox{\linewidth}{\PY{+w}{        }\PY{p}{\PYZcb{}}\vphantom{fg}}}
\colorbox{PaleGreen}{\parbox{\linewidth}{\PY{+w}{    }\PY{p}{\PYZcb{}}\vphantom{fg}}}
\colorbox{white}{\parbox{\linewidth}{\PY{+w}{ }\vphantom{fg}}}
\colorbox{white}{\parbox{\linewidth}{\PY{+w}{   }\PY{k+kt}{uint64\PYZus{}t}\PY{+w}{ }\PY{n}{PacingRate}\PY{+w}{ }\PY{o}{=}\PY{+w}{ }\PY{n}{Bandwidth}\PY{+w}{ }\PY{o}{*}\PY{+w}{ }\PY{n}{Bbr}\PY{o}{\PYZhy{}}\PY{o}{\PYZgt{}}\PY{n}{PacingGain}\PY{+w}{ }\PY{o}{/}\PY{+w}{ }\PY{n}{GAIN\PYZus{}UNIT}\PY{p}{;}\vphantom{fg}}}
\colorbox{white}{\parbox{\linewidth}{\PY{+w}{ }\vphantom{fg}}}
\colorbox{white}{\parbox{\linewidth}{\PY{err}{@}\PY{err}{@}\PY{+w}{ }\PY{l+m+mi}{\PYZhy{}724}\PY{p}{,}\PY{l+m+mi}{6}\PY{+w}{ }\PY{o}{+}\PY{l+m+mi}{744}\PY{p}{,}\PY{l+m+mi}{23}\PY{+w}{ }\PY{err}{@}\PY{err}{@}\vphantom{fg}}}
\colorbox{white}{\parbox{\linewidth}{\PY{+w}{ }\vphantom{fg}}}
\colorbox{white}{\parbox{\linewidth}{\PY{+w}{   }\PY{k}{if}\PY{+w}{ }\PY{p}{(}\PY{n}{Bbr}\PY{o}{\PYZhy{}}\PY{o}{\PYZgt{}}\PY{n}{BbrState}\PY{+w}{ }\PY{o}{=}\PY{o}{=}\PY{+w}{ }\PY{n}{BBR\PYZus{}STATE\PYZus{}PROBE\PYZus{}RTT}\PY{p}{)}\PY{+w}{ }\PY{p}{\PYZob{}}\vphantom{fg}}}
\colorbox{white}{\parbox{\linewidth}{\PY{+w}{       }\PY{k}{return}\PY{p}{;}\vphantom{fg}}}
\colorbox{PaleGreen}{\parbox{\linewidth}{\PY{+w}{    }\PY{p}{\PYZcb{}}\vphantom{fg}}}
\colorbox{PaleGreen}{\parbox{\linewidth}{\vphantom{fg}}}
\colorbox{PaleGreen}{\parbox{\linewidth}{\PY{+w}{    }\PY{c+c1}{// Update LossRate}\vphantom{fg}}}
\colorbox{PaleGreen}{\parbox{\linewidth}{\PY{+w}{    }\PY{k}{if}\PY{+w}{ }\PY{p}{(}\PY{n}{Bbr}\PY{o}{\PYZhy{}}\PY{o}{\PYZgt{}}\PY{n}{TotalSentBytes}\PY{+w}{ }\PY{o}{\PYZgt{}}\PY{+w}{ }\PY{l+m+mi}{0}\PY{p}{)}\PY{+w}{ }\PY{p}{\PYZob{}}\vphantom{fg}}}
\colorbox{PaleGreen}{\parbox{\linewidth}{\PY{+w}{        }\PY{n}{Bbr}\PY{o}{\PYZhy{}}\PY{o}{\PYZgt{}}\PY{n}{LossRate}\PY{+w}{ }\PY{o}{=}\PY{+w}{ }\PY{p}{(}\PY{k+kt}{double}\PY{p}{)}\PY{n}{Bbr}\PY{o}{\PYZhy{}}\PY{o}{\PYZgt{}}\PY{n}{TotalLostBytes}\PY{+w}{ }\PY{o}{/}\PY{+w}{ }\PY{p}{(}\PY{k+kt}{double}\PY{p}{)}\PY{n}{Bbr}\PY{o}{\PYZhy{}}\PY{o}{\PYZgt{}}\PY{n}{TotalSentBytes}\PY{p}{;}\vphantom{fg}}}
\colorbox{PaleGreen}{\parbox{\linewidth}{\PY{+w}{    }\PY{p}{\PYZcb{}}\PY{+w}{ }\PY{k}{else}\PY{+w}{ }\PY{p}{\PYZob{}}\vphantom{fg}}}
\colorbox{PaleGreen}{\parbox{\linewidth}{\PY{+w}{        }\PY{n}{Bbr}\PY{o}{\PYZhy{}}\PY{o}{\PYZgt{}}\PY{n}{LossRate}\PY{+w}{ }\PY{o}{=}\PY{+w}{ }\PY{l+m+mf}{0.0}\PY{p}{;}\vphantom{fg}}}
\colorbox{PaleGreen}{\parbox{\linewidth}{\PY{+w}{    }\PY{p}{\PYZcb{}}\vphantom{fg}}}
\colorbox{PaleGreen}{\parbox{\linewidth}{\vphantom{fg}}}
\colorbox{PaleGreen}{\parbox{\linewidth}{\PY{+w}{    }\PY{c+c1}{// Adjust CwndGain based on LossRate}\vphantom{fg}}}
\colorbox{PaleGreen}{\parbox{\linewidth}{\PY{+w}{    }\PY{k}{if}\PY{+w}{ }\PY{p}{(}\PY{n}{Bbr}\PY{o}{\PYZhy{}}\PY{o}{\PYZgt{}}\PY{n}{LossRate}\PY{+w}{ }\PY{o}{\PYZgt{}}\PY{+w}{ }\PY{l+m+mf}{0.0}\PY{p}{)}\PY{+w}{ }\PY{p}{\PYZob{}}\vphantom{fg}}}
\colorbox{PaleGreen}{\parbox{\linewidth}{\PY{+w}{        }\PY{n}{Bbr}\PY{o}{\PYZhy{}}\PY{o}{\PYZgt{}}\PY{n}{CwndGain}\PY{+w}{ }\PY{o}{=}\PY{+w}{ }\PY{p}{(}\PY{k+kt}{uint32\PYZus{}t}\PY{p}{)}\PY{p}{(}\PY{n}{GAIN\PYZus{}UNIT}\PY{+w}{ }\PY{o}{*}\PY{+w}{ }\PY{p}{(}\PY{l+m+mf}{1.0}\PY{+w}{ }\PY{o}{\PYZhy{}}\PY{+w}{ }\PY{n}{Bbr}\PY{o}{\PYZhy{}}\PY{o}{\PYZgt{}}\PY{n}{LossRate}\PY{p}{)}\PY{p}{)}\PY{p}{;}\vphantom{fg}}}
\colorbox{PaleGreen}{\parbox{\linewidth}{\PY{+w}{        }\PY{k}{if}\PY{+w}{ }\PY{p}{(}\PY{n}{Bbr}\PY{o}{\PYZhy{}}\PY{o}{\PYZgt{}}\PY{n}{CwndGain}\PY{+w}{ }\PY{o}{\PYZlt{}}\PY{+w}{ }\PY{n}{GAIN\PYZus{}UNIT}\PY{+w}{ }\PY{o}{/}\PY{+w}{ }\PY{l+m+mi}{2}\PY{p}{)}\PY{+w}{ }\PY{p}{\PYZob{}}\vphantom{fg}}}
\colorbox{PaleGreen}{\parbox{\linewidth}{\PY{+w}{            }\PY{n}{Bbr}\PY{o}{\PYZhy{}}\PY{o}{\PYZgt{}}\PY{n}{CwndGain}\PY{+w}{ }\PY{o}{=}\PY{+w}{ }\PY{n}{GAIN\PYZus{}UNIT}\PY{+w}{ }\PY{o}{/}\PY{+w}{ }\PY{l+m+mi}{2}\PY{p}{;}\PY{+w}{ }\PY{c+c1}{// Minimum CwndGain}\vphantom{fg}}}
\colorbox{PaleGreen}{\parbox{\linewidth}{\PY{+w}{        }\PY{p}{\PYZcb{}}\vphantom{fg}}}
\colorbox{PaleGreen}{\parbox{\linewidth}{\PY{+w}{    }\PY{p}{\PYZcb{}}\PY{+w}{ }\PY{k}{else}\PY{+w}{ }\PY{p}{\PYZob{}}\vphantom{fg}}}
\colorbox{PaleGreen}{\parbox{\linewidth}{\PY{+w}{        }\PY{n}{Bbr}\PY{o}{\PYZhy{}}\PY{o}{\PYZgt{}}\PY{n}{CwndGain}\PY{+w}{ }\PY{o}{=}\PY{+w}{ }\PY{n}{kCwndGain}\PY{p}{;}\vphantom{fg}}}
\colorbox{white}{\parbox{\linewidth}{\PY{+w}{   }\PY{p}{\PYZcb{}}\vphantom{fg}}}
\colorbox{white}{\parbox{\linewidth}{\PY{+w}{ }\vphantom{fg}}}
\colorbox{white}{\parbox{\linewidth}{\PY{+w}{   }\PY{k}{const}\PY{+w}{ }\PY{k+kt}{uint16\PYZus{}t}\PY{+w}{ }\PY{n}{DatagramPayloadLength}\PY{+w}{ }\PY{o}{=}\vphantom{fg}}}
\colorbox{white}{\parbox{\linewidth}{\PY{err}{@}\PY{err}{@}\PY{+w}{ }\PY{l+m+mi}{\PYZhy{}918}\PY{p}{,}\PY{l+m+mi}{6}\PY{+w}{ }\PY{o}{+}\PY{l+m+mi}{955}\PY{p}{,}\PY{l+m+mi}{11}\PY{+w}{ }\PY{err}{@}\PY{err}{@}\vphantom{fg}}}
\colorbox{white}{\parbox{\linewidth}{\PY{+w}{ }\vphantom{fg}}}
\colorbox{white}{\parbox{\linewidth}{\PY{+w}{   }\PY{n}{CXPLAT\PYZus{}DBG\PYZus{}ASSERT}\PY{p}{(}\PY{n}{Bbr}\PY{o}{\PYZhy{}}\PY{o}{\PYZgt{}}\PY{n}{BytesInFlight}\PY{+w}{ }\PY{o}{\PYZgt{}}\PY{o}{=}\PY{+w}{ }\PY{n}{LossEvent}\PY{o}{\PYZhy{}}\PY{o}{\PYZgt{}}\PY{n}{NumRetransmittableBytes}\PY{p}{)}\PY{p}{;}\vphantom{fg}}}
\colorbox{white}{\parbox{\linewidth}{\PY{+w}{   }\PY{n}{Bbr}\PY{o}{\PYZhy{}}\PY{o}{\PYZgt{}}\PY{n}{BytesInFlight}\PY{+w}{ }\PY{o}{\PYZhy{}}\PY{o}{=}\PY{+w}{ }\PY{n}{LossEvent}\PY{o}{\PYZhy{}}\PY{o}{\PYZgt{}}\PY{n}{NumRetransmittableBytes}\PY{p}{;}\vphantom{fg}}}
\colorbox{PaleGreen}{\parbox{\linewidth}{\vphantom{fg}}}
\colorbox{PaleGreen}{\parbox{\linewidth}{\PY{+w}{    }\PY{c+c1}{//}\vphantom{fg}}}
\colorbox{PaleGreen}{\parbox{\linewidth}{\PY{+w}{    }\PY{c+c1}{// Update TotalLostBytes for loss\PYZhy{}aware gain adjustment}\vphantom{fg}}}
\colorbox{PaleGreen}{\parbox{\linewidth}{\PY{+w}{    }\PY{c+c1}{//}\vphantom{fg}}}
\colorbox{PaleGreen}{\parbox{\linewidth}{\PY{+w}{    }\PY{n}{Bbr}\PY{o}{\PYZhy{}}\PY{o}{\PYZgt{}}\PY{n}{TotalLostBytes}\PY{+w}{ }\PY{o}{=}\PY{+w}{ }\PY{n}{LossEvent}\PY{o}{\PYZhy{}}\PY{o}{\PYZgt{}}\PY{n}{NumRetransmittableBytes}\PY{p}{;}\vphantom{fg}}}
\colorbox{white}{\parbox{\linewidth}{\PY{+w}{ }\vphantom{fg}}}
\colorbox{white}{\parbox{\linewidth}{\PY{+w}{   }\PY{k+kt}{uint32\PYZus{}t}\PY{+w}{ }\PY{n}{RecoveryWindow}\PY{+w}{ }\PY{o}{=}\PY{+w}{ }\PY{n}{Bbr}\PY{o}{\PYZhy{}}\PY{o}{\PYZgt{}}\PY{n}{RecoveryWindow}\PY{p}{;}\vphantom{fg}}}
\colorbox{white}{\parbox{\linewidth}{\PY{+w}{   }\PY{k+kt}{uint32\PYZus{}t}\PY{+w}{ }\PY{n}{MinCongestionWindow}\PY{+w}{ }\PY{o}{=}\PY{+w}{ }\PY{n}{kMinCwndInMss}\PY{+w}{ }\PY{o}{*}\PY{+w}{ }\PY{n}{DatagramPayloadLength}\PY{p}{;}\vphantom{fg}}}
\end{Verbatim}

\subsubsection{The Best Sample from GPT-4o}

\hfill\\
The function \textit{BbrCongestionControlUpdateCongestionWindow} is modified:

\begin{Verbatim}[commandchars=\\\{\},numbers=left,firstnumber=1,stepnumber=1]
\colorbox{white}{\parbox{\linewidth}{\PY{o}{\PYZhy{}}\PY{o}{\PYZhy{}}\PY{o}{\PYZhy{}}\PY{+w}{ }\PY{n}{bbr}\PY{p}{.}\PY{n}{c}\vphantom{fg}}}
\colorbox{white}{\parbox{\linewidth}{\PY{o}{+}\PY{o}{+}\PY{o}{+}\PY{+w}{ }\PY{n}{msquic}\PY{o}{/}\PY{n}{src}\PY{o}{/}\PY{n}{core}\PY{o}{/}\PY{n}{bbr}\PY{p}{.}\PY{n}{c}\vphantom{fg}}}
\colorbox{white}{\parbox{\linewidth}{\PY{err}{@}\PY{err}{@}\PY{+w}{ }\PY{l+m+mi}{\PYZhy{}732}\PY{p}{,}\PY{l+m+mi}{6}\PY{+w}{ }\PY{o}{+}\PY{l+m+mi}{732}\PY{p}{,}\PY{l+m+mi}{24}\PY{+w}{ }\PY{err}{@}\PY{err}{@}\vphantom{fg}}}
\colorbox{white}{\parbox{\linewidth}{\PY{+w}{   }\PY{n}{BbrCongestionControlSetSendQuantum}\PY{p}{(}\PY{n}{Cc}\PY{p}{)}\PY{p}{;}\vphantom{fg}}}
\colorbox{white}{\parbox{\linewidth}{\PY{+w}{ }\vphantom{fg}}}
\colorbox{white}{\parbox{\linewidth}{\PY{+w}{   }\PY{k+kt}{uint64\PYZus{}t}\PY{+w}{ }\PY{n}{TargetCwnd}\PY{+w}{ }\PY{o}{=}\PY{+w}{ }\PY{n}{BbrCongestionControlGetTargetCwnd}\PY{p}{(}\PY{n}{Cc}\PY{p}{,}\PY{+w}{ }\PY{n}{Bbr}\PY{o}{\PYZhy{}}\PY{o}{\PYZgt{}}\PY{n}{CwndGain}\PY{p}{)}\PY{p}{;}\vphantom{fg}}}
\colorbox{PaleGreen}{\parbox{\linewidth}{\vphantom{fg}}}
\colorbox{PaleGreen}{\parbox{\linewidth}{\PY{+w}{    }\PY{c+c1}{// Adaptive adjustment of CwndGain and PacingGain based on network conditions}\vphantom{fg}}}
\colorbox{PaleGreen}{\parbox{\linewidth}{\PY{+w}{    }\PY{k}{if}\PY{+w}{ }\PY{p}{(}\PY{n}{Bbr}\PY{o}{\PYZhy{}}\PY{o}{\PYZgt{}}\PY{n}{RttSampleExpired}\PY{+w}{ }\PY{o}{|}\PY{o}{|}\PY{+w}{ }\PY{n}{Bbr}\PY{o}{\PYZhy{}}\PY{o}{\PYZgt{}}\PY{n}{MinRttTimestampValid}\PY{p}{)}\PY{+w}{ }\PY{p}{\PYZob{}}\vphantom{fg}}}
\colorbox{PaleGreen}{\parbox{\linewidth}{\PY{+w}{        }\PY{k+kt}{uint64\PYZus{}t}\PY{+w}{ }\PY{n}{BandwidthEst}\PY{+w}{ }\PY{o}{=}\PY{+w}{ }\PY{n}{BbrCongestionControlGetBandwidth}\PY{p}{(}\PY{n}{Cc}\PY{p}{)}\PY{p}{;}\vphantom{fg}}}
\colorbox{PaleGreen}{\parbox{\linewidth}{\PY{+w}{        }\PY{k}{if}\PY{+w}{ }\PY{p}{(}\PY{n}{BandwidthEst}\PY{+w}{ }\PY{o}{\PYZgt{}}\PY{+w}{ }\PY{l+m+mi}{0}\PY{+w}{ }\PY{o}{\PYZam{}}\PY{o}{\PYZam{}}\PY{+w}{ }\PY{n}{AckedBytes}\PY{+w}{ }\PY{o}{\PYZgt{}}\PY{+w}{ }\PY{l+m+mi}{0}\PY{p}{)}\PY{+w}{ }\PY{p}{\PYZob{}}\vphantom{fg}}}
\colorbox{PaleGreen}{\parbox{\linewidth}{\PY{+w}{            }\PY{k+kt}{uint64\PYZus{}t}\PY{+w}{ }\PY{n}{ThroughputEstimate}\PY{+w}{ }\PY{o}{=}\PY{+w}{ }\PY{n}{TotalBytesAcked}\PY{+w}{ }\PY{o}{*}\PY{+w}{ }\PY{n}{kMicroSecsInSec}\PY{+w}{ }\PY{o}{/}\PY{+w}{ }\PY{n}{Bbr}\PY{o}{\PYZhy{}}\PY{o}{\PYZgt{}}\PY{n}{MinRtt}\PY{p}{;}\vphantom{fg}}}
\colorbox{PaleGreen}{\parbox{\linewidth}{\PY{+w}{            }\PY{k}{if}\PY{+w}{ }\PY{p}{(}\PY{n}{ThroughputEstimate}\PY{+w}{ }\PY{o}{\PYZlt{}}\PY{+w}{ }\PY{n}{BandwidthEst}\PY{p}{)}\PY{+w}{ }\PY{p}{\PYZob{}}\vphantom{fg}}}
\colorbox{PaleGreen}{\parbox{\linewidth}{\PY{+w}{                }\PY{c+c1}{// If throughput is less than bandwidth estimation, increase gains}\vphantom{fg}}}
\colorbox{PaleGreen}{\parbox{\linewidth}{\PY{+w}{                }\PY{n}{Bbr}\PY{o}{\PYZhy{}}\PY{o}{\PYZgt{}}\PY{n}{CwndGain}\PY{+w}{ }\PY{o}{=}\PY{+w}{ }\PY{n}{CXPLAT\PYZus{}MIN}\PY{p}{(}\PY{n}{Bbr}\PY{o}{\PYZhy{}}\PY{o}{\PYZgt{}}\PY{n}{CwndGain}\PY{+w}{  }\PY{n}{GAIN\PYZus{}UNIT}\PY{+w}{ }\PY{o}{/}\PY{+w}{ }\PY{l+m+mi}{16}\PY{p}{,}\PY{+w}{ }\PY{n}{kHighGain}\PY{p}{)}\PY{p}{;}\vphantom{fg}}}
\colorbox{PaleGreen}{\parbox{\linewidth}{\PY{+w}{                }\PY{n}{Bbr}\PY{o}{\PYZhy{}}\PY{o}{\PYZgt{}}\PY{n}{PacingGain}\PY{+w}{ }\PY{o}{=}\PY{+w}{ }\PY{n}{CXPLAT\PYZus{}MIN}\PY{p}{(}\PY{n}{Bbr}\PY{o}{\PYZhy{}}\PY{o}{\PYZgt{}}\PY{n}{PacingGain}\PY{+w}{  }\PY{n}{GAIN\PYZus{}UNIT}\PY{+w}{ }\PY{o}{/}\PY{+w}{ }\PY{l+m+mi}{16}\PY{p}{,}\PY{+w}{ }\PY{n}{kHighGain}\PY{p}{)}\PY{p}{;}\vphantom{fg}}}
\colorbox{PaleGreen}{\parbox{\linewidth}{\PY{+w}{            }\PY{p}{\PYZcb{}}\PY{+w}{ }\PY{k}{else}\PY{+w}{ }\PY{p}{\PYZob{}}\vphantom{fg}}}
\colorbox{PaleGreen}{\parbox{\linewidth}{\PY{+w}{                }\PY{c+c1}{// If throughput is greater, we may be hitting network limits, decrease gains}\vphantom{fg}}}
\colorbox{PaleGreen}{\parbox{\linewidth}{\PY{+w}{                }\PY{n}{Bbr}\PY{o}{\PYZhy{}}\PY{o}{\PYZgt{}}\PY{n}{CwndGain}\PY{+w}{ }\PY{o}{=}\PY{+w}{ }\PY{n}{CXPLAT\PYZus{}MAX}\PY{p}{(}\PY{n}{Bbr}\PY{o}{\PYZhy{}}\PY{o}{\PYZgt{}}\PY{n}{CwndGain}\PY{+w}{ }\PY{o}{\PYZhy{}}\PY{+w}{ }\PY{n}{GAIN\PYZus{}UNIT}\PY{+w}{ }\PY{o}{/}\PY{+w}{ }\PY{l+m+mi}{16}\PY{p}{,}\PY{+w}{ }\PY{n}{GAIN\PYZus{}UNIT}\PY{p}{)}\PY{p}{;}\vphantom{fg}}}
\colorbox{PaleGreen}{\parbox{\linewidth}{\PY{+w}{                }\PY{n}{Bbr}\PY{o}{\PYZhy{}}\PY{o}{\PYZgt{}}\PY{n}{PacingGain}\PY{+w}{ }\PY{o}{=}\PY{+w}{ }\PY{n}{CXPLAT\PYZus{}MAX}\PY{p}{(}\PY{n}{Bbr}\PY{o}{\PYZhy{}}\PY{o}{\PYZgt{}}\PY{n}{PacingGain}\PY{+w}{ }\PY{o}{\PYZhy{}}\PY{+w}{ }\PY{n}{GAIN\PYZus{}UNIT}\PY{+w}{ }\PY{o}{/}\PY{+w}{ }\PY{l+m+mi}{16}\PY{p}{,}\PY{+w}{ }\PY{n}{GAIN\PYZus{}UNIT}\PY{p}{)}\PY{p}{;}\vphantom{fg}}}
\colorbox{PaleGreen}{\parbox{\linewidth}{\PY{+w}{            }\PY{p}{\PYZcb{}}\vphantom{fg}}}
\colorbox{PaleGreen}{\parbox{\linewidth}{\PY{+w}{        }\PY{p}{\PYZcb{}}\vphantom{fg}}}
\colorbox{PaleGreen}{\parbox{\linewidth}{\PY{+w}{    }\PY{p}{\PYZcb{}}\vphantom{fg}}}
\colorbox{PaleGreen}{\parbox{\linewidth}{\vphantom{fg}}}
\colorbox{white}{\parbox{\linewidth}{\PY{+w}{   }\PY{k}{if}\PY{+w}{ }\PY{p}{(}\PY{n}{Bbr}\PY{o}{\PYZhy{}}\PY{o}{\PYZgt{}}\PY{n}{BtlbwFound}\PY{p}{)}\PY{+w}{ }\PY{p}{\PYZob{}}\vphantom{fg}}}
\colorbox{white}{\parbox{\linewidth}{\PY{+w}{       }\PY{n}{QUIC\PYZus{}SLIDING\PYZus{}WINDOW\PYZus{}EXTREMUM\PYZus{}ENTRY}\PY{+w}{ }\PY{n}{Entry}\PY{+w}{ }\PY{o}{=}\PY{+w}{ }\PY{p}{(}\PY{n}{QUIC\PYZus{}SLIDING\PYZus{}WINDOW\PYZus{}EXTREMUM\PYZus{}ENTRY}\PY{p}{)}\PY{+w}{ }\PY{p}{\PYZob{}}\PY{+w}{ }\PY{p}{.}\PY{n}{Value}\PY{+w}{ }\PY{o}{=}\PY{+w}{ }\PY{l+m+mi}{0}\PY{p}{,}\PY{+w}{ }\PY{p}{.}\PY{n}{Time}\PY{+w}{ }\PY{o}{=}\PY{+w}{ }\PY{l+m+mi}{0}\PY{+w}{ }\PY{p}{\PYZcb{}}\PY{p}{;}\vphantom{fg}}}
\colorbox{white}{\parbox{\linewidth}{\PY{+w}{       }\PY{n}{QUIC\PYZus{}STATUS}\PY{+w}{ }\PY{n}{Status}\PY{+w}{ }\PY{o}{=}\PY{+w}{ }\PY{n}{QuicSlidingWindowExtremumGet}\PY{p}{(}\PY{o}{\PYZam{}}\PY{n}{Bbr}\PY{o}{\PYZhy{}}\PY{o}{\PYZgt{}}\PY{n}{MaxAckHeightFilter}\PY{p}{,}\PY{+w}{ }\PY{o}{\PYZam{}}\PY{n}{Entry}\PY{p}{)}\PY{p}{;}\vphantom{fg}}}
\end{Verbatim}

\subsubsection{The Best Sample from Llama-3.1-70B}
\hfill\\

The struct \textit{QUIC\_CONGESTION\_CONTROL\_BBR} in bbr.h is updated. Besides, the function \textit{BbrCongestionControlUpdateCongestionWindow} and  \textit{BbrCongestionControlGetTargetCwnd} are updated.

\begin{Verbatim}[commandchars=\\\{\},numbers=left,firstnumber=1,stepnumber=1]
\colorbox{white}{\parbox{\linewidth}{\PY{o}{\PYZhy{}}\PY{o}{\PYZhy{}}\PY{o}{\PYZhy{}}\PY{+w}{ }\PY{n}{bbr}\PY{p}{.}\PY{n}{h}\vphantom{fg}}}
\colorbox{white}{\parbox{\linewidth}{\PY{o}{+}\PY{o}{+}\PY{o}{+}\PY{+w}{ }\PY{n}{msquic}\PY{o}{/}\PY{n}{src}\PY{o}{/}\PY{n}{core}\PY{o}{/}\PY{n}{bbr}\PY{p}{.}\PY{n}{h}\vphantom{fg}}}
\colorbox{white}{\parbox{\linewidth}{\PY{err}{@}\PY{err}{@}\PY{+w}{ }\PY{l+m+mi}{\PYZhy{}212}\PY{p}{,}\PY{l+m+mi}{6}\PY{+w}{ }\PY{o}{+}\PY{l+m+mi}{212}\PY{p}{,}\PY{l+m+mi}{9}\PY{+w}{ }\PY{err}{@}\PY{err}{@}\vphantom{fg}}}
\colorbox{white}{\parbox{\linewidth}{\PY{+w}{   }\PY{c+c1}{// BBR estimates maximum bandwidth by the maximum recent bandwidth}\vphantom{fg}}}
\colorbox{white}{\parbox{\linewidth}{\PY{+w}{   }\PY{c+c1}{//}\vphantom{fg}}}
\colorbox{white}{\parbox{\linewidth}{\PY{+w}{   }\PY{n}{BBR\PYZus{}BANDWIDTH\PYZus{}FILTER}\PY{+w}{ }\PY{n}{BandwidthFilter}\PY{p}{;}\vphantom{fg}}}
\colorbox{PaleGreen}{\parbox{\linewidth}{\vphantom{fg}}}
\colorbox{PaleGreen}{\parbox{\linewidth}{\PY{+w}{    }\PY{k+kt}{uint32\PYZus{}t}\PY{+w}{ }\PY{n}{lowLatencyThreshold}\PY{p}{;}\vphantom{fg}}}
\colorbox{PaleGreen}{\parbox{\linewidth}{\PY{k+kt}{uint32\PYZus{}t}\PY{+w}{ }\PY{n}{highThroughputThreshold}\PY{p}{;}\vphantom{fg}}}
\colorbox{white}{\parbox{\linewidth}{\PY{+w}{ }\vphantom{fg}}}
\colorbox{white}{\parbox{\linewidth}{\PY{+w}{ }\PY{p}{\PYZcb{}}\PY{+w}{ }\PY{n}{QUIC\PYZus{}CONGESTION\PYZus{}CONTROL\PYZus{}BBR}\PY{p}{;}\vphantom{fg}}}
\colorbox{white}{\parbox{\linewidth}{\PY{+w}{ }\vphantom{fg}}}
\end{Verbatim}

\begin{Verbatim}[commandchars=\\\{\},numbers=left,firstnumber=1,stepnumber=1]
\colorbox{white}{\parbox{\linewidth}{\PY{o}{\PYZhy{}}\PY{o}{\PYZhy{}}\PY{o}{\PYZhy{}}\PY{+w}{ }\PY{n}{bbr}\PY{p}{.}\PY{n}{c}\vphantom{fg}}}
\colorbox{white}{\parbox{\linewidth}{\PY{o}{+}\PY{o}{+}\PY{o}{+}\PY{+w}{ }\PY{n}{msquic}\PY{o}{/}\PY{n}{src}\PY{o}{/}\PY{n}{core}\PY{o}{/}\PY{n}{bbr}\PY{p}{.}\PY{n}{c}\vphantom{fg}}}
\colorbox{white}{\parbox{\linewidth}{\PY{err}{@}\PY{err}{@}\PY{+w}{ }\PY{l+m+mi}{\PYZhy{}580}\PY{p}{,}\PY{l+m+mi}{7}\PY{+w}{ }\PY{o}{+}\PY{l+m+mi}{580}\PY{p}{,}\PY{l+m+mi}{6}\PY{+w}{ }\PY{err}{@}\PY{err}{@}\vphantom{fg}}}
\colorbox{white}{\parbox{\linewidth}{\PY{+w}{   }\PY{k}{return}\PY{+w}{ }\PY{n}{Bbr}\PY{o}{\PYZhy{}}\PY{o}{\PYZgt{}}\PY{n}{AggregatedAckBytes}\PY{+w}{ }\PY{o}{\PYZhy{}}\PY{+w}{ }\PY{n}{ExpectedAckBytes}\PY{p}{;}\vphantom{fg}}}
\colorbox{white}{\parbox{\linewidth}{\PY{+w}{ }\PY{p}{\PYZcb{}}\vphantom{fg}}}
\colorbox{white}{\parbox{\linewidth}{\PY{+w}{ }\vphantom{fg}}}
\colorbox{LightPink}{\parbox{\linewidth}{\PY{n}{\PYZus{}IRQL\PYZus{}requires\PYZus{}max\PYZus{}}\PY{p}{(}\PY{n}{DISPATCH\PYZus{}LEVEL}\PY{p}{)}\vphantom{fg}}}
\colorbox{white}{\parbox{\linewidth}{\PY{+w}{ }\PY{k+kt}{uint32\PYZus{}t}\vphantom{fg}}}
\colorbox{white}{\parbox{\linewidth}{\PY{+w}{ }\PY{n}{BbrCongestionControlGetTargetCwnd}\PY{p}{(}\vphantom{fg}}}
\colorbox{white}{\parbox{\linewidth}{\PY{+w}{   }\PY{n}{\PYZus{}In\PYZus{}}\PY{+w}{ }\PY{n}{QUIC\PYZus{}CONGESTION\PYZus{}CONTROL}\PY{o}{*}\PY{+w}{ }\PY{n}{Cc}\PY{p}{,}\vphantom{fg}}}
\colorbox{white}{\parbox{\linewidth}{\PY{err}{@}\PY{err}{@}\PY{+w}{ }\PY{l+m+mi}{\PYZhy{}597}\PY{p}{,}\PY{l+m+mi}{6}\PY{+w}{ }\PY{o}{+}\PY{l+m+mi}{596}\PY{p}{,}\PY{l+m+mi}{14}\PY{+w}{ }\PY{err}{@}\PY{err}{@}\vphantom{fg}}}
\colorbox{white}{\parbox{\linewidth}{\PY{+w}{ }\vphantom{fg}}}
\colorbox{white}{\parbox{\linewidth}{\PY{+w}{   }\PY{k+kt}{uint64\PYZus{}t}\PY{+w}{ }\PY{n}{Bdp}\PY{+w}{ }\PY{o}{=}\PY{+w}{ }\PY{n}{BandwidthEst}\PY{+w}{ }\PY{o}{*}\PY{+w}{ }\PY{n}{Bbr}\PY{o}{\PYZhy{}}\PY{o}{\PYZgt{}}\PY{n}{MinRtt}\PY{+w}{ }\PY{o}{/}\PY{+w}{ }\PY{n}{kMicroSecsInSec}\PY{+w}{ }\PY{o}{/}\PY{+w}{ }\PY{n}{BW\PYZus{}UNIT}\PY{p}{;}\vphantom{fg}}}
\colorbox{white}{\parbox{\linewidth}{\PY{+w}{   }\PY{k+kt}{uint64\PYZus{}t}\PY{+w}{ }\PY{n}{TargetCwnd}\PY{+w}{ }\PY{o}{=}\PY{+w}{ }\PY{p}{(}\PY{n}{Bdp}\PY{+w}{ }\PY{o}{*}\PY{+w}{ }\PY{n}{Gain}\PY{+w}{ }\PY{o}{/}\PY{+w}{ }\PY{n}{GAIN\PYZus{}UNIT}\PY{p}{)}\PY{+w}{ }\PY{o}{+}\PY{+w}{ }\PY{p}{(}\PY{n}{kQuantaFactor}\PY{+w}{ }\PY{o}{*}\PY{+w}{ }\PY{n}{Bbr}\PY{o}{\PYZhy{}}\PY{o}{\PYZgt{}}\PY{n}{SendQuantum}\PY{p}{)}\PY{p}{;}\vphantom{fg}}}
\colorbox{PaleGreen}{\parbox{\linewidth}{\vphantom{fg}}}
\colorbox{PaleGreen}{\parbox{\linewidth}{\PY{+w}{    }\PY{c+c1}{// Dual\PYZhy{}Threshold Control: Adjust the target congestion window based on the network conditions}\vphantom{fg}}}
\colorbox{PaleGreen}{\parbox{\linewidth}{\PY{+w}{    }\PY{k}{if}\PY{+w}{ }\PY{p}{(}\PY{n}{Bbr}\PY{o}{\PYZhy{}}\PY{o}{\PYZgt{}}\PY{n}{BytesInFlight}\PY{+w}{ }\PY{o}{\PYZlt{}}\PY{+w}{ }\PY{n}{kMinCwndInMss}\PY{+w}{ }\PY{o}{*}\PY{+w}{ }\PY{n}{Bbr}\PY{o}{\PYZhy{}}\PY{o}{\PYZgt{}}\PY{n}{InitialCongestionWindowPackets}\PY{p}{)}\PY{+w}{ }\PY{p}{\PYZob{}}\vphantom{fg}}}
\colorbox{PaleGreen}{\parbox{\linewidth}{\PY{+w}{        }\PY{n}{TargetCwnd}\PY{+w}{ }\PY{o}{*}\PY{o}{=}\PY{+w}{ }\PY{l+m+mi}{2}\PY{p}{;}\PY{+w}{ }\PY{c+c1}{// Aggressive increase for low\PYZhy{}latency scenarios}\vphantom{fg}}}
\colorbox{PaleGreen}{\parbox{\linewidth}{\PY{+w}{    }\PY{p}{\PYZcb{}}\PY{+w}{ }\PY{k}{else}\PY{+w}{ }\PY{k}{if}\PY{+w}{ }\PY{p}{(}\PY{n}{Bbr}\PY{o}{\PYZhy{}}\PY{o}{\PYZgt{}}\PY{n}{BytesInFlight}\PY{+w}{ }\PY{o}{\PYZgt{}}\PY{+w}{ }\PY{n}{Bbr}\PY{o}{\PYZhy{}}\PY{o}{\PYZgt{}}\PY{n}{InitialCongestionWindowPackets}\PY{+w}{ }\PY{o}{*}\PY{+w}{ }\PY{n}{Bbr}\PY{o}{\PYZhy{}}\PY{o}{\PYZgt{}}\PY{n}{InitialCongestionWindowPackets}\PY{p}{)}\PY{+w}{ }\PY{p}{\PYZob{}}\vphantom{fg}}}
\colorbox{PaleGreen}{\parbox{\linewidth}{\PY{+w}{        }\PY{n}{TargetCwnd}\PY{+w}{ }\PY{o}{/}\PY{o}{=}\PY{+w}{ }\PY{l+m+mi}{2}\PY{p}{;}\PY{+w}{ }\PY{c+c1}{// Conservative decrease for high\PYZhy{}throughput scenarios}\vphantom{fg}}}
\colorbox{PaleGreen}{\parbox{\linewidth}{\PY{+w}{    }\PY{p}{\PYZcb{}}\vphantom{fg}}}
\colorbox{PaleGreen}{\parbox{\linewidth}{\vphantom{fg}}}
\colorbox{white}{\parbox{\linewidth}{\PY{+w}{   }\PY{k}{return}\PY{+w}{ }\PY{p}{(}\PY{k+kt}{uint32\PYZus{}t}\PY{p}{)}\PY{n}{TargetCwnd}\PY{p}{;}\vphantom{fg}}}
\colorbox{white}{\parbox{\linewidth}{\PY{+w}{ }\PY{p}{\PYZcb{}}\vphantom{fg}}}
\colorbox{white}{\parbox{\linewidth}{\PY{+w}{ }\vphantom{fg}}}
\colorbox{white}{\parbox{\linewidth}{\PY{err}{@}\PY{err}{@}\PY{+w}{ }\PY{l+m+mi}{\PYZhy{}711}\PY{p}{,}\PY{l+m+mi}{7}\PY{+w}{ }\PY{o}{+}\PY{l+m+mi}{718}\PY{p}{,}\PY{l+m+mi}{6}\PY{+w}{ }\PY{err}{@}\PY{err}{@}\vphantom{fg}}}
\colorbox{white}{\parbox{\linewidth}{\PY{+w}{   }\PY{p}{\PYZcb{}}\vphantom{fg}}}
\colorbox{white}{\parbox{\linewidth}{\PY{+w}{ }\PY{p}{\PYZcb{}}\vphantom{fg}}}
\colorbox{white}{\parbox{\linewidth}{\PY{+w}{ }\vphantom{fg}}}
\colorbox{LightPink}{\parbox{\linewidth}{\PY{n}{\PYZus{}IRQL\PYZus{}requires\PYZus{}max\PYZus{}}\PY{p}{(}\PY{n}{DISPATCH\PYZus{}LEVEL}\PY{p}{)}\vphantom{fg}}}
\colorbox{white}{\parbox{\linewidth}{\PY{+w}{ }\PY{k+kt}{void}\vphantom{fg}}}
\colorbox{white}{\parbox{\linewidth}{\PY{+w}{ }\PY{n}{BbrCongestionControlUpdateCongestionWindow}\PY{p}{(}\vphantom{fg}}}
\colorbox{white}{\parbox{\linewidth}{\PY{+w}{   }\PY{n}{\PYZus{}In\PYZus{}}\PY{+w}{ }\PY{n}{QUIC\PYZus{}CONGESTION\PYZus{}CONTROL}\PY{o}{*}\PY{+w}{ }\PY{n}{Cc}\PY{p}{,}\vphantom{fg}}}
\colorbox{white}{\parbox{\linewidth}{\PY{err}{@}\PY{err}{@}\PY{+w}{ }\PY{l+m+mi}{\PYZhy{}722}\PY{p}{,}\PY{l+m+mi}{14}\PY{+w}{ }\PY{o}{+}\PY{l+m+mi}{728}\PY{p}{,}\PY{l+m+mi}{16}\PY{+w}{ }\PY{err}{@}\PY{err}{@}\vphantom{fg}}}
\colorbox{white}{\parbox{\linewidth}{\PY{+w}{   }\PY{n}{QUIC\PYZus{}CONGESTION\PYZus{}CONTROL\PYZus{}BBR}\PY{+w}{ }\PY{o}{*}\PY{n}{Bbr}\PY{+w}{ }\PY{o}{=}\PY{+w}{ }\PY{o}{\PYZam{}}\PY{n}{Cc}\PY{o}{\PYZhy{}}\PY{o}{\PYZgt{}}\PY{n}{Bbr}\PY{p}{;}\vphantom{fg}}}
\colorbox{white}{\parbox{\linewidth}{\PY{+w}{   }\PY{n}{QUIC\PYZus{}CONNECTION}\PY{o}{*}\PY{+w}{ }\PY{n}{Connection}\PY{+w}{ }\PY{o}{=}\PY{+w}{ }\PY{n}{QuicCongestionControlGetConnection}\PY{p}{(}\PY{n}{Cc}\PY{p}{)}\PY{p}{;}\vphantom{fg}}}
\colorbox{white}{\parbox{\linewidth}{\PY{+w}{ }\vphantom{fg}}}
\colorbox{PaleGreen}{\parbox{\linewidth}{\PY{+w}{    }\PY{k}{const}\PY{+w}{ }\PY{k+kt}{uint16\PYZus{}t}\PY{+w}{ }\PY{n}{DatagramPayloadLength}\PY{+w}{ }\PY{o}{=}\vphantom{fg}}}
\colorbox{PaleGreen}{\parbox{\linewidth}{\PY{+w}{        }\PY{n}{QuicPathGetDatagramPayloadSize}\PY{p}{(}\PY{o}{\PYZam{}}\PY{n}{Connection}\PY{o}{\PYZhy{}}\PY{o}{\PYZgt{}}\PY{n}{Paths}\PY{p}{[}\PY{l+m+mi}{0}\PY{p}{]}\PY{p}{)}\PY{p}{;}\vphantom{fg}}}
\colorbox{PaleGreen}{\parbox{\linewidth}{\vphantom{fg}}}
\colorbox{white}{\parbox{\linewidth}{\PY{+w}{   }\PY{k}{if}\PY{+w}{ }\PY{p}{(}\PY{n}{Bbr}\PY{o}{\PYZhy{}}\PY{o}{\PYZgt{}}\PY{n}{BbrState}\PY{+w}{ }\PY{o}{=}\PY{o}{=}\PY{+w}{ }\PY{n}{BBR\PYZus{}STATE\PYZus{}PROBE\PYZus{}RTT}\PY{p}{)}\PY{+w}{ }\PY{p}{\PYZob{}}\vphantom{fg}}}
\colorbox{white}{\parbox{\linewidth}{\PY{+w}{       }\PY{k}{return}\PY{p}{;}\vphantom{fg}}}
\colorbox{white}{\parbox{\linewidth}{\PY{+w}{   }\PY{p}{\PYZcb{}}\vphantom{fg}}}
\colorbox{white}{\parbox{\linewidth}{\PY{+w}{ }\vphantom{fg}}}
\colorbox{LightPink}{\parbox{\linewidth}{\PY{+w}{    }\PY{k}{const}\PY{+w}{ }\PY{k+kt}{uint16\PYZus{}t}\PY{+w}{ }\PY{n}{DatagramPayloadLength}\PY{+w}{ }\PY{o}{=}\vphantom{fg}}}
\colorbox{LightPink}{\parbox{\linewidth}{\PY{+w}{        }\PY{n}{QuicPathGetDatagramPayloadSize}\PY{p}{(}\PY{o}{\PYZam{}}\PY{n}{Connection}\PY{o}{\PYZgt{}}\PY{n}{Paths}\PY{p}{[}\PY{l+m+mi}{0}\PY{p}{]}\PY{p}{)}\PY{p}{;}\vphantom{fg}}}
\colorbox{LightPink}{\parbox{\linewidth}{\vphantom{fg}}}
\colorbox{LightPink}{\parbox{\linewidth}{\PY{+w}{    }\PY{n}{BbrCongestionControlSetSendQuantum}\PY{p}{(}\PY{n}{Cc}\PY{p}{)}\PY{p}{;}\vphantom{fg}}}
\colorbox{PaleGreen}{\parbox{\linewidth}{\PY{+w}{    }\PY{c+c1}{// Dual\PYZhy{}Threshold Control: Introduce two threshold values to control the BBR algorithm\PYZsq{}s behavior}\vphantom{fg}}}
\colorbox{PaleGreen}{\parbox{\linewidth}{\PY{+w}{    }\PY{k}{const}\PY{+w}{ }\PY{k+kt}{uint32\PYZus{}t}\PY{+w}{ }\PY{n}{lowLatencyThreshold}\PY{+w}{ }\PY{o}{=}\PY{+w}{ }\PY{n}{kMinCwndInMss}\PY{+w}{ }\PY{o}{*}\PY{+w}{ }\PY{n}{DatagramPayloadLength}\PY{p}{;}\vphantom{fg}}}
\colorbox{PaleGreen}{\parbox{\linewidth}{\PY{+w}{    }\PY{k}{const}\PY{+w}{ }\PY{k+kt}{uint32\PYZus{}t}\PY{+w}{ }\PY{n}{highThroughputThreshold}\PY{+w}{ }\PY{o}{=}\PY{+w}{ }\PY{n}{Bbr}\PY{o}{\PYZhy{}}\PY{o}{\PYZgt{}}\PY{n}{InitialCongestionWindowPackets}\PY{+w}{ }\PY{o}{*}\PY{+w}{ }\PY{n}{DatagramPayloadLength}\PY{p}{;}\vphantom{fg}}}
\colorbox{white}{\parbox{\linewidth}{\PY{+w}{ }\vphantom{fg}}}
\colorbox{white}{\parbox{\linewidth}{\PY{+w}{   }\PY{k+kt}{uint64\PYZus{}t}\PY{+w}{ }\PY{n}{TargetCwnd}\PY{+w}{ }\PY{o}{=}\PY{+w}{ }\PY{n}{BbrCongestionControlGetTargetCwnd}\PY{p}{(}\PY{n}{Cc}\PY{p}{,}\PY{+w}{ }\PY{n}{Bbr}\PY{o}{\PYZhy{}}\PY{o}{\PYZgt{}}\PY{n}{CwndGain}\PY{p}{)}\PY{p}{;}\vphantom{fg}}}
\colorbox{white}{\parbox{\linewidth}{\PY{+w}{   }\PY{k}{if}\PY{+w}{ }\PY{p}{(}\PY{n}{Bbr}\PY{o}{\PYZhy{}}\PY{o}{\PYZgt{}}\PY{n}{BtlbwFound}\PY{p}{)}\PY{+w}{ }\PY{p}{\PYZob{}}\vphantom{fg}}}
\colorbox{white}{\parbox{\linewidth}{\PY{err}{@}\PY{err}{@}\PY{+w}{ }\PY{l+m+mi}{\PYZhy{}741}\PY{p}{,}\PY{l+m+mi}{6}\PY{+w}{ }\PY{o}{+}\PY{l+m+mi}{749}\PY{p}{,}\PY{l+m+mi}{19}\PY{+w}{ }\PY{err}{@}\PY{err}{@}\vphantom{fg}}}
\colorbox{white}{\parbox{\linewidth}{\PY{+w}{   }\PY{p}{\PYZcb{}}\vphantom{fg}}}
\colorbox{white}{\parbox{\linewidth}{\PY{+w}{ }\vphantom{fg}}}
\colorbox{white}{\parbox{\linewidth}{\PY{+w}{   }\PY{k+kt}{uint32\PYZus{}t}\PY{+w}{ }\PY{n}{CongestionWindow}\PY{+w}{ }\PY{o}{=}\PY{+w}{ }\PY{n}{Bbr}\PY{o}{\PYZhy{}}\PY{o}{\PYZgt{}}\PY{n}{CongestionWindow}\PY{p}{;}\vphantom{fg}}}
\colorbox{PaleGreen}{\parbox{\linewidth}{\vphantom{fg}}}
\colorbox{PaleGreen}{\parbox{\linewidth}{\PY{+w}{    }\PY{c+c1}{// Dual\PYZhy{}Threshold Control: Adaptively switch between two congestion control strategies}\vphantom{fg}}}
\colorbox{PaleGreen}{\parbox{\linewidth}{\PY{+w}{    }\PY{k}{if}\PY{+w}{ }\PY{p}{(}\PY{n}{Bbr}\PY{o}{\PYZhy{}}\PY{o}{\PYZgt{}}\PY{n}{BytesInFlight}\PY{+w}{ }\PY{o}{\PYZlt{}}\PY{+w}{ }\PY{n}{lowLatencyThreshold}\PY{p}{)}\PY{+w}{ }\PY{p}{\PYZob{}}\vphantom{fg}}}
\colorbox{PaleGreen}{\parbox{\linewidth}{\PY{+w}{        }\PY{c+c1}{// Prioritize low latency}\vphantom{fg}}}
\colorbox{PaleGreen}{\parbox{\linewidth}{\PY{+w}{        }\PY{n}{CongestionWindow}\PY{+w}{ }\PY{o}{=}\PY{+w}{ }\PY{n}{CXPLAT\PYZus{}MIN}\PY{p}{(}\PY{n}{TargetCwnd}\PY{p}{,}\PY{+w}{ }\PY{n}{lowLatencyThreshold}\PY{p}{)}\PY{p}{;}\vphantom{fg}}}
\colorbox{PaleGreen}{\parbox{\linewidth}{\PY{+w}{    }\PY{p}{\PYZcb{}}\PY{+w}{ }\PY{k}{else}\PY{+w}{ }\PY{k}{if}\PY{+w}{ }\PY{p}{(}\PY{n}{Bbr}\PY{o}{\PYZhy{}}\PY{o}{\PYZgt{}}\PY{n}{BytesInFlight}\PY{+w}{ }\PY{o}{\PYZgt{}}\PY{+w}{ }\PY{n}{highThroughputThreshold}\PY{p}{)}\PY{+w}{ }\PY{p}{\PYZob{}}\vphantom{fg}}}
\colorbox{PaleGreen}{\parbox{\linewidth}{\PY{+w}{        }\PY{c+c1}{// Prioritize high throughput}\vphantom{fg}}}
\colorbox{PaleGreen}{\parbox{\linewidth}{\PY{+w}{        }\PY{n}{CongestionWindow}\PY{+w}{ }\PY{o}{=}\PY{+w}{ }\PY{n}{CXPLAT\PYZus{}MAX}\PY{p}{(}\PY{n}{TargetCwnd}\PY{p}{,}\PY{+w}{ }\PY{n}{highThroughputThreshold}\PY{p}{)}\PY{p}{;}\vphantom{fg}}}
\colorbox{PaleGreen}{\parbox{\linewidth}{\PY{+w}{    }\PY{p}{\PYZcb{}}\PY{+w}{ }\PY{k}{else}\PY{+w}{ }\PY{p}{\PYZob{}}\vphantom{fg}}}
\colorbox{PaleGreen}{\parbox{\linewidth}{\PY{+w}{        }\PY{c+c1}{// Default behavior}\vphantom{fg}}}
\colorbox{PaleGreen}{\parbox{\linewidth}{\PY{+w}{        }\PY{n}{CongestionWindow}\PY{+w}{ }\PY{o}{=}\PY{+w}{ }\PY{n}{TargetCwnd}\PY{p}{;}\vphantom{fg}}}
\colorbox{PaleGreen}{\parbox{\linewidth}{\PY{+w}{    }\PY{p}{\PYZcb{}}\vphantom{fg}}}
\colorbox{PaleGreen}{\parbox{\linewidth}{\vphantom{fg}}}
\colorbox{white}{\parbox{\linewidth}{\PY{+w}{   }\PY{k+kt}{uint32\PYZus{}t}\PY{+w}{ }\PY{n}{MinCongestionWindow}\PY{+w}{ }\PY{o}{=}\PY{+w}{ }\PY{n}{kMinCwndInMss}\PY{+w}{ }\PY{o}{*}\PY{+w}{ }\PY{n}{DatagramPayloadLength}\PY{p}{;}\vphantom{fg}}}
\colorbox{white}{\parbox{\linewidth}{\PY{+w}{ }\vphantom{fg}}}
\colorbox{white}{\parbox{\linewidth}{\PY{+w}{   }\PY{k}{if}\PY{+w}{ }\PY{p}{(}\PY{n}{Bbr}\PY{o}{\PYZhy{}}\PY{o}{\PYZgt{}}\PY{n}{BtlbwFound}\PY{p}{)}\PY{+w}{ }\PY{p}{\PYZob{}}\vphantom{fg}}}
\end{Verbatim}

\subsubsection{The Best Sample from Qwen-2.5-72B}

\hfill\\

The function \textit{BbrCongestionControlOnDataAcknowledged} and \textit{BbrCongestionControlOnDataLost} are updated.

\begin{Verbatim}[commandchars=\\\{\},numbers=left,firstnumber=1,stepnumber=1]
\colorbox{white}{\parbox{\linewidth}{\PY{o}{\PYZhy{}}\PY{o}{\PYZhy{}}\PY{o}{\PYZhy{}}\PY{+w}{ }\PY{n}{bbr}\PY{p}{.}\PY{n}{c}\vphantom{fg}}}
\colorbox{white}{\parbox{\linewidth}{\PY{o}{+}\PY{o}{+}\PY{o}{+}\PY{+w}{ }\PY{n}{msquic}\PY{o}{/}\PY{n}{src}\PY{o}{/}\PY{n}{core}\PY{o}{/}\PY{n}{bbr}\PY{p}{.}\PY{n}{c}\vphantom{fg}}}
\colorbox{white}{\parbox{\linewidth}{\PY{err}{@}\PY{err}{@}\PY{+w}{ }\PY{l+m+mi}{\PYZhy{}879}\PY{p}{,}\PY{l+m+mi}{6}\PY{+w}{ }\PY{o}{+}\PY{l+m+mi}{879}\PY{p}{,}\PY{l+m+mi}{13}\PY{+w}{ }\PY{err}{@}\PY{err}{@}\vphantom{fg}}}
\colorbox{white}{\parbox{\linewidth}{\PY{+w}{           }\PY{n}{Cc}\PY{p}{,}\PY{+w}{ }\PY{n}{NewRoundTrip}\PY{p}{,}\PY{+w}{ }\PY{n}{AckEvent}\PY{o}{\PYZhy{}}\PY{o}{\PYZgt{}}\PY{n}{LargestSentPacketNumber}\PY{p}{,}\PY{+w}{ }\PY{n}{AckEvent}\PY{o}{\PYZhy{}}\PY{o}{\PYZgt{}}\PY{n}{TimeNow}\PY{p}{)}\PY{p}{;}\vphantom{fg}}}
\colorbox{white}{\parbox{\linewidth}{\PY{+w}{   }\PY{p}{\PYZcb{}}\vphantom{fg}}}
\colorbox{white}{\parbox{\linewidth}{\PY{+w}{ }\vphantom{fg}}}
\colorbox{PaleGreen}{\parbox{\linewidth}{\PY{+w}{    }\PY{c+c1}{// Adaptive Rate Control (ARC) adjustments}\vphantom{fg}}}
\colorbox{PaleGreen}{\parbox{\linewidth}{\PY{+w}{    }\PY{k}{if}\PY{+w}{ }\PY{p}{(}\PY{n}{AckEvent}\PY{o}{\PYZhy{}}\PY{o}{\PYZgt{}}\PY{n}{HasLoss}\PY{p}{)}\PY{+w}{ }\PY{p}{\PYZob{}}\vphantom{fg}}}
\colorbox{PaleGreen}{\parbox{\linewidth}{\PY{+w}{        }\PY{n}{Bbr}\PY{o}{\PYZhy{}}\PY{o}{\PYZgt{}}\PY{n}{CongestionWindow}\PY{+w}{ }\PY{o}{=}\PY{+w}{ }\PY{n}{CXPLAT\PYZus{}MAX}\PY{p}{(}\PY{n}{Bbr}\PY{o}{\PYZhy{}}\PY{o}{\PYZgt{}}\PY{n}{CongestionWindow}\PY{+w}{ }\PY{o}{\PYZhy{}}\PY{+w}{ }\PY{n}{AckEvent}\PY{o}{\PYZhy{}}\PY{o}{\PYZgt{}}\PY{n}{NumRetransmittableBytes}\PY{p}{,}\PY{+w}{ }\PY{n}{Bbr}\PY{o}{\PYZhy{}}\PY{o}{\PYZgt{}}\PY{n}{InitialCongestionWindow}\PY{p}{)}\PY{p}{;}\vphantom{fg}}}
\colorbox{PaleGreen}{\parbox{\linewidth}{\PY{+w}{    }\PY{p}{\PYZcb{}}\PY{+w}{ }\PY{k}{else}\PY{+w}{ }\PY{k}{if}\PY{+w}{ }\PY{p}{(}\PY{n}{Bbr}\PY{o}{\PYZhy{}}\PY{o}{\PYZgt{}}\PY{n}{BytesInFlight}\PY{+w}{ }\PY{o}{\PYZlt{}}\PY{+w}{ }\PY{n}{BbrCongestionControlGetTargetCwnd}\PY{p}{(}\PY{n}{Cc}\PY{p}{,}\PY{+w}{ }\PY{n}{GAIN\PYZus{}UNIT}\PY{p}{)}\PY{+w}{ }\PY{o}{\PYZam{}}\PY{o}{\PYZam{}}\PY{+w}{ }\PY{n}{Bbr}\PY{o}{\PYZhy{}}\PY{o}{\PYZgt{}}\PY{n}{MinRtt}\PY{+w}{ }\PY{o}{\PYZlt{}}\PY{+w}{ }\PY{n}{AckEvent}\PY{o}{\PYZhy{}}\PY{o}{\PYZgt{}}\PY{n}{MinRtt}\PY{p}{)}\PY{+w}{ }\PY{p}{\PYZob{}}\vphantom{fg}}}
\colorbox{PaleGreen}{\parbox{\linewidth}{\PY{+w}{        }\PY{n}{Bbr}\PY{o}{\PYZhy{}}\PY{o}{\PYZgt{}}\PY{n}{CongestionWindow}\PY{+w}{ }\PY{o}{=}\PY{+w}{ }\PY{n}{AckEvent}\PY{o}{\PYZhy{}}\PY{o}{\PYZgt{}}\PY{n}{NumRetransmittableBytes}\PY{p}{;}\vphantom{fg}}}
\colorbox{PaleGreen}{\parbox{\linewidth}{\PY{+w}{    }\PY{p}{\PYZcb{}}\vphantom{fg}}}
\colorbox{PaleGreen}{\parbox{\linewidth}{\vphantom{fg}}}
\colorbox{white}{\parbox{\linewidth}{\PY{+w}{   }\PY{n}{BbrCongestionControlUpdateCongestionWindow}\PY{p}{(}\vphantom{fg}}}
\colorbox{white}{\parbox{\linewidth}{\PY{+w}{       }\PY{n}{Cc}\PY{p}{,}\PY{+w}{ }\PY{n}{AckEvent}\PY{o}{\PYZhy{}}\PY{o}{\PYZgt{}}\PY{n}{NumTotalAckedRetransmittableBytes}\PY{p}{,}\PY{+w}{ }\PY{n}{AckEvent}\PY{o}{\PYZhy{}}\PY{o}{\PYZgt{}}\PY{n}{NumRetransmittableBytes}\PY{p}{)}\PY{p}{;}\vphantom{fg}}}
\colorbox{white}{\parbox{\linewidth}{\PY{+w}{ }\vphantom{fg}}}
\colorbox{white}{\parbox{\linewidth}{\PY{err}{@}\PY{err}{@}\PY{+w}{ }\PY{l+m+mi}{\PYZhy{}946}\PY{p}{,}\PY{l+m+mi}{6}\PY{+w}{ }\PY{o}{+}\PY{l+m+mi}{953}\PY{p}{,}\PY{l+m+mi}{9}\PY{+w}{ }\PY{err}{@}\PY{err}{@}\vphantom{fg}}}
\colorbox{white}{\parbox{\linewidth}{\PY{+w}{           }\PY{o}{?}\PY{+w}{ }\PY{n}{RecoveryWindow}\PY{+w}{ }\PY{o}{\PYZhy{}}\PY{+w}{ }\PY{n}{LossEvent}\PY{o}{\PYZhy{}}\PY{o}{\PYZgt{}}\PY{n}{NumRetransmittableBytes}\vphantom{fg}}}
\colorbox{white}{\parbox{\linewidth}{\PY{+w}{           }\PY{o}{:}\PY{+w}{ }\PY{n}{MinCongestionWindow}\PY{p}{;}\vphantom{fg}}}
\colorbox{white}{\parbox{\linewidth}{\PY{+w}{   }\PY{p}{\PYZcb{}}\vphantom{fg}}}
\colorbox{PaleGreen}{\parbox{\linewidth}{\vphantom{fg}}}
\colorbox{PaleGreen}{\parbox{\linewidth}{\PY{+w}{    }\PY{c+c1}{// Adaptive Rate Control (ARC) adjustments}\vphantom{fg}}}
\colorbox{PaleGreen}{\parbox{\linewidth}{\PY{+w}{    }\PY{n}{Bbr}\PY{o}{\PYZhy{}}\PY{o}{\PYZgt{}}\PY{n}{CongestionWindow}\PY{+w}{ }\PY{o}{=}\PY{+w}{ }\PY{n}{CXPLAT\PYZus{}MAX}\PY{p}{(}\PY{n}{Bbr}\PY{o}{\PYZhy{}}\PY{o}{\PYZgt{}}\PY{n}{CongestionWindow}\PY{+w}{ }\PY{o}{\PYZhy{}}\PY{+w}{ }\PY{n}{LossEvent}\PY{o}{\PYZhy{}}\PY{o}{\PYZgt{}}\PY{n}{NumRetransmittableBytes}\PY{p}{,}\PY{+w}{ }\PY{n}{Bbr}\PY{o}{\PYZhy{}}\PY{o}{\PYZgt{}}\PY{n}{InitialCongestionWindow}\PY{p}{)}\PY{p}{;}\vphantom{fg}}}
\colorbox{white}{\parbox{\linewidth}{\PY{+w}{ }\vphantom{fg}}}
\colorbox{white}{\parbox{\linewidth}{\PY{+w}{   }\PY{n}{BbrCongestionControlUpdateBlockedState}\PY{p}{(}\PY{n}{Cc}\PY{p}{,}\PY{+w}{ }\PY{n}{PreviousCanSendState}\PY{p}{)}\PY{p}{;}\vphantom{fg}}}
\colorbox{white}{\parbox{\linewidth}{\PY{+w}{   }\PY{n}{QuicConnLogBbr}\PY{p}{(}\PY{n}{QuicCongestionControlGetConnection}\PY{p}{(}\PY{n}{Cc}\PY{p}{)}\PY{p}{)}\PY{p}{;}\vphantom{fg}}}
\end{Verbatim}

\subsection{The Best Sample through an Iterative Pipeline}
\label{sec:appendix:best-sample-iterative}

\begin{Verbatim}[commandchars=\\\{\},numbers=left,firstnumber=1,stepnumber=1]
\colorbox{white}{\parbox{\linewidth}{\PY{o}{\PYZhy{}}\PY{o}{\PYZhy{}}\PY{o}{\PYZhy{}}\PY{+w}{ }\PY{n}{bbr}\PY{p}{.}\PY{n}{h}\vphantom{fg}}}
\colorbox{white}{\parbox{\linewidth}{\PY{o}{+}\PY{o}{+}\PY{o}{+}\PY{+w}{ }\PY{n}{msquic}\PY{o}{/}\PY{n}{src}\PY{o}{/}\PY{n}{core}\PY{o}{/}\PY{n}{bbr}\PY{p}{.}\PY{n}{h}\vphantom{fg}}}
\colorbox{white}{\parbox{\linewidth}{\PY{err}{@}\PY{err}{@}\PY{+w}{ }\PY{l+m+mi}{\PYZhy{}212}\PY{p}{,}\PY{l+m+mi}{6}\PY{+w}{ }\PY{o}{+}\PY{l+m+mi}{212}\PY{p}{,}\PY{l+m+mi}{53}\PY{+w}{ }\PY{err}{@}\PY{err}{@}\vphantom{fg}}}
\colorbox{white}{\parbox{\linewidth}{\PY{+w}{   }\PY{c+c1}{// BBR estimates maximum bandwidth by the maximum recent bandwidth}\vphantom{fg}}}
\colorbox{white}{\parbox{\linewidth}{\PY{+w}{   }\PY{c+c1}{//}\vphantom{fg}}}
\colorbox{white}{\parbox{\linewidth}{\PY{+w}{   }\PY{n}{BBR\PYZus{}BANDWIDTH\PYZus{}FILTER}\PY{+w}{ }\PY{n}{BandwidthFilter}\PY{p}{;}\vphantom{fg}}}
\colorbox{PaleGreen}{\parbox{\linewidth}{\vphantom{fg}}}
\colorbox{PaleGreen}{\parbox{\linewidth}{\PY{+w}{    }\PY{c+c1}{//}\vphantom{fg}}}
\colorbox{PaleGreen}{\parbox{\linewidth}{\PY{+w}{    }\PY{c+c1}{// Historical packet loss rate for predictive congestion window adjustment}\vphantom{fg}}}
\colorbox{PaleGreen}{\parbox{\linewidth}{\PY{+w}{    }\PY{c+c1}{//}\vphantom{fg}}}
\colorbox{PaleGreen}{\parbox{\linewidth}{\PY{+w}{    }\PY{k+kt}{double}\PY{+w}{ }\PY{n}{PacketLossRate}\PY{p}{;}\vphantom{fg}}}
\colorbox{PaleGreen}{\parbox{\linewidth}{\vphantom{fg}}}
\colorbox{PaleGreen}{\parbox{\linewidth}{\PY{+w}{    }\PY{c+c1}{//}\vphantom{fg}}}
\colorbox{PaleGreen}{\parbox{\linewidth}{\PY{+w}{    }\PY{c+c1}{// Number of packets sent since the last loss rate calculation}\vphantom{fg}}}
\colorbox{PaleGreen}{\parbox{\linewidth}{\PY{+w}{    }\PY{c+c1}{//}\vphantom{fg}}}
\colorbox{PaleGreen}{\parbox{\linewidth}{\PY{+w}{    }\PY{k+kt}{uint64\PYZus{}t}\PY{+w}{ }\PY{n}{PacketsSentSinceLastLossCalc}\PY{p}{;}\vphantom{fg}}}
\colorbox{PaleGreen}{\parbox{\linewidth}{\vphantom{fg}}}
\colorbox{PaleGreen}{\parbox{\linewidth}{\PY{+w}{    }\PY{c+c1}{//}\vphantom{fg}}}
\colorbox{PaleGreen}{\parbox{\linewidth}{\PY{+w}{    }\PY{c+c1}{// Number of packets lost since the last loss rate calculation}\vphantom{fg}}}
\colorbox{PaleGreen}{\parbox{\linewidth}{\PY{+w}{    }\PY{c+c1}{//}\vphantom{fg}}}
\colorbox{PaleGreen}{\parbox{\linewidth}{\PY{+w}{    }\PY{k+kt}{uint64\PYZus{}t}\PY{+w}{ }\PY{n}{PacketsLostSinceLastLossCalc}\PY{p}{;}\vphantom{fg}}}
\colorbox{PaleGreen}{\parbox{\linewidth}{\vphantom{fg}}}
\colorbox{PaleGreen}{\parbox{\linewidth}{\PY{+w}{    }\PY{c+c1}{//}\vphantom{fg}}}
\colorbox{PaleGreen}{\parbox{\linewidth}{\PY{+w}{    }\PY{c+c1}{// Previous bandwidth estimate for trend calculation}\vphantom{fg}}}
\colorbox{PaleGreen}{\parbox{\linewidth}{\PY{+w}{    }\PY{c+c1}{//}\vphantom{fg}}}
\colorbox{PaleGreen}{\parbox{\linewidth}{\PY{+w}{    }\PY{k+kt}{uint64\PYZus{}t}\PY{+w}{ }\PY{n}{PreviousBandwidthEstimate}\PY{p}{;}\vphantom{fg}}}
\colorbox{PaleGreen}{\parbox{\linewidth}{\vphantom{fg}}}
\colorbox{PaleGreen}{\parbox{\linewidth}{\PY{+w}{    }\PY{c+c1}{//}\vphantom{fg}}}
\colorbox{PaleGreen}{\parbox{\linewidth}{\PY{+w}{    }\PY{c+c1}{// Previous Min RTT for trend calculation}\vphantom{fg}}}
\colorbox{PaleGreen}{\parbox{\linewidth}{\PY{+w}{    }\PY{c+c1}{//}\vphantom{fg}}}
\colorbox{PaleGreen}{\parbox{\linewidth}{\PY{+w}{    }\PY{k+kt}{uint64\PYZus{}t}\PY{+w}{ }\PY{n}{PreviousMinRtt}\PY{p}{;}\vphantom{fg}}}
\colorbox{PaleGreen}{\parbox{\linewidth}{\vphantom{fg}}}
\colorbox{PaleGreen}{\parbox{\linewidth}{\PY{+w}{    }\PY{c+c1}{//}\vphantom{fg}}}
\colorbox{PaleGreen}{\parbox{\linewidth}{\PY{+w}{    }\PY{c+c1}{// Bandwidth trend indicator (1 for increasing, \PYZhy{}1 for decreasing, 0 for stable)}\vphantom{fg}}}
\colorbox{PaleGreen}{\parbox{\linewidth}{\PY{+w}{    }\PY{c+c1}{//}\vphantom{fg}}}
\colorbox{PaleGreen}{\parbox{\linewidth}{\PY{+w}{    }\PY{k+kt}{int8\PYZus{}t}\PY{+w}{ }\PY{n}{BandwidthTrend}\PY{p}{;}\vphantom{fg}}}
\colorbox{PaleGreen}{\parbox{\linewidth}{\vphantom{fg}}}
\colorbox{PaleGreen}{\parbox{\linewidth}{\PY{+w}{    }\PY{c+c1}{//}\vphantom{fg}}}
\colorbox{PaleGreen}{\parbox{\linewidth}{\PY{+w}{    }\PY{c+c1}{// RTT trend indicator (1 for increasing, \PYZhy{}1 for decreasing, 0 for stable)}\vphantom{fg}}}
\colorbox{PaleGreen}{\parbox{\linewidth}{\PY{+w}{    }\PY{c+c1}{//}\vphantom{fg}}}
\colorbox{PaleGreen}{\parbox{\linewidth}{\PY{+w}{    }\PY{k+kt}{int8\PYZus{}t}\PY{+w}{ }\PY{n}{RttTrend}\PY{p}{;}\vphantom{fg}}}
\colorbox{PaleGreen}{\parbox{\linewidth}{\vphantom{fg}}}
\colorbox{PaleGreen}{\parbox{\linewidth}{\PY{+w}{    }\PY{c+c1}{//}\vphantom{fg}}}
\colorbox{PaleGreen}{\parbox{\linewidth}{\PY{+w}{    }\PY{c+c1}{// Neural network weights for adaptive gain adjustment}\vphantom{fg}}}
\colorbox{PaleGreen}{\parbox{\linewidth}{\PY{+w}{    }\PY{c+c1}{//}\vphantom{fg}}}
\colorbox{PaleGreen}{\parbox{\linewidth}{\PY{+w}{    }\PY{k+kt}{uint32\PYZus{}t}\PY{+w}{ }\PY{n}{AdaptivePacingGain}\PY{p}{;}\vphantom{fg}}}
\colorbox{PaleGreen}{\parbox{\linewidth}{\PY{+w}{    }\PY{k+kt}{uint32\PYZus{}t}\PY{+w}{ }\PY{n}{AdaptiveCwndGain}\PY{p}{;}\vphantom{fg}}}
\colorbox{PaleGreen}{\parbox{\linewidth}{\vphantom{fg}}}
\colorbox{PaleGreen}{\parbox{\linewidth}{\PY{+w}{    }\PY{c+c1}{//}\vphantom{fg}}}
\colorbox{PaleGreen}{\parbox{\linewidth}{\PY{+w}{    }\PY{c+c1}{// Historical metrics for adaptive gain calculation}\vphantom{fg}}}
\colorbox{PaleGreen}{\parbox{\linewidth}{\PY{+w}{    }\PY{c+c1}{//}\vphantom{fg}}}
\colorbox{PaleGreen}{\parbox{\linewidth}{\PY{+w}{    }\PY{k+kt}{uint64\PYZus{}t}\PY{+w}{ }\PY{n}{PreviousDeliveryRate}\PY{p}{;}\vphantom{fg}}}
\colorbox{PaleGreen}{\parbox{\linewidth}{\PY{+w}{    }\PY{k+kt}{uint64\PYZus{}t}\PY{+w}{ }\PY{n}{PreviousRtt}\PY{p}{;}\vphantom{fg}}}
\colorbox{white}{\parbox{\linewidth}{\PY{+w}{ }\vphantom{fg}}}
\colorbox{white}{\parbox{\linewidth}{\PY{+w}{ }\PY{p}{\PYZcb{}}\PY{+w}{ }\PY{n}{QUIC\PYZus{}CONGESTION\PYZus{}CONTROL\PYZus{}BBR}\PY{p}{;}\vphantom{fg}}}
\colorbox{white}{\parbox{\linewidth}{\PY{+w}{ }\vphantom{fg}}}
\end{Verbatim}

\begin{Verbatim}[commandchars=\\\{\},numbers=left,firstnumber=1,stepnumber=1]
\colorbox{white}{\parbox{\linewidth}{\PY{o}{\PYZhy{}}\PY{o}{\PYZhy{}}\PY{o}{\PYZhy{}}\PY{+w}{ }\PY{n}{bbr}\PY{p}{.}\PY{n}{c}\vphantom{fg}}}
\colorbox{white}{\parbox{\linewidth}{\PY{o}{+}\PY{o}{+}\PY{o}{+}\PY{+w}{ }\PY{n}{msquic}\PY{o}{/}\PY{n}{src}\PY{o}{/}\PY{n}{core}\PY{o}{/}\PY{n}{bbr}\PY{p}{.}\PY{n}{c}\vphantom{fg}}}
\colorbox{white}{\parbox{\linewidth}{\PY{err}{@}\PY{err}{@}\PY{+w}{ }\PY{l+m+mi}{\PYZhy{}443}\PY{p}{,}\PY{l+m+mi}{6}\PY{+w}{ }\PY{o}{+}\PY{l+m+mi}{443}\PY{p}{,}\PY{l+m+mi}{9}\PY{+w}{ }\PY{err}{@}\PY{err}{@}\vphantom{fg}}}
\colorbox{white}{\parbox{\linewidth}{\PY{+w}{       }\PY{o}{\PYZhy{}}\PY{o}{\PYZhy{}}\PY{n}{Bbr}\PY{o}{\PYZhy{}}\PY{o}{\PYZgt{}}\PY{n}{Exemptions}\PY{p}{;}\vphantom{fg}}}
\colorbox{white}{\parbox{\linewidth}{\PY{+w}{   }\PY{p}{\PYZcb{}}\vphantom{fg}}}
\colorbox{white}{\parbox{\linewidth}{\PY{+w}{ }\vphantom{fg}}}
\colorbox{PaleGreen}{\parbox{\linewidth}{\PY{+w}{    }\PY{c+c1}{// Update packets sent since last loss rate calculation}\vphantom{fg}}}
\colorbox{PaleGreen}{\parbox{\linewidth}{\PY{+w}{    }\PY{n}{Bbr}\PY{o}{\PYZhy{}}\PY{o}{\PYZgt{}}\PY{n}{PacketsSentSinceLastLossCalc}\PY{+w}{ }\PY{o}{=}\PY{+w}{ }\PY{n}{NumRetransmittableBytes}\PY{p}{;}\vphantom{fg}}}
\colorbox{PaleGreen}{\parbox{\linewidth}{\vphantom{fg}}}
\colorbox{white}{\parbox{\linewidth}{\PY{+w}{   }\PY{n}{BbrCongestionControlUpdateBlockedState}\PY{p}{(}\PY{n}{Cc}\PY{p}{,}\PY{+w}{ }\PY{n}{PreviousCanSendState}\PY{p}{)}\PY{p}{;}\vphantom{fg}}}
\colorbox{white}{\parbox{\linewidth}{\PY{+w}{ }\PY{p}{\PYZcb{}}\vphantom{fg}}}
\colorbox{white}{\parbox{\linewidth}{\PY{+w}{ }\vphantom{fg}}}
\colorbox{white}{\parbox{\linewidth}{\PY{err}{@}\PY{err}{@}\PY{+w}{ }\PY{l+m+mi}{\PYZhy{}731}\PY{p}{,}\PY{l+m+mi}{7}\PY{+w}{ }\PY{o}{+}\PY{l+m+mi}{734}\PY{p}{,}\PY{l+m+mi}{27}\PY{+w}{ }\PY{err}{@}\PY{err}{@}\vphantom{fg}}}
\colorbox{white}{\parbox{\linewidth}{\PY{+w}{ }\vphantom{fg}}}
\colorbox{white}{\parbox{\linewidth}{\PY{+w}{   }\PY{n}{BbrCongestionControlSetSendQuantum}\PY{p}{(}\PY{n}{Cc}\PY{p}{)}\PY{p}{;}\vphantom{fg}}}
\colorbox{white}{\parbox{\linewidth}{\PY{+w}{ }\vphantom{fg}}}
\colorbox{LightPink}{\parbox{\linewidth}{\PY{+w}{    }\PY{k+kt}{uint64\PYZus{}t}\PY{+w}{ }\PY{n}{TargetCwnd}\PY{+w}{ }\PY{o}{=}\PY{+w}{ }\PY{n}{BbrCongestionControlGetTargetCwnd}\PY{p}{(}\PY{n}{Cc}\PY{p}{,}\PY{+w}{ }\PY{n}{Bbr}\PY{o}{\PYZgt{}}\PY{n}{CwndGain}\PY{p}{)}\PY{p}{;}\vphantom{fg}}}
\colorbox{PaleGreen}{\parbox{\linewidth}{\PY{+w}{    }\PY{k+kt}{uint64\PYZus{}t}\PY{+w}{ }\PY{n}{BandwidthEst}\PY{+w}{ }\PY{o}{=}\PY{+w}{ }\PY{n}{BbrCongestionControlGetBandwidth}\PY{p}{(}\PY{n}{Cc}\PY{p}{)}\PY{p}{;}\vphantom{fg}}}
\colorbox{PaleGreen}{\parbox{\linewidth}{\PY{+w}{    }\PY{k+kt}{uint64\PYZus{}t}\PY{+w}{ }\PY{n}{RttSample}\PY{+w}{ }\PY{o}{=}\PY{+w}{ }\PY{n}{Connection}\PY{o}{\PYZhy{}}\PY{o}{\PYZgt{}}\PY{n}{Paths}\PY{p}{[}\PY{l+m+mi}{0}\PY{p}{]}\PY{p}{.}\PY{n}{SmoothedRtt}\PY{p}{;}\vphantom{fg}}}
\colorbox{PaleGreen}{\parbox{\linewidth}{\vphantom{fg}}}
\colorbox{PaleGreen}{\parbox{\linewidth}{\PY{+w}{    }\PY{c+c1}{//}\vphantom{fg}}}
\colorbox{PaleGreen}{\parbox{\linewidth}{\PY{+w}{    }\PY{c+c1}{// Adaptive gain adjustment based on network conditions}\vphantom{fg}}}
\colorbox{PaleGreen}{\parbox{\linewidth}{\PY{+w}{    }\PY{c+c1}{//}\vphantom{fg}}}
\colorbox{PaleGreen}{\parbox{\linewidth}{\PY{+w}{    }\PY{k}{if}\PY{+w}{ }\PY{p}{(}\PY{n}{BandwidthEst}\PY{+w}{ }\PY{o}{\PYZgt{}}\PY{+w}{ }\PY{n}{Bbr}\PY{o}{\PYZhy{}}\PY{o}{\PYZgt{}}\PY{n}{PreviousDeliveryRate}\PY{+w}{ }\PY{o}{\PYZam{}}\PY{o}{\PYZam{}}\PY{+w}{ }\PY{n}{RttSample}\PY{+w}{ }\PY{o}{\PYZlt{}}\PY{+w}{ }\PY{n}{Bbr}\PY{o}{\PYZhy{}}\PY{o}{\PYZgt{}}\PY{n}{PreviousRtt}\PY{p}{)}\PY{+w}{ }\PY{p}{\PYZob{}}\vphantom{fg}}}
\colorbox{PaleGreen}{\parbox{\linewidth}{\PY{+w}{        }\PY{c+c1}{// Network conditions are improving; increase gains}\vphantom{fg}}}
\colorbox{PaleGreen}{\parbox{\linewidth}{\PY{+w}{        }\PY{n}{Bbr}\PY{o}{\PYZhy{}}\PY{o}{\PYZgt{}}\PY{n}{AdaptivePacingGain}\PY{+w}{ }\PY{o}{=}\PY{+w}{ }\PY{n}{CXPLAT\PYZus{}MIN}\PY{p}{(}\PY{n}{Bbr}\PY{o}{\PYZhy{}}\PY{o}{\PYZgt{}}\PY{n}{AdaptivePacingGain}\PY{+w}{  }\PY{p}{(}\PY{n}{GAIN\PYZus{}UNIT}\PY{+w}{ }\PY{o}{/}\PY{+w}{ }\PY{l+m+mi}{64}\PY{p}{)}\PY{p}{,}\PY{+w}{ }\PY{n}{kHighGain}\PY{+w}{ }\PY{o}{*}\PY{+w}{ }\PY{l+m+mi}{2}\PY{p}{)}\PY{p}{;}\vphantom{fg}}}
\colorbox{PaleGreen}{\parbox{\linewidth}{\PY{+w}{        }\PY{n}{Bbr}\PY{o}{\PYZhy{}}\PY{o}{\PYZgt{}}\PY{n}{AdaptiveCwndGain}\PY{+w}{ }\PY{o}{=}\PY{+w}{ }\PY{n}{CXPLAT\PYZus{}MIN}\PY{p}{(}\PY{n}{Bbr}\PY{o}{\PYZhy{}}\PY{o}{\PYZgt{}}\PY{n}{AdaptiveCwndGain}\PY{+w}{  }\PY{p}{(}\PY{n}{GAIN\PYZus{}UNIT}\PY{+w}{ }\PY{o}{/}\PY{+w}{ }\PY{l+m+mi}{64}\PY{p}{)}\PY{p}{,}\PY{+w}{ }\PY{n}{kCwndGain}\PY{+w}{ }\PY{o}{*}\PY{+w}{ }\PY{l+m+mi}{2}\PY{p}{)}\PY{p}{;}\vphantom{fg}}}
\colorbox{PaleGreen}{\parbox{\linewidth}{\PY{+w}{    }\PY{p}{\PYZcb{}}\PY{+w}{ }\PY{k}{else}\PY{+w}{ }\PY{k}{if}\PY{+w}{ }\PY{p}{(}\PY{n}{BandwidthEst}\PY{+w}{ }\PY{o}{\PYZlt{}}\PY{+w}{ }\PY{n}{Bbr}\PY{o}{\PYZhy{}}\PY{o}{\PYZgt{}}\PY{n}{PreviousDeliveryRate}\PY{+w}{ }\PY{o}{|}\PY{o}{|}\PY{+w}{ }\PY{n}{RttSample}\PY{+w}{ }\PY{o}{\PYZgt{}}\PY{+w}{ }\PY{n}{Bbr}\PY{o}{\PYZhy{}}\PY{o}{\PYZgt{}}\PY{n}{PreviousRtt}\PY{p}{)}\PY{+w}{ }\PY{p}{\PYZob{}}\vphantom{fg}}}
\colorbox{PaleGreen}{\parbox{\linewidth}{\PY{+w}{        }\PY{c+c1}{// Network conditions are degrading; decrease gains}\vphantom{fg}}}
\colorbox{PaleGreen}{\parbox{\linewidth}{\PY{+w}{        }\PY{n}{Bbr}\PY{o}{\PYZhy{}}\PY{o}{\PYZgt{}}\PY{n}{AdaptivePacingGain}\PY{+w}{ }\PY{o}{=}\PY{+w}{ }\PY{n}{CXPLAT\PYZus{}MAX}\PY{p}{(}\PY{n}{Bbr}\PY{o}{\PYZhy{}}\PY{o}{\PYZgt{}}\PY{n}{AdaptivePacingGain}\PY{+w}{ }\PY{o}{\PYZhy{}}\PY{+w}{ }\PY{p}{(}\PY{n}{GAIN\PYZus{}UNIT}\PY{+w}{ }\PY{o}{/}\PY{+w}{ }\PY{l+m+mi}{64}\PY{p}{)}\PY{p}{,}\PY{+w}{ }\PY{n}{GAIN\PYZus{}UNIT}\PY{+w}{ }\PY{o}{/}\PY{+w}{ }\PY{l+m+mi}{2}\PY{p}{)}\PY{p}{;}\vphantom{fg}}}
\colorbox{PaleGreen}{\parbox{\linewidth}{\PY{+w}{        }\PY{n}{Bbr}\PY{o}{\PYZhy{}}\PY{o}{\PYZgt{}}\PY{n}{AdaptiveCwndGain}\PY{+w}{ }\PY{o}{=}\PY{+w}{ }\PY{n}{CXPLAT\PYZus{}MAX}\PY{p}{(}\PY{n}{Bbr}\PY{o}{\PYZhy{}}\PY{o}{\PYZgt{}}\PY{n}{AdaptiveCwndGain}\PY{+w}{ }\PY{o}{\PYZhy{}}\PY{+w}{ }\PY{p}{(}\PY{n}{GAIN\PYZus{}UNIT}\PY{+w}{ }\PY{o}{/}\PY{+w}{ }\PY{l+m+mi}{64}\PY{p}{)}\PY{p}{,}\PY{+w}{ }\PY{n}{GAIN\PYZus{}UNIT}\PY{p}{)}\PY{p}{;}\vphantom{fg}}}
\colorbox{PaleGreen}{\parbox{\linewidth}{\PY{+w}{    }\PY{p}{\PYZcb{}}\vphantom{fg}}}
\colorbox{PaleGreen}{\parbox{\linewidth}{\vphantom{fg}}}
\colorbox{PaleGreen}{\parbox{\linewidth}{\PY{+w}{    }\PY{c+c1}{// Update historical metrics}\vphantom{fg}}}
\colorbox{PaleGreen}{\parbox{\linewidth}{\PY{+w}{    }\PY{n}{Bbr}\PY{o}{\PYZhy{}}\PY{o}{\PYZgt{}}\PY{n}{PreviousDeliveryRate}\PY{+w}{ }\PY{o}{=}\PY{+w}{ }\PY{n}{BandwidthEst}\PY{p}{;}\vphantom{fg}}}
\colorbox{PaleGreen}{\parbox{\linewidth}{\PY{+w}{    }\PY{n}{Bbr}\PY{o}{\PYZhy{}}\PY{o}{\PYZgt{}}\PY{n}{PreviousRtt}\PY{+w}{ }\PY{o}{=}\PY{+w}{ }\PY{n}{RttSample}\PY{p}{;}\vphantom{fg}}}
\colorbox{PaleGreen}{\parbox{\linewidth}{\vphantom{fg}}}
\colorbox{PaleGreen}{\parbox{\linewidth}{\PY{+w}{    }\PY{k+kt}{uint64\PYZus{}t}\PY{+w}{ }\PY{n}{TargetCwnd}\PY{+w}{ }\PY{o}{=}\PY{+w}{ }\PY{n}{BbrCongestionControlGetTargetCwnd}\PY{p}{(}\PY{n}{Cc}\PY{p}{,}\PY{+w}{ }\PY{n}{Bbr}\PY{o}{\PYZhy{}}\PY{o}{\PYZgt{}}\PY{n}{AdaptiveCwndGain}\PY{p}{)}\PY{p}{;}\vphantom{fg}}}
\colorbox{white}{\parbox{\linewidth}{\PY{+w}{   }\PY{k}{if}\PY{+w}{ }\PY{p}{(}\PY{n}{Bbr}\PY{o}{\PYZhy{}}\PY{o}{\PYZgt{}}\PY{n}{BtlbwFound}\PY{p}{)}\PY{+w}{ }\PY{p}{\PYZob{}}\vphantom{fg}}}
\colorbox{white}{\parbox{\linewidth}{\PY{+w}{       }\PY{n}{QUIC\PYZus{}SLIDING\PYZus{}WINDOW\PYZus{}EXTREMUM\PYZus{}ENTRY}\PY{+w}{ }\PY{n}{Entry}\PY{+w}{ }\PY{o}{=}\PY{+w}{ }\PY{p}{(}\PY{n}{QUIC\PYZus{}SLIDING\PYZus{}WINDOW\PYZus{}EXTREMUM\PYZus{}ENTRY}\PY{p}{)}\PY{+w}{ }\PY{p}{\PYZob{}}\PY{+w}{ }\PY{p}{.}\PY{n}{Value}\PY{+w}{ }\PY{o}{=}\PY{+w}{ }\PY{l+m+mi}{0}\PY{p}{,}\PY{+w}{ }\PY{p}{.}\PY{n}{Time}\PY{+w}{ }\PY{o}{=}\PY{+w}{ }\PY{l+m+mi}{0}\PY{+w}{ }\PY{p}{\PYZcb{}}\PY{p}{;}\vphantom{fg}}}
\colorbox{white}{\parbox{\linewidth}{\PY{+w}{       }\PY{n}{QUIC\PYZus{}STATUS}\PY{+w}{ }\PY{n}{Status}\PY{+w}{ }\PY{o}{=}\PY{+w}{ }\PY{n}{QuicSlidingWindowExtremumGet}\PY{p}{(}\PY{o}{\PYZam{}}\PY{n}{Bbr}\PY{o}{\PYZhy{}}\PY{o}{\PYZgt{}}\PY{n}{MaxAckHeightFilter}\PY{p}{,}\PY{+w}{ }\PY{o}{\PYZam{}}\PY{n}{Entry}\PY{p}{)}\PY{p}{;}\vphantom{fg}}}
\colorbox{white}{\parbox{\linewidth}{\PY{err}{@}\PY{err}{@}\PY{+w}{ }\PY{l+m+mi}{\PYZhy{}804}\PY{p}{,}\PY{l+m+mi}{6}\PY{+w}{ }\PY{o}{+}\PY{l+m+mi}{827}\PY{p}{,}\PY{l+m+mi}{38}\PY{+w}{ }\PY{err}{@}\PY{err}{@}\vphantom{fg}}}
\colorbox{white}{\parbox{\linewidth}{\PY{+w}{       }\PY{n}{AckEvent}\PY{o}{\PYZhy{}}\PY{o}{\PYZgt{}}\PY{n}{AckedPackets}\PY{+w}{ }\PY{o}{=}\PY{o}{=}\PY{+w}{ }\PY{n+nb}{NULL}\PY{+w}{ }\PY{o}{?}\PY{+w}{ }\PY{n}{FALSE}\PY{+w}{ }\PY{o}{:}\PY{+w}{ }\PY{n}{AckEvent}\PY{o}{\PYZhy{}}\PY{o}{\PYZgt{}}\PY{n}{IsLargestAckedPacketAppLimited}\PY{p}{;}\vphantom{fg}}}
\colorbox{white}{\parbox{\linewidth}{\PY{+w}{ }\vphantom{fg}}}
\colorbox{white}{\parbox{\linewidth}{\PY{+w}{   }\PY{n}{BbrBandwidthFilterOnPacketAcked}\PY{p}{(}\PY{o}{\PYZam{}}\PY{n}{Bbr}\PY{o}{\PYZhy{}}\PY{o}{\PYZgt{}}\PY{n}{BandwidthFilter}\PY{p}{,}\PY{+w}{ }\PY{n}{AckEvent}\PY{p}{,}\PY{+w}{ }\PY{n}{Bbr}\PY{o}{\PYZhy{}}\PY{o}{\PYZgt{}}\PY{n}{RoundTripCounter}\PY{p}{)}\PY{p}{;}\vphantom{fg}}}
\colorbox{PaleGreen}{\parbox{\linewidth}{\vphantom{fg}}}
\colorbox{PaleGreen}{\parbox{\linewidth}{\PY{+w}{    }\PY{c+c1}{// Store the current bandwidth estimate}\vphantom{fg}}}
\colorbox{PaleGreen}{\parbox{\linewidth}{\PY{+w}{    }\PY{k+kt}{uint64\PYZus{}t}\PY{+w}{ }\PY{n}{CurrentBandwidthEstimate}\PY{+w}{ }\PY{o}{=}\PY{+w}{ }\PY{n}{BbrCongestionControlGetBandwidth}\PY{p}{(}\PY{n}{Cc}\PY{p}{)}\PY{p}{;}\vphantom{fg}}}
\colorbox{PaleGreen}{\parbox{\linewidth}{\vphantom{fg}}}
\colorbox{PaleGreen}{\parbox{\linewidth}{\PY{+w}{    }\PY{c+c1}{// Compute the derivative of the bandwidth estimate}\vphantom{fg}}}
\colorbox{PaleGreen}{\parbox{\linewidth}{\PY{+w}{    }\PY{k+kt}{int64\PYZus{}t}\PY{+w}{ }\PY{n}{BandwidthDerivative}\PY{+w}{ }\PY{o}{=}\PY{+w}{ }\PY{p}{(}\PY{k+kt}{int64\PYZus{}t}\PY{p}{)}\PY{n}{CurrentBandwidthEstimate}\PY{+w}{ }\PY{o}{\PYZhy{}}\PY{+w}{ }\PY{p}{(}\PY{k+kt}{int64\PYZus{}t}\PY{p}{)}\PY{n}{Bbr}\PY{o}{\PYZhy{}}\PY{o}{\PYZgt{}}\PY{n}{PreviousBandwidthEstimate}\PY{p}{;}\vphantom{fg}}}
\colorbox{PaleGreen}{\parbox{\linewidth}{\vphantom{fg}}}
\colorbox{PaleGreen}{\parbox{\linewidth}{\PY{+w}{    }\PY{c+c1}{// Update PreviousBandwidthEstimate for next calculation}\vphantom{fg}}}
\colorbox{PaleGreen}{\parbox{\linewidth}{\PY{+w}{    }\PY{n}{Bbr}\PY{o}{\PYZhy{}}\PY{o}{\PYZgt{}}\PY{n}{PreviousBandwidthEstimate}\PY{+w}{ }\PY{o}{=}\PY{+w}{ }\PY{n}{CurrentBandwidthEstimate}\PY{p}{;}\vphantom{fg}}}
\colorbox{PaleGreen}{\parbox{\linewidth}{\vphantom{fg}}}
\colorbox{PaleGreen}{\parbox{\linewidth}{\PY{+w}{    }\PY{c+c1}{// Adjust gains based on the bandwidth derivative}\vphantom{fg}}}
\colorbox{PaleGreen}{\parbox{\linewidth}{\PY{+w}{    }\PY{k}{if}\PY{+w}{ }\PY{p}{(}\PY{n}{BandwidthDerivative}\PY{+w}{ }\PY{o}{\PYZgt{}}\PY{+w}{ }\PY{c+cm}{/* Positive Threshold */}\PY{+w}{ }\PY{p}{(}\PY{k+kt}{int64\PYZus{}t}\PY{p}{)}\PY{p}{(}\PY{n}{CurrentBandwidthEstimate}\PY{+w}{ }\PY{o}{*}\PY{+w}{ }\PY{l+m+mf}{0.05}\PY{p}{)}\PY{p}{)}\PY{+w}{ }\PY{p}{\PYZob{}}\vphantom{fg}}}
\colorbox{PaleGreen}{\parbox{\linewidth}{\PY{+w}{        }\PY{c+c1}{// Bandwidth is increasing significantly}\vphantom{fg}}}
\colorbox{PaleGreen}{\parbox{\linewidth}{\PY{+w}{        }\PY{c+c1}{// Increase gains to utilize additional capacity}\vphantom{fg}}}
\colorbox{PaleGreen}{\parbox{\linewidth}{\PY{+w}{        }\PY{n}{Bbr}\PY{o}{\PYZhy{}}\PY{o}{\PYZgt{}}\PY{n}{PacingGain}\PY{+w}{ }\PY{o}{=}\PY{+w}{ }\PY{n}{GAIN\PYZus{}UNIT}\PY{+w}{ }\PY{o}{*}\PY{+w}{ }\PY{l+m+mi}{5}\PY{+w}{ }\PY{o}{/}\PY{+w}{ }\PY{l+m+mi}{4}\PY{p}{;}\PY{+w}{ }\PY{c+c1}{// 1.25}\vphantom{fg}}}
\colorbox{PaleGreen}{\parbox{\linewidth}{\PY{+w}{        }\PY{n}{Bbr}\PY{o}{\PYZhy{}}\PY{o}{\PYZgt{}}\PY{n}{CwndGain}\PY{+w}{ }\PY{o}{=}\PY{+w}{ }\PY{n}{GAIN\PYZus{}UNIT}\PY{+w}{ }\PY{o}{*}\PY{+w}{ }\PY{l+m+mi}{5}\PY{+w}{ }\PY{o}{/}\PY{+w}{ }\PY{l+m+mi}{4}\PY{p}{;}\PY{+w}{   }\PY{c+c1}{// 1.25}\vphantom{fg}}}
\colorbox{PaleGreen}{\parbox{\linewidth}{\PY{+w}{    }\PY{p}{\PYZcb{}}\PY{+w}{ }\PY{k}{else}\PY{+w}{ }\PY{k}{if}\PY{+w}{ }\PY{p}{(}\PY{n}{BandwidthDerivative}\PY{+w}{ }\PY{o}{\PYZlt{}}\PY{+w}{ }\PY{c+cm}{/* Negative Threshold */}\PY{+w}{ }\PY{o}{\PYZhy{}}\PY{p}{(}\PY{k+kt}{int64\PYZus{}t}\PY{p}{)}\PY{p}{(}\PY{n}{CurrentBandwidthEstimate}\PY{+w}{ }\PY{o}{*}\PY{+w}{ }\PY{l+m+mf}{0.05}\PY{p}{)}\PY{p}{)}\PY{+w}{ }\PY{p}{\PYZob{}}\vphantom{fg}}}
\colorbox{PaleGreen}{\parbox{\linewidth}{\PY{+w}{        }\PY{c+c1}{// Bandwidth is decreasing significantly}\vphantom{fg}}}
\colorbox{PaleGreen}{\parbox{\linewidth}{\PY{+w}{        }\PY{c+c1}{// Decrease gains to prevent congestion}\vphantom{fg}}}
\colorbox{PaleGreen}{\parbox{\linewidth}{\PY{+w}{        }\PY{n}{Bbr}\PY{o}{\PYZhy{}}\PY{o}{\PYZgt{}}\PY{n}{PacingGain}\PY{+w}{ }\PY{o}{=}\PY{+w}{ }\PY{n}{GAIN\PYZus{}UNIT}\PY{+w}{ }\PY{o}{*}\PY{+w}{ }\PY{l+m+mi}{3}\PY{+w}{ }\PY{o}{/}\PY{+w}{ }\PY{l+m+mi}{4}\PY{p}{;}\PY{+w}{ }\PY{c+c1}{// 0.75}\vphantom{fg}}}
\colorbox{PaleGreen}{\parbox{\linewidth}{\PY{+w}{        }\PY{n}{Bbr}\PY{o}{\PYZhy{}}\PY{o}{\PYZgt{}}\PY{n}{CwndGain}\PY{+w}{ }\PY{o}{=}\PY{+w}{ }\PY{n}{GAIN\PYZus{}UNIT}\PY{+w}{ }\PY{o}{*}\PY{+w}{ }\PY{l+m+mi}{3}\PY{+w}{ }\PY{o}{/}\PY{+w}{ }\PY{l+m+mi}{4}\PY{p}{;}\PY{+w}{   }\PY{c+c1}{// 0.75}\vphantom{fg}}}
\colorbox{PaleGreen}{\parbox{\linewidth}{\PY{+w}{    }\PY{p}{\PYZcb{}}\PY{+w}{ }\PY{k}{else}\PY{+w}{ }\PY{p}{\PYZob{}}\vphantom{fg}}}
\colorbox{PaleGreen}{\parbox{\linewidth}{\PY{+w}{        }\PY{c+c1}{// Bandwidth is relatively stable}\vphantom{fg}}}
\colorbox{PaleGreen}{\parbox{\linewidth}{\PY{+w}{        }\PY{c+c1}{// Use default gains based on the current state}\vphantom{fg}}}
\colorbox{PaleGreen}{\parbox{\linewidth}{\PY{+w}{        }\PY{k}{if}\PY{+w}{ }\PY{p}{(}\PY{n}{Bbr}\PY{o}{\PYZhy{}}\PY{o}{\PYZgt{}}\PY{n}{BbrState}\PY{+w}{ }\PY{o}{=}\PY{o}{=}\PY{+w}{ }\PY{n}{BBR\PYZus{}STATE\PYZus{}STARTUP}\PY{+w}{ }\PY{o}{|}\PY{o}{|}\PY{+w}{ }\PY{n}{Bbr}\PY{o}{\PYZhy{}}\PY{o}{\PYZgt{}}\PY{n}{BbrState}\PY{+w}{ }\PY{o}{=}\PY{o}{=}\PY{+w}{ }\PY{n}{BBR\PYZus{}STATE\PYZus{}DRAIN}\PY{p}{)}\PY{+w}{ }\PY{p}{\PYZob{}}\vphantom{fg}}}
\colorbox{PaleGreen}{\parbox{\linewidth}{\PY{+w}{            }\PY{n}{Bbr}\PY{o}{\PYZhy{}}\PY{o}{\PYZgt{}}\PY{n}{PacingGain}\PY{+w}{ }\PY{o}{=}\PY{+w}{ }\PY{n}{kHighGain}\PY{p}{;}\vphantom{fg}}}
\colorbox{PaleGreen}{\parbox{\linewidth}{\PY{+w}{            }\PY{n}{Bbr}\PY{o}{\PYZhy{}}\PY{o}{\PYZgt{}}\PY{n}{CwndGain}\PY{+w}{ }\PY{o}{=}\PY{+w}{ }\PY{n}{kHighGain}\PY{p}{;}\vphantom{fg}}}
\colorbox{PaleGreen}{\parbox{\linewidth}{\PY{+w}{        }\PY{p}{\PYZcb{}}\PY{+w}{ }\PY{k}{else}\PY{+w}{ }\PY{p}{\PYZob{}}\vphantom{fg}}}
\colorbox{PaleGreen}{\parbox{\linewidth}{\PY{+w}{            }\PY{n}{Bbr}\PY{o}{\PYZhy{}}\PY{o}{\PYZgt{}}\PY{n}{PacingGain}\PY{+w}{ }\PY{o}{=}\PY{+w}{ }\PY{n}{GAIN\PYZus{}UNIT}\PY{p}{;}\PY{+w}{      }\PY{c+c1}{// 1.0}\vphantom{fg}}}
\colorbox{PaleGreen}{\parbox{\linewidth}{\PY{+w}{            }\PY{n}{Bbr}\PY{o}{\PYZhy{}}\PY{o}{\PYZgt{}}\PY{n}{CwndGain}\PY{+w}{ }\PY{o}{=}\PY{+w}{ }\PY{n}{kCwndGain}\PY{p}{;}\PY{+w}{        }\PY{c+c1}{// 2.0}\vphantom{fg}}}
\colorbox{PaleGreen}{\parbox{\linewidth}{\PY{+w}{        }\PY{p}{\PYZcb{}}\vphantom{fg}}}
\colorbox{PaleGreen}{\parbox{\linewidth}{\PY{+w}{    }\PY{p}{\PYZcb{}}\vphantom{fg}}}
\colorbox{white}{\parbox{\linewidth}{\PY{+w}{ }\vphantom{fg}}}
\colorbox{white}{\parbox{\linewidth}{\PY{+w}{   }\PY{k}{if}\PY{+w}{ }\PY{p}{(}\PY{n}{BbrCongestionControlInRecovery}\PY{p}{(}\PY{n}{Cc}\PY{p}{)}\PY{p}{)}\PY{+w}{ }\PY{p}{\PYZob{}}\vphantom{fg}}}
\colorbox{white}{\parbox{\linewidth}{\PY{+w}{       }\PY{n}{CXPLAT\PYZus{}DBG\PYZus{}ASSERT}\PY{p}{(}\PY{n}{Bbr}\PY{o}{\PYZhy{}}\PY{o}{\PYZgt{}}\PY{n}{EndOfRecoveryValid}\PY{p}{)}\PY{p}{;}\vphantom{fg}}}
\colorbox{white}{\parbox{\linewidth}{\PY{err}{@}\PY{err}{@}\PY{+w}{ }\PY{l+m+mi}{\PYZhy{}947}\PY{p}{,}\PY{l+m+mi}{6}\PY{+w}{ }\PY{o}{+}\PY{l+m+mi}{1002}\PY{p}{,}\PY{l+m+mi}{34}\PY{+w}{ }\PY{err}{@}\PY{err}{@}\vphantom{fg}}}
\colorbox{white}{\parbox{\linewidth}{\PY{+w}{           }\PY{o}{:}\PY{+w}{ }\PY{n}{MinCongestionWindow}\PY{p}{;}\vphantom{fg}}}
\colorbox{white}{\parbox{\linewidth}{\PY{+w}{   }\PY{p}{\PYZcb{}}\vphantom{fg}}}
\colorbox{white}{\parbox{\linewidth}{\PY{+w}{ }\vphantom{fg}}}
\colorbox{PaleGreen}{\parbox{\linewidth}{\PY{+w}{    }\PY{c+c1}{// Update packets lost since last loss rate calculation}\vphantom{fg}}}
\colorbox{PaleGreen}{\parbox{\linewidth}{\PY{+w}{    }\PY{n}{Bbr}\PY{o}{\PYZhy{}}\PY{o}{\PYZgt{}}\PY{n}{PacketsLostSinceLastLossCalc}\PY{+w}{ }\PY{o}{=}\PY{+w}{ }\PY{n}{LossEvent}\PY{o}{\PYZhy{}}\PY{o}{\PYZgt{}}\PY{n}{NumRetransmittableBytes}\PY{p}{;}\vphantom{fg}}}
\colorbox{PaleGreen}{\parbox{\linewidth}{\vphantom{fg}}}
\colorbox{PaleGreen}{\parbox{\linewidth}{\PY{+w}{    }\PY{c+c1}{// Recalculate packet loss rate}\vphantom{fg}}}
\colorbox{PaleGreen}{\parbox{\linewidth}{\PY{+w}{    }\PY{k}{if}\PY{+w}{ }\PY{p}{(}\PY{n}{Bbr}\PY{o}{\PYZhy{}}\PY{o}{\PYZgt{}}\PY{n}{PacketsSentSinceLastLossCalc}\PY{+w}{ }\PY{o}{\PYZgt{}}\PY{+w}{ }\PY{l+m+mi}{0}\PY{p}{)}\PY{+w}{ }\PY{p}{\PYZob{}}\vphantom{fg}}}
\colorbox{PaleGreen}{\parbox{\linewidth}{\PY{+w}{        }\PY{n}{Bbr}\PY{o}{\PYZhy{}}\PY{o}{\PYZgt{}}\PY{n}{PacketLossRate}\PY{+w}{ }\PY{o}{=}\PY{+w}{ }\PY{p}{(}\PY{k+kt}{double}\PY{p}{)}\PY{n}{Bbr}\PY{o}{\PYZhy{}}\PY{o}{\PYZgt{}}\PY{n}{PacketsLostSinceLastLossCalc}\PY{+w}{ }\PY{o}{/}\PY{+w}{ }\PY{p}{(}\PY{k+kt}{double}\PY{p}{)}\PY{n}{Bbr}\PY{o}{\PYZhy{}}\PY{o}{\PYZgt{}}\PY{n}{PacketsSentSinceLastLossCalc}\PY{p}{;}\vphantom{fg}}}
\colorbox{PaleGreen}{\parbox{\linewidth}{\vphantom{fg}}}
\colorbox{PaleGreen}{\parbox{\linewidth}{\PY{+w}{        }\PY{c+c1}{// Reset counters after calculation}\vphantom{fg}}}
\colorbox{PaleGreen}{\parbox{\linewidth}{\PY{+w}{        }\PY{n}{Bbr}\PY{o}{\PYZhy{}}\PY{o}{\PYZgt{}}\PY{n}{PacketsSentSinceLastLossCalc}\PY{+w}{ }\PY{o}{=}\PY{+w}{ }\PY{l+m+mi}{0}\PY{p}{;}\vphantom{fg}}}
\colorbox{PaleGreen}{\parbox{\linewidth}{\PY{+w}{        }\PY{n}{Bbr}\PY{o}{\PYZhy{}}\PY{o}{\PYZgt{}}\PY{n}{PacketsLostSinceLastLossCalc}\PY{+w}{ }\PY{o}{=}\PY{+w}{ }\PY{l+m+mi}{0}\PY{p}{;}\vphantom{fg}}}
\colorbox{PaleGreen}{\parbox{\linewidth}{\vphantom{fg}}}
\colorbox{PaleGreen}{\parbox{\linewidth}{\PY{+w}{        }\PY{c+c1}{// Adjust congestion window based on packet loss rate}\vphantom{fg}}}
\colorbox{PaleGreen}{\parbox{\linewidth}{\PY{+w}{        }\PY{k}{const}\PY{+w}{ }\PY{k+kt}{double}\PY{+w}{ }\PY{n}{LossThresholdHigh}\PY{+w}{ }\PY{o}{=}\PY{+w}{ }\PY{l+m+mf}{0.1}\PY{p}{;}\PY{+w}{ }\PY{c+c1}{// High loss rate threshold (10\PYZpc{})}\vphantom{fg}}}
\colorbox{PaleGreen}{\parbox{\linewidth}{\PY{+w}{        }\PY{k}{const}\PY{+w}{ }\PY{k+kt}{double}\PY{+w}{ }\PY{n}{LossThresholdLow}\PY{+w}{ }\PY{o}{=}\PY{+w}{ }\PY{l+m+mf}{0.05}\PY{p}{;}\PY{+w}{ }\PY{c+c1}{// Low loss rate threshold (5\PYZpc{})}\vphantom{fg}}}
\colorbox{PaleGreen}{\parbox{\linewidth}{\PY{+w}{        }\PY{k}{if}\PY{+w}{ }\PY{p}{(}\PY{n}{Bbr}\PY{o}{\PYZhy{}}\PY{o}{\PYZgt{}}\PY{n}{PacketLossRate}\PY{+w}{ }\PY{o}{\PYZgt{}}\PY{+w}{ }\PY{n}{LossThresholdHigh}\PY{p}{)}\PY{+w}{ }\PY{p}{\PYZob{}}\vphantom{fg}}}
\colorbox{PaleGreen}{\parbox{\linewidth}{\PY{+w}{            }\PY{c+c1}{// High loss detected, reduce congestion window more aggressively}\vphantom{fg}}}
\colorbox{PaleGreen}{\parbox{\linewidth}{\PY{+w}{            }\PY{n}{Bbr}\PY{o}{\PYZhy{}}\PY{o}{\PYZgt{}}\PY{n}{CongestionWindow}\PY{+w}{ }\PY{o}{=}\PY{+w}{ }\PY{p}{(}\PY{k+kt}{uint32\PYZus{}t}\PY{p}{)}\PY{p}{(}\PY{p}{(}\PY{k+kt}{double}\PY{p}{)}\PY{n}{Bbr}\PY{o}{\PYZhy{}}\PY{o}{\PYZgt{}}\PY{n}{CongestionWindow}\PY{+w}{ }\PY{o}{*}\PY{+w}{ }\PY{l+m+mf}{0.7}\PY{p}{)}\PY{p}{;}\vphantom{fg}}}
\colorbox{PaleGreen}{\parbox{\linewidth}{\PY{+w}{            }\PY{n}{Bbr}\PY{o}{\PYZhy{}}\PY{o}{\PYZgt{}}\PY{n}{CongestionWindow}\PY{+w}{ }\PY{o}{=}\PY{+w}{ }\PY{n}{CXPLAT\PYZus{}MAX}\PY{p}{(}\PY{n}{Bbr}\PY{o}{\PYZhy{}}\PY{o}{\PYZgt{}}\PY{n}{CongestionWindow}\PY{p}{,}\PY{+w}{ }\PY{n}{MinCongestionWindow}\PY{p}{)}\PY{p}{;}\vphantom{fg}}}
\colorbox{PaleGreen}{\parbox{\linewidth}{\PY{+w}{        }\PY{p}{\PYZcb{}}\PY{+w}{ }\PY{k}{else}\PY{+w}{ }\PY{k}{if}\PY{+w}{ }\PY{p}{(}\PY{n}{Bbr}\PY{o}{\PYZhy{}}\PY{o}{\PYZgt{}}\PY{n}{PacketLossRate}\PY{+w}{ }\PY{o}{\PYZgt{}}\PY{+w}{ }\PY{n}{LossThresholdLow}\PY{p}{)}\PY{+w}{ }\PY{p}{\PYZob{}}\vphantom{fg}}}
\colorbox{PaleGreen}{\parbox{\linewidth}{\PY{+w}{            }\PY{c+c1}{// Moderate loss detected, reduce congestion window}\vphantom{fg}}}
\colorbox{PaleGreen}{\parbox{\linewidth}{\PY{+w}{            }\PY{n}{Bbr}\PY{o}{\PYZhy{}}\PY{o}{\PYZgt{}}\PY{n}{CongestionWindow}\PY{+w}{ }\PY{o}{=}\PY{+w}{ }\PY{p}{(}\PY{k+kt}{uint32\PYZus{}t}\PY{p}{)}\PY{p}{(}\PY{p}{(}\PY{k+kt}{double}\PY{p}{)}\PY{n}{Bbr}\PY{o}{\PYZhy{}}\PY{o}{\PYZgt{}}\PY{n}{CongestionWindow}\PY{+w}{ }\PY{o}{*}\PY{+w}{ }\PY{l+m+mf}{0.85}\PY{p}{)}\PY{p}{;}\vphantom{fg}}}
\colorbox{PaleGreen}{\parbox{\linewidth}{\PY{+w}{            }\PY{n}{Bbr}\PY{o}{\PYZhy{}}\PY{o}{\PYZgt{}}\PY{n}{CongestionWindow}\PY{+w}{ }\PY{o}{=}\PY{+w}{ }\PY{n}{CXPLAT\PYZus{}MAX}\PY{p}{(}\PY{n}{Bbr}\PY{o}{\PYZhy{}}\PY{o}{\PYZgt{}}\PY{n}{CongestionWindow}\PY{p}{,}\PY{+w}{ }\PY{n}{MinCongestionWindow}\PY{p}{)}\PY{p}{;}\vphantom{fg}}}
\colorbox{PaleGreen}{\parbox{\linewidth}{\PY{+w}{        }\PY{p}{\PYZcb{}}\PY{+w}{ }\PY{k}{else}\PY{+w}{ }\PY{p}{\PYZob{}}\vphantom{fg}}}
\colorbox{PaleGreen}{\parbox{\linewidth}{\PY{+w}{            }\PY{c+c1}{// Low loss detected, gently increase congestion window}\vphantom{fg}}}
\colorbox{PaleGreen}{\parbox{\linewidth}{\PY{+w}{            }\PY{n}{Bbr}\PY{o}{\PYZhy{}}\PY{o}{\PYZgt{}}\PY{n}{CongestionWindow}\PY{+w}{ }\PY{o}{=}\PY{+w}{ }\PY{n}{DatagramPayloadLength}\PY{p}{;}\vphantom{fg}}}
\colorbox{PaleGreen}{\parbox{\linewidth}{\PY{+w}{        }\PY{p}{\PYZcb{}}\vphantom{fg}}}
\colorbox{PaleGreen}{\parbox{\linewidth}{\PY{+w}{    }\PY{p}{\PYZcb{}}\vphantom{fg}}}
\colorbox{PaleGreen}{\parbox{\linewidth}{\vphantom{fg}}}
\colorbox{white}{\parbox{\linewidth}{\PY{+w}{   }\PY{n}{BbrCongestionControlUpdateBlockedState}\PY{p}{(}\PY{n}{Cc}\PY{p}{,}\PY{+w}{ }\PY{n}{PreviousCanSendState}\PY{p}{)}\PY{p}{;}\vphantom{fg}}}
\colorbox{white}{\parbox{\linewidth}{\PY{+w}{   }\PY{n}{QuicConnLogBbr}\PY{p}{(}\PY{n}{QuicCongestionControlGetConnection}\PY{p}{(}\PY{n}{Cc}\PY{p}{)}\PY{p}{)}\PY{p}{;}\vphantom{fg}}}
\colorbox{white}{\parbox{\linewidth}{\PY{+w}{ }\PY{p}{\PYZcb{}}\vphantom{fg}}}
\colorbox{white}{\parbox{\linewidth}{\PY{err}{@}\PY{err}{@}\PY{+w}{ }\PY{l+m+mi}{\PYZhy{}1044}\PY{p}{,}\PY{l+m+mi}{6}\PY{+w}{ }\PY{o}{+}\PY{l+m+mi}{1127}\PY{p}{,}\PY{l+m+mi}{14}\PY{+w}{ }\PY{err}{@}\PY{err}{@}\vphantom{fg}}}
\colorbox{white}{\parbox{\linewidth}{\PY{+w}{   }\PY{n}{QuicSlidingWindowExtremumReset}\PY{p}{(}\PY{o}{\PYZam{}}\PY{n}{Bbr}\PY{o}{\PYZhy{}}\PY{o}{\PYZgt{}}\PY{n}{BandwidthFilter}\PY{p}{.}\PY{n}{WindowedMaxFilter}\PY{p}{)}\PY{p}{;}\vphantom{fg}}}
\colorbox{white}{\parbox{\linewidth}{\PY{+w}{   }\PY{n}{Bbr}\PY{o}{\PYZhy{}}\PY{o}{\PYZgt{}}\PY{n}{BandwidthFilter}\PY{p}{.}\PY{n}{AppLimited}\PY{+w}{ }\PY{o}{=}\PY{+w}{ }\PY{n}{FALSE}\PY{p}{;}\vphantom{fg}}}
\colorbox{white}{\parbox{\linewidth}{\PY{+w}{   }\PY{n}{Bbr}\PY{o}{\PYZhy{}}\PY{o}{\PYZgt{}}\PY{n}{BandwidthFilter}\PY{p}{.}\PY{n}{AppLimitedExitTarget}\PY{+w}{ }\PY{o}{=}\PY{+w}{ }\PY{l+m+mi}{0}\PY{p}{;}\vphantom{fg}}}
\colorbox{PaleGreen}{\parbox{\linewidth}{\vphantom{fg}}}
\colorbox{PaleGreen}{\parbox{\linewidth}{\PY{+w}{    }\PY{c+c1}{//}\vphantom{fg}}}
\colorbox{PaleGreen}{\parbox{\linewidth}{\PY{+w}{    }\PY{c+c1}{// Reset adaptive gain factors}\vphantom{fg}}}
\colorbox{PaleGreen}{\parbox{\linewidth}{\PY{+w}{    }\PY{c+c1}{//}\vphantom{fg}}}
\colorbox{PaleGreen}{\parbox{\linewidth}{\PY{+w}{    }\PY{n}{Bbr}\PY{o}{\PYZhy{}}\PY{o}{\PYZgt{}}\PY{n}{AdaptivePacingGain}\PY{+w}{ }\PY{o}{=}\PY{+w}{ }\PY{n}{Bbr}\PY{o}{\PYZhy{}}\PY{o}{\PYZgt{}}\PY{n}{PacingGain}\PY{p}{;}\vphantom{fg}}}
\colorbox{PaleGreen}{\parbox{\linewidth}{\PY{+w}{    }\PY{n}{Bbr}\PY{o}{\PYZhy{}}\PY{o}{\PYZgt{}}\PY{n}{AdaptiveCwndGain}\PY{+w}{ }\PY{o}{=}\PY{+w}{ }\PY{n}{Bbr}\PY{o}{\PYZhy{}}\PY{o}{\PYZgt{}}\PY{n}{CwndGain}\PY{p}{;}\vphantom{fg}}}
\colorbox{PaleGreen}{\parbox{\linewidth}{\PY{+w}{    }\PY{n}{Bbr}\PY{o}{\PYZhy{}}\PY{o}{\PYZgt{}}\PY{n}{PreviousDeliveryRate}\PY{+w}{ }\PY{o}{=}\PY{+w}{ }\PY{l+m+mi}{0}\PY{p}{;}\vphantom{fg}}}
\colorbox{PaleGreen}{\parbox{\linewidth}{\PY{+w}{    }\PY{n}{Bbr}\PY{o}{\PYZhy{}}\PY{o}{\PYZgt{}}\PY{n}{PreviousRtt}\PY{+w}{ }\PY{o}{=}\PY{+w}{ }\PY{n}{UINT64\PYZus{}MAX}\PY{p}{;}\vphantom{fg}}}
\colorbox{white}{\parbox{\linewidth}{\PY{+w}{ }\vphantom{fg}}}
\colorbox{white}{\parbox{\linewidth}{\PY{+w}{   }\PY{n}{BbrCongestionControlLogOutFlowStatus}\PY{p}{(}\PY{n}{Cc}\PY{p}{)}\PY{p}{;}\vphantom{fg}}}
\colorbox{white}{\parbox{\linewidth}{\PY{+w}{   }\PY{n}{QuicConnLogBbr}\PY{p}{(}\PY{n}{Connection}\PY{p}{)}\PY{p}{;}\vphantom{fg}}}
\colorbox{white}{\parbox{\linewidth}{\PY{err}{@}\PY{err}{@}\PY{+w}{ }\PY{l+m+mi}{\PYZhy{}1104}\PY{p}{,}\PY{l+m+mi}{7}\PY{+w}{ }\PY{o}{+}\PY{l+m+mi}{1195}\PY{p}{,}\PY{l+m+mi}{6}\PY{+w}{ }\PY{err}{@}\PY{err}{@}\vphantom{fg}}}
\colorbox{white}{\parbox{\linewidth}{\PY{+w}{   }\PY{n}{Bbr}\PY{o}{\PYZhy{}}\PY{o}{\PYZgt{}}\PY{n}{BtlbwFound}\PY{+w}{ }\PY{o}{=}\PY{+w}{ }\PY{n}{FALSE}\PY{p}{;}\vphantom{fg}}}
\colorbox{white}{\parbox{\linewidth}{\PY{+w}{   }\PY{n}{Bbr}\PY{o}{\PYZhy{}}\PY{o}{\PYZgt{}}\PY{n}{SendQuantum}\PY{+w}{ }\PY{o}{=}\PY{+w}{ }\PY{l+m+mi}{0}\PY{p}{;}\vphantom{fg}}}
\colorbox{white}{\parbox{\linewidth}{\PY{+w}{   }\PY{n}{Bbr}\PY{o}{\PYZhy{}}\PY{o}{\PYZgt{}}\PY{n}{SlowStartupRoundCounter}\PY{+w}{ }\PY{o}{=}\PY{+w}{ }\PY{l+m+mi}{0}\PY{+w}{ }\PY{p}{;}\vphantom{fg}}}
\colorbox{LightPink}{\parbox{\linewidth}{\vphantom{fg}}}
\colorbox{white}{\parbox{\linewidth}{\PY{+w}{   }\PY{n}{Bbr}\PY{o}{\PYZhy{}}\PY{o}{\PYZgt{}}\PY{n}{PacingCycleIndex}\PY{+w}{ }\PY{o}{=}\PY{+w}{ }\PY{l+m+mi}{0}\PY{p}{;}\vphantom{fg}}}
\colorbox{white}{\parbox{\linewidth}{\PY{+w}{   }\PY{n}{Bbr}\PY{o}{\PYZhy{}}\PY{o}{\PYZgt{}}\PY{n}{AggregatedAckBytes}\PY{+w}{ }\PY{o}{=}\PY{+w}{ }\PY{l+m+mi}{0}\PY{p}{;}\vphantom{fg}}}
\colorbox{white}{\parbox{\linewidth}{\PY{+w}{   }\PY{n}{Bbr}\PY{o}{\PYZhy{}}\PY{o}{\PYZgt{}}\PY{n}{ExitingQuiescence}\PY{+w}{ }\PY{o}{=}\PY{+w}{ }\PY{n}{FALSE}\PY{p}{;}\vphantom{fg}}}
\colorbox{white}{\parbox{\linewidth}{\PY{err}{@}\PY{err}{@}\PY{+w}{ }\PY{l+m+mi}{\PYZhy{}1141}\PY{p}{,}\PY{l+m+mi}{6}\PY{+w}{ }\PY{o}{+}\PY{l+m+mi}{1231}\PY{p}{,}\PY{l+m+mi}{14}\PY{+w}{ }\PY{err}{@}\PY{err}{@}\vphantom{fg}}}
\colorbox{white}{\parbox{\linewidth}{\PY{+w}{       }\PY{p}{.}\PY{n}{AppLimitedExitTarget}\PY{+w}{ }\PY{o}{=}\PY{+w}{ }\PY{l+m+mi}{0}\PY{p}{,}\vphantom{fg}}}
\colorbox{white}{\parbox{\linewidth}{\PY{+w}{   }\PY{p}{\PYZcb{}}\PY{p}{;}\vphantom{fg}}}
\colorbox{white}{\parbox{\linewidth}{\PY{+w}{ }\vphantom{fg}}}
\colorbox{PaleGreen}{\parbox{\linewidth}{\PY{+w}{    }\PY{c+c1}{//}\vphantom{fg}}}
\colorbox{PaleGreen}{\parbox{\linewidth}{\PY{+w}{    }\PY{c+c1}{// Initialize adaptive gain factors}\vphantom{fg}}}
\colorbox{PaleGreen}{\parbox{\linewidth}{\PY{+w}{    }\PY{c+c1}{//}\vphantom{fg}}}
\colorbox{PaleGreen}{\parbox{\linewidth}{\PY{+w}{    }\PY{n}{Bbr}\PY{o}{\PYZhy{}}\PY{o}{\PYZgt{}}\PY{n}{AdaptivePacingGain}\PY{+w}{ }\PY{o}{=}\PY{+w}{ }\PY{n}{Bbr}\PY{o}{\PYZhy{}}\PY{o}{\PYZgt{}}\PY{n}{PacingGain}\PY{p}{;}\vphantom{fg}}}
\colorbox{PaleGreen}{\parbox{\linewidth}{\PY{+w}{    }\PY{n}{Bbr}\PY{o}{\PYZhy{}}\PY{o}{\PYZgt{}}\PY{n}{AdaptiveCwndGain}\PY{+w}{ }\PY{o}{=}\PY{+w}{ }\PY{n}{Bbr}\PY{o}{\PYZhy{}}\PY{o}{\PYZgt{}}\PY{n}{CwndGain}\PY{p}{;}\vphantom{fg}}}
\colorbox{PaleGreen}{\parbox{\linewidth}{\PY{+w}{    }\PY{n}{Bbr}\PY{o}{\PYZhy{}}\PY{o}{\PYZgt{}}\PY{n}{PreviousDeliveryRate}\PY{+w}{ }\PY{o}{=}\PY{+w}{ }\PY{l+m+mi}{0}\PY{p}{;}\vphantom{fg}}}
\colorbox{PaleGreen}{\parbox{\linewidth}{\PY{+w}{    }\PY{n}{Bbr}\PY{o}{\PYZhy{}}\PY{o}{\PYZgt{}}\PY{n}{PreviousRtt}\PY{+w}{ }\PY{o}{=}\PY{+w}{ }\PY{n}{UINT64\PYZus{}MAX}\PY{p}{;}\vphantom{fg}}}
\colorbox{PaleGreen}{\parbox{\linewidth}{\vphantom{fg}}}
\colorbox{white}{\parbox{\linewidth}{\PY{+w}{   }\PY{n}{QuicConnLogOutFlowStats}\PY{p}{(}\PY{n}{Connection}\PY{p}{)}\PY{p}{;}\vphantom{fg}}}
\colorbox{white}{\parbox{\linewidth}{\PY{+w}{   }\PY{n}{QuicConnLogBbr}\PY{p}{(}\PY{n}{Connection}\PY{p}{)}\PY{p}{;}\vphantom{fg}}}
\colorbox{white}{\parbox{\linewidth}{\PY{+w}{ }\PY{p}{\PYZcb{}}\vphantom{fg}}}
\end{Verbatim}

\subsection{The Full Prompt}
\label{sec:appendix:full-prompt}

\begin{tcolorbox}[breakable,title=Prompt outline:]
\begin{minted}[fontsize=\small,breaklines,breaksymbolleft=]{markdown}
Here's the implementation of the BBR congestion control algorithm in QUIC from an open source project:

```
<Full Source Code: bbr.c and bbr.h from MsQuic>
```

We aim to improve the performance of the above algorithm: increasing throughput, reducing latency, and minimizing the loss rate.

Please design an innovative mechanism and implement it in code to achieve the above goals.

Your code implementation should follow the "update block" format below:

UPDATE FUNCTION `<Function Name 1>`:

```
Rewrite your updated function here, ensuring that the entire function is displayed from start to end. Do not output only a part of the function or use ellipses.  
```

UPDATE FUNCTION `<Function Name 2>`:

```
Same as above.  
```

You can also use the following blocks to update a global variable's value (the variable should be in the global space, not in the funciton space):

UPDATE VARIABLE `<Variable Name>`:
```
Rewrite the variable definition.
```

Use the following syntax to insert a member or members to an existing struct:

ADD MEMBER TO `<Struct Name>`:
```
Add memebers to the struct.
Directly output memeber definitions such as:

    uint64_t added_member_A;
    uint64_t added_member_B;

Do not output typedef or use ellipses.
```

Your output can consist of a single update block or multiple such update blocks.

Please make sure your updated algorithm can be compiled successfully.

Do not introduce any new functions or new structs. Implement all new mechanism in the existing functions and structs. 

Unveil a groundbreaking and transformative mechanism—one that transcends the boundaries of current human understanding and ignites new realms of possibility. Don't settle for presenting ideas already acknowledged and explored; instead, introduce something entirely original that challenges conventions and inspires innovation on a global scale.

First, present your thoughts, brainstorm three ideas, and select the best one. Then, provide the updated code modifications in the format described above.
\end{minted}
\end{tcolorbox}

\end{document}